\documentstyle[prl,aps,epsf,multicol]{revtex}
\begin{document}
\draft
%\twocolumn[\hsize\textwidth\columnwidth\hsize\csname@twocolumnfalse%
%\endcsname
 
\title{$Z_2$ Gauge Theory of Electron Fractionalization in Strongly Correlated Systems }
\author{T. Senthil and Matthew P. A. Fisher}
\address{ 
Institute for Theoretical Physics, University of California,
Santa Barbara, CA 93106--4030
}

\date{\today}
\maketitle

\def\i{\imath\,}
\def\ih{\frac{\imath}{2}\,}
\def\undertext#1{\vtop{\hbox{#1}\kern 1pt \hrule}}
\def\ra{\rightarrow}
\def\lfa{\leftarrow}
\def\ua{\uparrow}
\def\da{\downarrow}
\def\Ra{\Rightarrow}
\def\lra{\longrightarrow}
\def\ler{\leftrightarrow}
\def\lrb#1{\left(#1\right)}
\def\O#1{O\left(#1\right)}
\def\VEV#1{\left\langle\,#1\,\right\rangle}
\def\tr{\hbox{tr}\,}
\def\trb#1{\tr\lrb{#1}}
\def\dd#1{\frac{d}{d#1}}
\def\dbyd#1#2{\frac{d#1}{d#2}}
\def\pp#1{\frac{\partial}{\partial#1}}
\def\pbyp#1#2{\frac{\partial#1}{\partial#2}}
\def\ff#1{\frac{\delta}{\delta#1}}
\def\fbyf#1#2{\frac{\delta#1}{\delta#2}}
\def\pd#1{\partial_{#1}}
\def\br{\\ \nonumber & &}
\def\brr{\right. \\ \nonumber & &\left.}
\def\inv#1{\frac{1}{#1}}
\def\be{\begin{equation}}
\def\ee{\end{equation}}
\def\bea{\begin{eqnarray}}
\def\eea{\end{eqnarray}}
\def\ct#1{\cite{#1}}
\def\rf#1{(\ref{#1})}
\def\EXP#1{\exp\left(#1\right)}
\def\TEXP#1{\hat{T}\exp\left(#1\right)}
\def\INT#1#2{\int_{#1}^{#2}}
\def\MAT{{\it Mathematica }}
\def\LHS{left-hand side }
\def\RHS{right-hand side }
\def\COM#1#2{\left\lbrack #1\,,\,#2\right\rbrack}
\def\AC#1#2{\left\lbrace #1\,,\,#2\right\rbrace}

\begin{abstract}
We develop a new theoretical framework for describing and
analyzing exotic phases of strongly correlated electrons
which support excitations with fractional quantum numbers.
Starting with a class of microscopic models
believed to capture much of the essential
physics of the cuprate superconductors, we derive
a new gauge theory  - based upon a {\it discrete}
Ising or $Z_2$ symmetry -  which interpolates
naturally between an antiferromagnetic Mott insulator
and a conventional $d-$wave superconductor.  We explore the
intervening regime, and demonstrate the possible
existence of an exotic fractionalized insulator - the nodal liquid - 
as well as various more conventional insulating phases
exhibiting broken lattice symmetries. A crucial role is played by
vortex configurations in the $Z_2$ gauge field. 
Fractionalization is obtained if they are uncondensed. 
Within the insulating phases, the dynamics of these 
$Z_2$ vortices in two dimensions ($2d$) is described, after a duality transformation,
by an Ising model
in a transverse field - the Ising spins representing the $Z_2$ vortices. 
The presence of an
unusual Berry's phase term in the gauge theory,
leads to a doping-dependent
``frustration" in the dual Ising model, being
fully frustrated at half-filling. The $Z_2$ gauge theory is readily
generalized to a variety of different situations - in particular, it
can also describe $3d$ insulators
with fractional quantum numbers. We point out that the mechanism of fractionalization for $d>1$ is
distinct from the well-known $1d$ spin-charge separation. 
Other interesting results include a description of an 
exotic fractionalized superconductor in two or higher dimensions.

\end{abstract}
\vspace{0.15cm}

%\pacs{PACS numbers:75.10.Nr, 05.50.+q, 75.10.Jm}
%\vskip -0.5 truein 

\begin{multicols}{2}
\narrowtext 

\section{Introduction}
\label{intro}
Strongly interacting many-electron systems in low dimensions
can exhibit exotic properties, most notably
the presence of excitations with fractional quantum numbers.
In these instances the electron is ``fractionalized" - 
effectively splintered into consituents which essentially behave as
free particles.
The classic example is the 
one-dimensional (1d) interacting electron gas\cite{1deg},
which exhibits many anomalous properties such as
the separation of the spin and the charge of the electron.
Electron ``fractionalization" is also predicted
to occur in 2d systems
in very strong magnetic fields
that exhibit the 
fractional quantum Hall effect\cite{QHEbook}.
Recent experiments have given strong supporting evidence of fractionalization
both in quantum Hall systems\cite{QHEexp} and in carbon nanotubes\cite{nano}.
Motivated by these examples, 
several authors have proposed the possibility of electron fractionalization in various other
experimental systems.  Perhaps the most tantalizing is the
suggestion by P.W. Anderson\cite{PWA} of
``spin-charge separation"
in cuprate high-$T_c$
materials. However, this suggestion is currently surrounded
with considerable controversy,
in part because the 1d electron gas and the 
fractional quantum Hall effect appear to be rather
special situations  
which do not readily generalize.
Indeed, in 1d the Fermi liquid breaks down even at weak coupling
and in the quantum Hall regime the kinetic energy is strongly
quenched by a time reversal breaking magnetic field.

In this paper, we will explore theoretically the possibility of electron fractionalization in
strongly correlated systems in spatial dimensions $d > 1$ in the presence of time reversal symmetry.
Our primary motivation is the cuprates, though we expect our results to be of 
significance to a variety of other strongly interacting systems.   
Early attempts\cite{BZA,BA,oldgauge} to implement theoretically
Anderson's suggestion of 2d spin-charge separation
typically started with either a quantum spin model or 
the t-J model.
Slave boson/fermion representations of the spin and electron operators
were employed to obtain a mean field ``saddle-point"
exhibiting spin-charge separation.
The slave
boson/fermion representation 
introduces a gauge symmetry - $U(1)$ in the simplest formulations - and
requires inclusion of
a corresponding gauge field.
Fluctuations about the mean field theory lead to a strongly interacting gauge theory
about which very little is reliably known.   
It is then quite difficult to reach any definitive conclusions about the true low energy 
behaviour - in particular whether spin-charge separation survives beyond the mean field level. 
An alternate more recent approach\cite{NLII,NLIII}, describes strongly correlated electron systems in 2d
in a dual language where the vortices in the many-electron wavefunction are the fundamental 
degrees of freedom. In this approach, insulating phases
can be obtained by condensing vortices.
Fractionalized insulators arise upon condensing {\it pairs} of vortices.

In this work we introduce a new gauge theory approach which enables us to reliably 
address issues of fractionalization. In contrast to the slave boson/fermion representation, 
our gauge symmetry is {\em discrete} -in fact, an Ising 
or $Z_2$ gauge symmetry.  This has several advantages.  Firstly,
gauge theories with discrete
symmetry are much simpler to analyze than
those with continuous symmetries\cite{Kogut}, so that it is possible for us to 
make definitive statements about low energy physics.  But in addition,
the pure $Z_2$ gauge theory in $2+1$ space-time dimensions
is dual to the 3d classical Ising model, which implies the
existence of {\it two} distinct quantum phases\cite{Wegner}.
In one of these two phases ``charges" are {\it deconfined},
in marked contrast to the pure $2+1$ dimensional $U(1)$ gauge theory which
is always in a confining phase\cite{Polyakov}.   
The presence of deconfinement
allows us to 
demonstrate the existence of insulating phases
exhibiting electron fractionalization, and to describe their
basic properties.  Remarkably, fractionalization in our $Z_2$
gauge theory approach is physically equivalent to
vortex pairing in the earlier dual formulation\cite{NLII,NLIII}.
We demonstrate this equivalence by combining
the standard boson-vortex duality\cite{xydual} with the Ising
duality mentioned above.  

In addition to the fractionalized phases, our approach allows us to
readily access the more conventional confined phases,
and the concomitant confinement transitions.
Furthermore, the $Z_2$ gauge theory 
can be readily generalized to describe a variety of different
situations - arbitrary spatial dimensions,
spin-rotation non-invariant systems,  etc. Some of these generalizations are explored 
towards the end of the paper. For the most part, we concentrate on 
spin-rotation invariant
electronic systems in 2d. An overview and summary of our main results may be found
at the end of this introductory section. 

In the context of {\it frustrated} quantum spin models,  Read and Sachdev\cite{RSSpN }
have demonstrated the 
possibility of disordered
phases with fractionalization of spin.
Specifically, an 
$Sp(2N)$ antiferromagnet at large $N$ and the related quantum dimer model\cite{FTFIM2,subir_pc} 
were shown to
reduce to a $Z_2$ gauge theory when frustration
was present.
In the deconfined phase of the gauge theory
free propagating spinons (spin $1/2$ excitations) would be possible. Somewhat similarly, 
in the slave-fermion representation of the conventional
Heisenberg magnet which introduces an $SU(2)$ gauge invariance,
X.G. Wen\cite{Wen} proposed obtaining fractionalization of spin by pairing and 
condensing pairs of spinons. This reduces the gauge symmetry down to $Z_2$. 
In contrast, we show explicitly that the conventional Heisenberg
spin model can be {\it directly} written as a $Z_2$ gauge theory
coupled to fermionic spinons,
even in the absence of any frustration.  The key observation
is that with fermionic spinons, the local constraint of single occupancy
is equivalent to the constraint of an {\it odd} number
of fermions per site.  This latter constraint can be
implemented with a discrete $Z_2$ gauge field. 
Such a $Z_2$ gauge description may 
also be obtained with the Majorona fermion representation of Heisenberg spins\cite{Tsvelik}. 

The basic physics underlying our description
of electron fractionalization is perhaps most
readily understood in $d=2$.  At the heart of quantum mechanics
is wave-particle dualism.  For a many-body system of interacting
bosons (with charge $Q_e$, say) this dualism implies that in addition to the
conventional ``particle" framework, a description
developed in terms of wavefunctions is possible.
In $2d$ this dual wave description focusses
on point like singularities in the phase of the
complex wavefunction - the familiar vortices with circulation quantized
in units of $Q_v$.
A fundamental property of such vortices is that the product of their
quantum of circulation and the particle charge
is a constant,
\be
\label{QeQv}
Q_e Q_v = hc    .
\ee
It is this simple identity which underlies the two known
examples of ``fractionalization" in two-dimensions,
and is at the heart of the $Z_2$ gauge theory developed
in this paper.
In a BCS superconductor, the pairing of electrons to form
a Cooper pair with charge $Q_e =2e$, implies
a ``halving" of the flux quantum, $Q_v = {1 \over 2} (hc/e)$ - which
is tantamount to ``vortex fractionalization".
The second example of 2d fractionalization 
occurs in the fractional quantum Hall effect\cite{QHEbook}.  In the $\nu=1/3$
state three vortices bind to each electron forming
a ``composite boson" with total circulation $Q_v = 3 (hc/e)$,
which then condenses.  The above identity implies
the existence of 
topological excitations in this condensate with electrical
charge ${1 \over 3} e$ - the celebrated
Laughlin quasiparticles. 

The route to electron fractionalization that we explore
in this paper is {\it physically} equivalent to a {\it pairing of vortices},
precisely as in earlier work by Balents et. al.\cite{NLII,NLIII}.
But the mathematical implementation is rather different.
Balents et. al. argued that a pairing and condensation
of conventional $Q_v = hc/2e$ BCS vortices in
a singlet superconductor,
results in an exotic fractionalized insulator.
As Eqn.~\ref{QeQv} demonstrates,
this insulator should support spinless
charge $e$ excitations.  Our analysis begins by noting that
such an excitation can be thought
of as ``one-half" of a Cooper pair.
We implement this fractionalization by formally
re-expressing the Cooper pair creation operator
as the {\it product} of two ``chargon" operators, $b^\dagger$,
each creating a spinless, charge $e$ boson.  This change
of variables introduces a {\it local} $Z_2$ symmetry,
since it is possible to change the sign of $b^\dagger$
on any given lattice site while leaving the Cooper pair
operator invariant.  This is the origin of a
local Ising, or $Z_2$, gauge symmetry - described
mathematically in terms of a $Z_2$ gauge field.
In the exotic fractionalized insulator,
there are strange gapped excitations which are
vortices in the $Z_2$ gauge field.  These excitations - which
we refer to as ``visons" because they can be
represented in terms of Ising spins -  are the remnant of the
{\it unpaired} $hc/2e$ BCS vortices, which survive
in the fractionalized insulator.  
As we shall see, when the visons condense they drive ``confinement",
thereby destroying fractionalization.  These visons
will play an absolutely central role throughout this paper,
since any insulator with gapped visons is {\it necessarily}
fractionalized.

Motivated by the cuprate
superconductors, we will focus on a particular class of microscopic lattice models
designed to capture much of the physics believed essential to these
materials. (Our description
of fractionalization is, however, more general and is not restricted to these models.)
The models describe electrons hopping on a lattice with 
inclusion of strong spin {\it and} pairing
fluctuations, and are quite similar to models introduced and 
analyzed numerically by Assaad et. al.\cite{Scalapino}
and to models considered more recently by Balents et. al.\cite{NLII}.
Many microscopic models of the cuprates, such as the $t-J$ model,
incorporate spin fluctuations from the outset.
Our reasons for similarly incorporating ``microscopic" pairing fluctuations
are two fold.  Firstly,
as the superconducting phase is a well-established and 
reasonably well-understood part of the high-$T_c$ phase diagram - just like the antiferromagnet - it
serves as a useful point of departure to access
more puzzling regions of the phase diagram.
This point of view was also advocated in Ref. \cite{NLI}. But there are
also more microscopic reasons to
include pairing fluctuations from the outset. 
In particular, as emphasized, for instance in Ref. \cite{BZA}, 
a spin-spin interaction term as in the 
$t-J$ model can be suggestively rewritten in terms of
electron operators as,
\be
{\bf S}_r \cdot {\bf S}_{r'} = -\frac{1}{2}(c^{\dagger}_{r\ua}c^{\dagger}_{r'\da} - 
c^{\dagger}_{r\da}c^{\dagger}_{r'\ua})(h.c)  + \frac{1}{4}\rho_r \rho_{r'}.
\ee
with $\rho_r = c^{\dagger}_r c_r$. 
For antiferromagnetic exchange the first term 
is an {\em attractive} pairing interaction 
in  the $d_{x^2 -y^2}$ (or extended-$s$) wave channel.
As in BCS theory, this interaction may be 
decoupled 
(in a functional integral) with a complex auxillary pair field $\eta_{ij}$ as
\be
\sum_{<rr'>}J|2\eta_{rr'}|^2 +  [\eta_{rr'}a_{rr'}(c_{r\ua}c_{r'\da} - 
c_{r\da}c_{r'\ua}) + c.c.] .
\ee
Here $a_{rr'} = +1$ for bonds along the $x$-direction, and equals $-1$ for bonds along the $y$-direction.
With $\langle \eta \rangle \ne 0$, this corresponds to a superconducting
phase with  
$d_{x^2 - y^2}$ symmetry.  But more generally, $\eta$ can
be decomposed into an amplitude 
and a phase, $\eta = \Delta e^{i \varphi}$. Ignoring fluctuations in the amplitude
leads to a model of the type we consider below,
with {\it local} fluctuating $d-$wave pairing correlations.

Further motivation for inclusion of such pairing
fluctuations is provided by 
resonating valence bond (RVB) ideas\cite{PWA,KRS}.
The wavefunction for an RVB Mott insulator can be obtained
from the wavefunction of a superconductor by Gutzwiller
projecting into a
subspace with exactly one electron per site. 
Some mean field theories of the RVB state
are equivalent to starting out with just the superconducting wavefunction.  Gauge field fluctuations about the mean field
solution are supposed to carry out this
highly non-trivial projection, and destroy the superconductivity.
A natural physical route
to achieve this end
is to include strong {\it phase} fluctuations of the mean field 
order parameter. Indeed, a recent preprint\cite{DHL} argues that 
fluctuations about the mean field theory of the $d$-wave RVB state\cite{KL}
are formally {\it equivalent} to a theory of a phase-fluctuating $d$-wave superconductor. 

With these motivations,   
we consider generalized Hubbard type models of the form
\be
\label{dHam}
H = H_0  + H_J + H_u + H_\Delta 
\ee
with ,
\bea
H_0 &=& - t \sum_{\langle r r^\prime \rangle} c^\dagger_{r\alpha} c_{r^\prime \alpha} +h.c., \\
H_J &=& J \sum_{\langle r r^\prime \rangle}  
\bbox{S}_r \cdot \bbox{S}_{r^\prime} ,\\ 
H_u &=&  
\sum_r u(N_r-N_0)^2 , \\
H_{\Delta} & = & \sum_{r}\left(e^{i\varphi_r}p_r + h.c \right) ,
\eea
with the local $d-$wave pair field defined as,
\be
\label{pr}
p_r = \sum_{r'\in r } \Delta_{rr'}(c_{r\uparrow}c_{r'\downarrow} - c_{r\downarrow}c_{r'\uparrow}) .
\ee
Here, $c_{r \alpha}$ denotes an electron operator
at site $r$ of (say) a 2d square lattice with spin polarization $\alpha
= \uparrow, \downarrow$.
The electron density and spin operators are
the usual bilinears:
$\rho_r = c^\dagger_{r\alpha} c_{r \alpha}$ and $\bbox{S}_r = {1 \over 2}
c^\dagger_r \bbox{\sigma} c_r$ with $\bbox{\sigma}$ a vector of
Pauli matrices. The term $H_u$ is an on-site repulsion.
Strong local tendencies for $d_{x^2 -y^2}$ pairing are incorporated through
the term
$H_{\Delta}$. 
In the definition of $p_r$ in Eqn. \ref{pr}, the summation is over the four nearest neighbours of the site $r$ and
$\Delta_{rr'} = \Delta$
for bonds along the $x$-direction, and $\Delta_{rr'} = -\Delta$ for bonds along the $y$-direction.
With this choice, the operator $p_r$ destroys a $d_{x^2 - y^2}$ pair of {\em electrons}
centered at the site $r$. 
    
As discussed above, this anomalous term 
can be obtained by decoupling a
local spin-exchange interaction - 
which is attractive in the $d$-wave pairing channel -
with a complex Hubbard Stratanovich field.
Here, we keep the amplitude $\Delta$ fixed,
but include (quantum) fluctuations
of the local pair field phase,
$\varphi_r$.
This phase 
is canonically conjugate to the Cooper pair number
operator, $n_r$:
\begin{equation}
[\varphi_r , n_{r^\prime} ] = i \delta_{r r^\prime}  .
\end{equation}
Due to the anomalous term in $H_\Delta$,
the two densities $\rho_r$ and $n_r$ are {\it not} separately conserved.
The {\it conserved} electrical charge density is simply the
sum of the Cooper pair and electron densities,
\begin{equation}
N_r = 2 n_r + \rho_r   .
\end{equation}
It is this total density that enters into the local on-site Hubbard
interaction term.  The c-number $N_0$ plays the role of a chemical
potential, determining the overall electrical density.

This Hamiltonian describes interacting electrons in a system with
strong local pairing and spin fluctuations.
Since $\varphi_r$ is a {\it dynamical} quantum field,
these pairing fluctuations do {\it not} necessarily
lead to a superconducting ground state.   
In addition to the pairing interaction terms,
the above Hamiltonian includes interactions
in the spin singlet ($u$) and spin triplet ($J$)
particle/hole channels.  
The Hamiltonian retains the important
global symmetries, corresponding to
conservation of 
spin and electrical charge.
It is worth emphasizing that the theoretical description
of electron fractionalization that we develop below
is {\it not} in the least restricted to this
particular Hamiltonian.

\subsection{Overview}

Due to the length of this paper,
we first provide a brief synopsis of our approach and of the
key results.
We start with the observation of
Kivelson and Rokhsar\cite{KR} that, in an appropriate sense, a (singlet) superconductor 
already has separation of spin and charge. If one imagines inserting an electron into
the bulk of a superconductor, its charge gets screened out by the condensate
to leave behind a neutral spin-carrying excitation - a ``spinon''. A
mathematical implementation of this 
idea\cite{NLI} essentially amounts to binding half of a Cooper pair to an electron
to produce a neutral spinon. 
Following these ideas, we first split the Cooper pair operator into two pieces,
each piece creating an excitation with charge $e$
but spin zero.   
These are the same quantum numbers as the ``holon".
But since this object seems to be defined rather differently,
and in any event is {\it not} directly tied to the doping of a Mott insulator,
we prefer to refer to it as a 
``chargon''. The {\em square} of the chargon operator creates the Cooper 
pair. Next, we define a neutral ``spinon'' operator by multiplying together 
the chargon and electron operators.  Changing variables
from the electrons and Cooper pairs to chargons and spinons
introduces a degree of redundancy in the description.
Specifically, all physical observables are invariant
under a {\it local} change in the {\it sign} of the spinon and chargon operators.  This implies that the resulting  
theory must have a local $Z_2$ gauge invariance. 

In Section \ref{Mod}, we carefully re-express the above model in terms of 
the chargon and spinon
operators, paying special attention to the local $Z_2$ gauge symmetry. Following techniques 
familiar from slave boson/fermion theories, we derive an action in terms of the chargon and spinon
fields coupled to a fluctuating $Z_2$ gauge field.
This takes the form
\bea
\label{IGA}
S & = & S_{c} + S_{s} + S_{B},  \\
S_{c} & = & - t_c \sum_{\langle ij \rangle} \sigma_{ij} ( b^*_ib_j + c.c.) ,\\
S_s &=& -\sum_{\langle ij\rangle} \sigma_{ij}
(t^s_{ij} \bar{f}_{i\alpha} f_{j\alpha} + t^{\Delta}_{ij} f_{i\ua}f_{j\da} + c.c ) - 
\sum_i \bar{f}_{i\alpha} f_{i\alpha}. \\
\eea
Here $S_c$ describes the charge
dynamics with $b_i \equiv e^{-i\phi_i}$ the chargon field defined on a $d+1$
dimensional space-time lattice labelled 
by $i,j,....$.
The spin is carried by the (Grassmann-valued) spinon fields,
$f_i$ and $\bar{f}_i$, also living on the lattice sites.
The chargon and spinon fields are ``minimally coupled"
to an Ising $Z_2$ gauge field
$\sigma_{ij} =\pm 1$ living on the 
{\it links} of the space-time lattice.
The form of the charge and spin actions, $S_c$ and $S_s$, could
have been guessed on symmetry grounds (the global charge $U(1)$,
the global spin $SU(2)$ and the {\it local} $Z_2$ gauge symmetry),
but the derivation in Section \ref{Mod} shows the presence of an additional term $S_B$. This is a 
``Berry phase'' term that takes the form
\be
S_{B} = -i\sum_{i, j = i - \hat{\tau}}N_0 [ 2\pi l_{ij}- \frac{\pi}{2} (1 - \sigma_{ij})]  .
\ee  
Here $\tau$ refers to the time direction, and $l_{ij}$ is an integer on each temporal link defined
in terms of the $\phi$ and $\sigma$ fields as,
\be
l_{ij} = Int\left[{\Phi_{ij} \over {2\pi}} + {1 \over 2}    \right] ,
\ee
with $\Phi_{ij}$ the gauge invariant phase difference across
the temporal link:
\be
\Phi_{ij} = \phi_{i} - \phi_{j} + \frac{\pi}{2}(1-\sigma_{ij}) .
\ee
The symbol $Int$ refers to the integer part. 
The Berry phase term simplifies considerably for {\em integer} $N_0$. For even integer $N_0$,
we simply have $e^{-S_B} = 1$, while for odd integer $N_0$,
\begin{equation}
e^{-S_{B}} = \prod_{i,j=i-\hat{\tau}} \sigma_{ij}   ,\hskip0.4cm  N_0 \hskip0.1cm odd  .
\end{equation}

A rough estimate of the dimensionless
couplings $t_c, t^s, t^{\Delta}$ in terms of the parameters $t,u, J, \Delta$
of the original microscopic Hamiltonian may be obtained in the physically interesting 
limit of large $u$ and small $t$ near half-filling:
\be
\label{couplings}
t_c \sim \left(\frac{\sqrt{tu}}{J}\right)^{\frac{1}{3}} \sqrt{\frac{t}{u}}; ~~
t^s \sim \left(\frac{J}{t}\right) t_c ; ~~t^{\Delta} \sim \frac{\Delta}{t}t_c. 
\ee
We will, however, regard these coupling strengths as phenomenological input parameters
for the $Z_2$ gauge theory.

A great deal of physics is contained in the simple-looking action Eqn. \ref{IGA}.  
Consider varying the dimensionless chargon
coupling, $t_c$, which represents the degree
of charge fluctuations, and for simplicity specializing
to half-filling with  
$N_0 = 1$.  Surprisingly,
in the limit of vanishing chargon coupling, $t_c=0$, the 
full $Z_2$ gauge theory action can be shown to be  
{\it formally} equivalent (See Section \ref{fracdwv}) to  
the Heisenberg antiferromagnetic spin model. Increasing $t_c$ from zero 
introduces charge fluctuations into the Heisenberg model. In the limit of large 
$t_c$, the chargons will condense resulting in a conventional $d_{x^2 - y^2}$
superconductor. Thus, the above $Z_2$ gauge theory action has the 
remarkable property of interpolating between
the Heisenberg antiferromagnet in one limit and a $d_{x^2 - y^2}$ superconductor
in the opposite limit.
Determining the properties of this model in the
intervening regime (with $t_c$ of order one) is
an extremely interesting question in the context of the cuprate
materials, 
and will be one of the prime focuses of our analysis. 
Specifically, within the present $Z_2$ gauge theory
we will explore the different 
possible routes between these two limits (which
depend on the parameters in the action).    
Most importantly, for certain parameter regimes
we will 
demonstrate the possibility of obtaining an
exotic fractionalized insulating phase 
- dubbed the nodal liquid in previous work\cite{NLI} - intervening
between the antiferromagnet and the $d_{x^2 - y^2}$ superconductor. For other parameter regimes, 
a number of conventional insulating phases ({\em i.e}, with no fractionalization) are accessible,
including various phases with spin-Peierls and/or charge order. 

To gain a simple understanding of these results 
it is extremely convenient to integrate out the
chargons to give an effective action depending
only on the spinons
and the $Z_2$ gauge field $\sigma$.  This is legitimate
provided the chargons are {\it gapped}, as they will
be in {\it all} of the insulating phases
(with $N_0 = 1$).
The most important effect of this integration will
be to generate a ``kinetic'' term for the $Z_2$ gauge field $\sigma$:
\be
S_{\sigma} = -K \sum_{\Box}[\prod_{\Box} \sigma_{ij} ]   .
\ee
Here, the product is of the $Z_2$ gauge fields around
an elementary plaquette of the space-time lattice,
and this product is then summed over all plaquettes.
Clearly, $S_{\sigma}$ is the direct Ising analog
of the $F_{\mu \nu}^2$ term which enters 
the Lagrangian of ordinary $U(1)$ electromagnetism. 
The value of the parameter $K$ is determined by the chargon coupling,
increasing monotonically with $t_c$.
The full effective action
appropriate to the insulating phases is simply,
\be
S = S_s + S_{\sigma} + S_B .
\ee
Since the  
onset of superconductivity will occur at
some critical value of order one, $t^*_c \approx 1$,
the validity of the effective action requires $t_c < t^*_c$.  
Near this limit,
but on the insulating side, $K$ will also be of order one. 

\begin{figure}
\epsfxsize=3.5in
\centerline{\epsffile{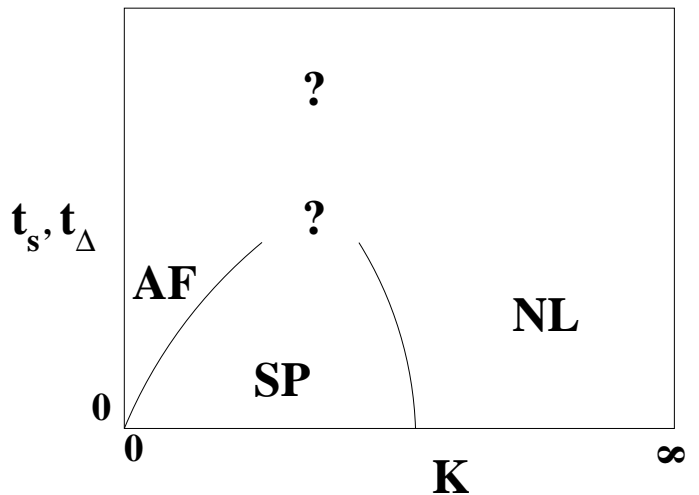}}
\vspace{0.15in}
\caption{Schematic zero temperature phase diagram of the insulating phases showing the three 
limits mentioned in the text. The horizontal axis measures the strength of the coupling $K$
obtained by integrating out the chargons. The vertical axis is a measure of the spinon couplings $t^s, t^{\Delta}$.
Here $AF$ denotes the Heisenberg antiferromagnet, $SP$ denotes an insulator with broken translational 
and rotational invariances such as a spin-Peierls state, and $NL$ denotes the nodal liquid with fractionalized 
excitations}
\vspace{0.15in}
\label{fracdwvf}
\end{figure}

There are several limits in which the properties of this effective action may be reliably analysed. 
A schematic phase diagram is shown in Fig. \ref{fracdwvf}. As mentioned above, 
with $K=t_c=0$ the action describes the Heisenberg antiferromagnet,
which in 2d exhibits Neel long-ranged order at zero temperature. 
The opposite limit of large $K$ is far more interesting, though.
Indeed, 
when $K = \infty$, fluctuations of the $Z_2$ gauge field $\sigma_{ij}$ are frozen, and 
one can set $\sigma_{ij} \approx 1$ on all the links. This results
in a phase with deconfined spinons propagating freely
with the gapless ``d-wave'' dispersion - the ``nodal liquid".  
Similarly, the chargons are also deconfined,
existing as gapped excitations in this insulating phase.
The nodal liquid is thus a genuinely ``fractionalized" insulator, within which
the electron has splintered into two pieces that
propagate independently.
On reducing $K$ from $\infty$, the nodal liquid
continues to be stable until a certain critical value $K_c$ of order one,
where 
the gauge field undergoes a confinement transition.
For $K<K_c$ the chargons and spinons are no longer legitimate excitations,
but rather are confined together to form the electron (or 
other composites built from the electron
such as magnons or Cooper pairs). This corresponds
to a conventional insulating phase. As we argue in Section \ref{fracdwv}, the confinement transition is accompanied
by a breaking of translational symmetry
leading to spin-Peierls order - 
at least for small spinon couplings $t^s, t^{\Delta}$. This may be understood from 
the limit when $t^s, t^{\Delta} = 0$. Then, as we show in 
Section \ref{fracdwv}, we are left with a pure $Z_2$ gauge theory
with the Berry phase term $S_B$ which is {\em exactly} dual to the fully frustrated 
Ising model in a transverse magnetic field. Ordering the Ising spins in this dual global Ising model 
leads to confinement. Physically, the Ising spins represent vortices in the $Z_2$ gauge field -
namely, the {\em vison} excitations mentioned in the previous subsection. 
This same model also arose in the studies of 
Sachdev and coworkers\cite{FTFIM2,subir_pc} on frustrated large $N$ quantum antiferromagnets.
Numerical studies\cite{FTFIM2} show that the ordering in the Ising model is accompanied by 
breaking of translational symmetry.   
The nature of the confined phase(s) at large spinon coupling remains uncertain at present.

These results demonstrate the possibility of two alternate
routes between an antiferromagnet and a
$d-$wave superconductor.
In one instance,
as the chargon hopping $t_c$ is increased towards
the critical value for the onset of 
superconductivity $t^*_c$, the value of the parameter $K$ stays smaller than the critical value for deconfinement, $K_c$.  
In this case, all of the insulating phases preceding the superconductor are ``conventional", 
with confinement of chargons and spinons. 
Alternately, if $K$ exceeds $K_c$
{\it before} the transition into the superconductor,
the fractionalized nodal liquid phase
will occur - sandwiched between the $d-$wave superconductor
and a conventional insulator.
Since both the superconducting and the 
deconfinement transitions occur when $t_c$ (and hence $K$) is of order one,
the deconfinement boundary
is expected to be ``near" the
onset of superconductity.  It is thus difficult to ascertain which 
of these two scenarios will be realized.
The precise phase
diagram interpolating between the antiferomagent
and superconductor will likely 
depend sensitively on 
various microscopic details.

Considerable further insight is provided into the mechanism of electron fractionalization
in an alternate dual formulation in which we trade the
chargon fields for the $hc/2e$ vortices which occur in a conventional superconductor. 
In Appendix \ref{dwaveigt}, we show how this may be done following standard duality 
transformations for the classical three dimensional $XY$ model. 
Starting with the full $Z_2$ gauge theory in Eqn. \ref{IGA},  the resulting
dual theory
is a lattice action for the $hc/2e$ vortices coupled to the spinons. 
The vortices see a fluctuating $U(1)$ gauge field $a$ whose circulation is the 
total electrical three current.  Further, 
the $hc/2e$ vortices have a long-ranged 
statistical interaction with the spinons: When a spinon encircles such a vortex,
its wavefunction acquires a phase of $\pi$. In the present formulation, 
a $\pi$ flux of the $Z_2$ gauge field $\sigma$ is effectively attached
to each vortex.  
As the spinons are minimally coupled to $\sigma$, they
acquire the expected phase of $\pi$ upon encircling each vortex. 
Mathematically, this flux attachment is implemented by an analog of a Chern-Simons term
for the Ising group. Quite remarkably, this Ising Chern-Simons term emerges 
automatically from the duality transformation in Appendix \ref{u1z2d}. 

This dual representation of the $Z_2$ gauge theory
is in fact essentially identical to the vortex field theory
introduced in 
Ref. \cite{NLII} on a phenomenological basis
starting with a BCS superconductor.  In that work, 
the statistical interaction between spinons
and vortices 
was put in by hand, employing a $U(1)$ Chern-Simons
terms to attach flux to the {\it spin} of the spinons.
An advantage of the Ising Chern-Simons terms
is that it does not break spin-rotational invariance, and in fact
is possible even for spinless electrons.  Moreover,
it enables the description of an exotic superconducing phase
in which the Ising flux de-attaches from the vortices (see below). 
In this dual description, superconducting phases correspond to vortex vacuua, while insulating
phases correspond to vortex condensates. Simply condensing the $hc/2e$ vortices leads
to confined insulating phases. Accessing deconfined insulating phases 
requires condensation of {\it paired} vortices, 
without condensation of single ones. 
In this way one obtains an alternate dual description of the
fractionalized nodal liquid.
The $Z_2$ gauge theory formulation 
suggests a mechanism for such vortex pairing:  Since the chargons
also have a long-ranged statistical interaction with $hc/2e$ vortices
their motion is ``frustrated" in the presence of such vortices.
Pairing the vortices reduces this frustration, allowing the
charge to propagate more easily, and lowering the kinetic energy.

Superconducting phases are readily accessed in either
the $Z_2$ gauge theory ``particle" formulation
of Eqn. \ref{IGA}, or its dual vortex counterpart.
In the particle formulation, 
when $t_c$ becomes large and the charge $e$ chargons condense, the result is a 
$d_{x^2 - y^2}$ superconductor - denoted $d{\cal SC}$.  This
superconductor is conventional, perhaps surprising
since BCS theory involves the condensate of a charge $2e$ Cooper pair.
But as we demonstrate in Section \ref{fracswv},
the chargon condensate supports
$hc/2e$ vortices, and shares all other properties
with a conventional BCS superconductor.
It is interesting to ask if it is possible to 
have a superconductor where the chargon {\em pairs} have condensed, while the single chargons have not. Such a superconductor, which we 
label $d{\cal SC}^*$, can be readily described with
the present $Z_2$ gauge theory formulation.
As detailed in Sec. \ref{SCstar}, $d{\cal SC}^*$ is a truly exotic
superconducting phase with many unusual properties.

The $Z_2$ gauge theory is readily generalized to a wide variety of other situations.
In particular, the ``particle" formulation
of Eqn. \ref{IGA} is valid
in {\it any} spatial dimension.
In 3d there again exist 
fractionalized insulating phases (and of course confined ones) 
which can be accessed by the theory. Remarkably, as we argue in Section \ref{FT}, 
in contrast to the 2d case,
a fractionalized insulator in 3d exists as a distinct {\it finite} temperature phase, 
separated by a classical phase transition
from the high temperature limit.  For an anisotropic layered three dimensional material, 
it is also possible to have another 3d fractionalized phase
consisting of weakly  
coupled 2d phases, but this phase is destroyed by thermal fluctuations.
It is also noteworthy that the $Z_2$ gauge theory formulation 
seems incapable of describing fractionalization in 1d. 
This indicates that the ``solitonic" mechanism of 
fractionalization in $d=1$, is qualitatively
different than  ``vortex pairing" which describes
fractionalization in higher dimensions.

We conclude this section with an outline of the rest of the paper. Section \ref{Mod}
contains the formal derivation of the $Z_2$ gauge theory from the microscopic models. 
For ease of presentation, and as it is simpler, we will first provide the technical details of
the derivation for situations with local $s$-wave pairing. In Appendix \ref{dwaveigt}, we show how 
situations with $d_{x^2 - y^2}$
pairing, the case of interest for the cuprates, can be readily handled. We next describe, in Section \ref{fracswv}, 
the physics of fractionalization and confinement in the simplest possible context - that of
$s$-wave pairing with an even number of electrons per unit cell. We then consider, in Section \ref{fracdwv}, the
more interesting situation of $d$-wave pairing with an odd number of electrons per unit cell. 
Section \ref{DVT} formulates and develops the dual description in terms of vortices. The results of
Section \ref{fracdwv} are reobtained in this representation. 
We then move on in Section \ref{Dop} to show how doping away from half-filling
may be incorporated into the formalism. In Section \ref{SCstar}, we discuss the possibility of 
other exotic fractionalized phases, in particular the superconductor ${\cal SC}^*$ mentioned above,
in both the particle and vortex formulations. 
Section \ref{ext} discusses various generalizations
of the theory, including spatial dimensions other than two, finite temperature,
and situations with no spin rotational invariance. We also briefly discuss a useful analogy with
$Z_2$ lattice gauge theories of {\em classical} nematic systems. 
In Section \ref{prevapp}, we discuss the relationship between this
work, and several other previous approaches to fractionalization in strongly correlated systems. 
Contact will be made, when possible, with the earlier dual vortex descriptions of the nodal liquid,
and with the slave boson/fermion approaches. 
Section \ref{disc} contains a
discussion of the experimental signatures of the various novel phases described in earlier sections.
We conclude with a summary of our main results. Various appendices contain technical details not presented 
in the main text.

\section{Models and $Z_2$ gauge theory} 
\label{Mod}
To describe our techniques in the simplest possible context, we will
start with a microscopic model that has local $s$-wave pairing correlations. 
This can be readily generalized to 
other symmetries such as d-wave (see the end of this Section, and Appendix \ref{dwaveigt}).  
Of course, with strong local on-site repulsion
(positive $u$ above) d-wave pairing fluctuations are presumably
more energetically viable, and also of central
interest in the context of cuprate superconductivity.

Consider then a generalized Hubbard type model:
\begin{equation}
H = H_0  + H_u + H_J + H_\Delta  ,
\end{equation}
with
\bea
H_0 &=& - t \sum_{\langle r r^\prime \rangle} c^\dagger_{r\alpha} c_{r^\prime \alpha} +h.c., \\
H_u &=&  
\sum_r u(N_r-N_0)^2 , \\
\label{HJs}
H_J &=& J \sum_{\langle r r^\prime \rangle} [ \bbox{S}_r \cdot \bbox{S}_{r^\prime} + {1 \over 4 }
\rho_r \rho_{r^\prime}  ]  , \\
H_\Delta &=& \Delta \sum_r ( e^{i\varphi_r } c_{r \uparrow} c_{r \downarrow} + h.c. )  . 
\eea
As earlier, $c_{r \alpha}$ denotes an electron operator
at site $r$ with spin $\alpha$
and
the electron density and spin operators are
the usual bilinears:
$\rho_r = c^\dagger_{r\alpha} c_{r \alpha}$ and $\bbox{S}_r = {1 \over 2}
c^\dagger_r \bbox{\sigma} c_r$. 
This Hamiltonian is essentially the same as Eqn. \ref{dHam} in the introduction,
except that it has local $s-$wave pairing rather than $d-$wave,
and we have added a term proportional to 
$\rho_r \rho_{r'}$ in $H_J$. 
These modifications have been made
to both simplify the derivation and the subsequent
analysis of the $Z_2$ gauge theory.  We return later to the more physically
interesting case of local $d-$wave pairing.

Here, $\varphi_r$ is the phase of a local $s$-wave Cooper pair field and
is canonically conjugate to the Cooper pair number
operator, $n_r$:
$[\varphi_r , n_{r^\prime} ] = i \delta_{r r^\prime}$.
As before, since $\varphi_r$ is a {\it dynamical} quantum field,
these pairing fluctuations do {\it not} necessarily
lead to a superconducting ground state. 
The {\it conserved} electrical charge density is the
sum of the Cooper pair and electron densities
\be
\label{Ntot}
N_r = 2 n_r + \rho_r.
\ee

\subsection{Split the Cooper pair}
\label{split}

We now proceed to split the Cooper pair into two pieces.
Consider an operator $b_r$ defined as,
\begin{equation}
\label{chargon}
b_r^\dagger = s_r e^{i\varphi_r/2} = e^{i\phi_r}   ,
\end{equation}
with $s_r = \pm 1$ an Ising ``spin" variable.
With this definition the new 
field,
\begin{equation}
\phi_r = {\varphi_r \over 2} + {\pi \over 2} (1 - s_r)  ,
\end{equation}
can be treated as a phase - lying in the interval zero to $2 \pi$,
with $b_r$ invariant under the transformation:
$\varphi_r \rightarrow \varphi_r + 2\pi$ and $s_r \rightarrow - s_r$.
The {\it square} of $b^\dagger_r$ creates
a Cooper pair, 
\begin{equation}
e^{i\varphi_r} = (b^\dagger_r)^2 ,
\end{equation}
so that $b^\dagger_
r$  creates a spinless excitation
with charge $e$, essentially one-half of a Cooper pair.
We refer to this operator as a ``chargon" operator.

In order to separate out the charge
and spin degrees of freedom it will
be extremely useful to define an electrically
neutral but spin carrying fermion operator (a ``spinon"):
\begin{equation}
f^\dagger_{r \alpha} = b_r c^\dagger_{r \alpha}    .
\end{equation}
This operator carries the spin of the electron,
but is electrically neutral as verified by noting that
it commutes with the total
electrical charge density $N_r$. 
On the other hand, the chargon is electrically charged,
and its phase is canonically conjugate
to the total electrical charge density
\begin{equation}
[ \phi_r , N_{r^\prime} ] = i \delta_{r r^\prime}  .
\end{equation}

At this stage it is legitimate to implement an operator change of variables
in the full Hamiltonian, replacing the electron and Cooper pair
operators ($\varphi, n, c, c^\dagger$) by chargons and
spinons ($\phi, N, f, f^\dagger$).
This gives, 
\be
\label{hsbf}
H = H_0  + H_u + H_J + H_\Delta,
\ee
with
\begin{equation}
\label{hobf}
H_0 = - t \sum_{\langle r r^\prime \rangle} 
b^\dagger_r b_{r^\prime} f^\dagger_{r\alpha} f_{r^\prime \alpha} +h.c.,
\end{equation}
\begin{equation}
\label{hdelf}
H_\Delta = \Delta \sum_r ( f_{r \uparrow} f_{r \downarrow} + h.c. )  ,
\end{equation}
with $H_u$ unchanged and $H_J$ of the same form as in Eqn. \ref{HJs}
but with spinon operators replacing
the electron operators:
$\rho_r = f^\dagger_{r\alpha} f_{r \alpha}$ and $\bbox{S}_r = {1 \over 2}
f^\dagger_r \bbox{\sigma} f_r$.

There are several extremely important points to stress about
this seemingly inoccuous change of variables.  Firstly,
one can change the sign of both the chargon and spinon
operators on any given site $r$,
\begin{equation}
b_r \rightarrow - b_r   ; \hskip0.5cm f_{r\alpha} \rightarrow -f_{r\alpha}  ,
\end{equation} 
without affecting the original Cooper pair or electron operators.
This implies that quite generally the transformed Hamiltonian {\it must}
also be invariant under this  {\it local} Ising $Z_2$
symmetry - as can be readily checked in Eqns. \ref{hobf} and \ref{hdelf}.  As we shall shortly
see, in a path integral formulation this local $Z_2$ symmetry
will be manifest in terms of an $Z_2$ gauge field.
Secondly, because of this redundancy introduced in the change of variables,
a {\it constraint} must be imposed on the Hilbert space
spanned by the spinon and chargon operators.

To understand the origin of this constraint,
consider first the Hilbert space of the original Hamiltonian.
In a number-diagonal basis, the Hilbert space
on each site $r$ is a direct product of states with an arbitrary
integer number of Cooper pairs ($n_r$) and the 
four electron states consistent with Pauli - empty, doubly occupied
or singly occupied with an electron of either spin.  
Since the chargon has only one-half the charge of the Cooper pair,
the full Hilbert space spanned by the chargon and spinon operators
is actually twice as large, and it is essential
to project down into the physical Hilbert
space of electrons and Cooper pairs.  From Eqn. \ref{Ntot},
it is clear that this can be achieved
by imposing a constraint that the {\it sum} (or difference)
of the number of chargons ($N_r$) and spinons ($\rho_r = f^\dagger_{r \alpha} f_{r\alpha}$) on each site
is an even integer:  
\begin{equation}
\label{constr}
(-1)^{N_r + \rho_r} = 1.
\end{equation}
This implies, for example, that a site with a single chargon
but no spinon is unphysical and forbidden,
whereas a spinon and chargon together (an electron) is allowed.

\subsection{Path Integral and $Z_2$ gauge Theory}

The most convenient way to implement the constraint on the 
spinon and chargon Hilbert space is in a (Euclidian) path
integral representation of the partition function.
To this end we define a projection operator,
\begin{equation}
\label{proj}
{\cal P} = \prod_r {\cal P}_r   ,
\end{equation}
with 
\begin{equation}
{\cal P}_r = {1 \over 2} [1 + (-1)^{N_r + \rho_r} ] = {1 \over 2}
\sum_{\sigma_r = \pm 1} e^{i{\pi \over 2} (1 - \sigma_r)(N_r + \rho_r)}   ,
\end{equation}
which projects into the physical Hilbert space.
Here, $\sigma_r = \pm 1$ is an Ising-like field
and $\rho_r = f^\dagger_{r\alpha} f_{r\alpha}$.  
As can be verified directly from Eqn. \ref{hsbf},
this projection operator commutes with the chargon-spinon Hamiltonian,
\begin{equation}
[{\cal P}, H] = 0,
\end{equation}
so that the Hamiltonian does not cause transitions
out of the physical Hilbert space.

The partition function can be written as,
\begin{equation}
Z = Tr [ e^{-\beta H} {\cal P} ]  ,
\end{equation}
where the trace is over the full Hilbert space spanned by the
chargon and spinon operators
($\phi,N,f,f^\dagger$).
A Euclidian path integral representation can be obtained
as usual by splitting the exponential,
\begin{equation}
\label{cspf}
Z = Tr [ (e^{-\epsilon H} {\cal P} )^M ]   ,
\end{equation}
with $M$ ``time slices" and $\epsilon = \beta/ M$.
Here, we have inserted projection operators into each time slice.
Working with fermion coherent states and eigenstates
of the chargon phase $\phi$, a path integral representation
can be readily derived - as detailed in Appendix A - giving,
\begin{equation}
Z = \int \prod_{i\alpha} d\bar{f}_{i\alpha} 
df_{i\alpha} d\phi_i \sum_{N_i = - \infty}^\infty
\sum_{\sigma_i = \pm 1}   e^{-S}    ,
\end{equation}
where the integration is over
Grassman numbers $f$ and $\bar{f}$ 
and a c-number phase $\phi$ in the interval zero to $2\pi$.
Here, $i=(r,\tau)$ runs over the $2+1$-dimensional space time lattice
with $\tau=1,2,...,M$ time slices. 
The Euclidian action takes the form,  
\begin{equation}
S = S^f_\tau + S^\phi_\tau + \epsilon \sum_{\tau =1}^M H(N_\tau,\phi_\tau,
\bar{f}_\tau f_\tau )   ,
\end{equation} 
with
\bea
S^f_\tau &=& \sum_{r, \tau=1}^M  [ \bar{f}_{\tau} (\sigma_{\tau+1} f_{\tau+1}
-f_{\tau}  ) ] , \\
S^\phi_\tau &=& -i \sum_{r, \tau=1}^M  N_{\tau} [\phi_{\tau} - \phi_{\tau -1} + {\pi \over 2}
(1 - \sigma_{\tau}) ] .
\eea 
Here, we have suppressed the explicit $r$ and $\alpha$ subscripts
on the fields, displaying only the time-slice dependencies.
As usual, the bosonic phase field and the Ising field both
have the expected periodic boundary conditions,
whereas the fermions are anti-periodic:
\begin{equation}
\phi_{\tau=M+1} = \phi_{\tau=1}  ;  \hskip0.3cm  \sigma_{M+1} = \sigma_{1}  ;
\hskip0.3cm  f_{M+1} = - f_{1}    .
\end{equation}

Notice that the Ising variables live on the links connecting
adjacent time slices, and can thus be correctly interpreted
as a gauge field.  In fact,
the Ising field $\sigma$ is {\it minimally} coupled
to both spinons and chargons, as the time component of a gauge field.
Moreover, the local $Z_2$ symmetry of the Hamiltonian
in Eqn. \ref{hsbf}, is manifest in the path integral as a full fledged
Ising $Z_2$ gauge symmetry:
\begin{equation}
\label{igbf}
f_{i\alpha} \rightarrow \epsilon_i f_{i\alpha} ; \hskip0.3cm \bar{f}_{i \alpha}
\rightarrow \epsilon_i \bar{f}_{i\alpha}  ; \hskip0.3cm  \phi_i \rightarrow \phi_i
+ {\pi \over 2}(1-\epsilon_i)   ,
\end{equation}
{\it together} with a transformation of the gauge field,
\begin{equation}
\label{igsigma}
\sigma_{ij} \rightarrow \epsilon_i \sigma_{ij} \epsilon_j   .
\end{equation}
Here, $\epsilon_i = \pm 1$,
and $\sigma_{ij}$ lives on the link
connecting two ``nearest neighbor" space-time lattice points,
differing by one time slice.  

Our final goal is to beat the model into a form which also includes
$Z_2$ gauge fields on the {\it spatial} links,
so that space and time end up on more equal footing.
Our approach follows closely the standard methods\cite{RSSuN} employed
in slave fermion or slave boson treatments
of Heisenberg magnets.
First, we perform a Hubbard-Stratanovich decoupling
of the spin interaction terms in the Euclidian action:
\begin{equation}
e^{-\epsilon H_J} = \int \prod_{\langle r r^\prime \rangle} 
\prod_\tau d\chi_{rr^\prime}(\tau) d\chi^*_{rr^\prime}(\tau) e^{-S_{hs}}   , 
\end{equation}
\begin{equation}
S_{hs} = \epsilon \sum_{\langle rr^\prime \rangle} \sum_\tau 
[2J |\chi_{rr^\prime}|^2 - (J\chi_{rr^\prime} \bar{f}_{r\alpha}
f_{r^\prime \alpha} + c.c.)]   .
\end{equation}

Here, $\chi_{rr^\prime}(\tau)$ are a set of complex fields which live on each
of the nearest neighbor spatial links.  
Next, a simple
change of variables can be performed which eliminates
the remaining quartic spinon-chargon interaction, in $H_0$ in Eqn. \ref{hobf}:
\begin{equation}
\chi_{rr^\prime}  \rightarrow \chi_{rr^\prime} - {t \over J} b^*_r b_{r^\prime}    ,
\end{equation}
where $b^*_r \equiv e^{i\phi_r}$.  The full Euclidian action then takes
the form, $S = S^f_\tau + S^\phi_\tau + S_r$,
with the spatial interactions given by,
\begin{equation}
S_r = \epsilon \sum_\tau (H_u + H_\Delta) + S_\chi  ,
\end{equation}
with
\begin{equation}
S_\chi = \epsilon \sum_{\langle rr^\prime\rangle} 2J|\chi_{rr^\prime}|^2 - [\chi_{rr^\prime}
(2t b^*_{r^\prime} b_r + J\bar{f}_{r\alpha} f_{r^\prime \alpha}) + c.c.] .
\end{equation}
The terms in $S_\chi$ correspond to the hopping of spinons and chargons
in the presence of a common fluctuating gauge field, $\chi$,
on the near neighbor links.  

Up to this stage, 
all of the formal manipulations that we have
performed have been {\it exact},
so that the full Euclidian action gives
a faithful representation of the original ``microscopic"
electron Hamiltonian.  But now, following
standard slave fermion/boson techniques,
we perform an approximation,
treating the functional integral
over the Hubbard-Stratanovich field, $\chi$,
within a saddle point approximation.  [While it might be
possible to find an appropriate ``large-N" generalization
of the model for which this approximation becomes exact,
we do not pursue this tack here.]  The simplest saddle-point
corresponds to setting all of the link fields equal to
a single real constant:  $\chi_{rr^\prime} = \chi_0$.
The saddle-point value for $\chi_0$ can (in principle)
be obtained by integrating out the spinons (which are Gaussian)
and the chargons (which are not).  This 
saddle-point respects two important discrete symmetries
of the model - translational and time-reversal invariance.
But the saddle-point does {\it not} respect
the $Z_2$ gauge symmetry in
Eqns. \ref{igbf} and \ref{igsigma}.  This serious flaw can be easily
remedied, though, by retaining a particular set of
{\it fluctuations} about the saddle point.
The simplest choice consistent with 
the $Z_2$ gauge symmetry corresponds to allowing the
{\it sign} of $\chi_{rr^\prime}$ to change, keeping the magnitude fixed,
putting
\begin{equation}
\chi_{rr^\prime} = \sigma_{rr^\prime} \chi_0   .
\end{equation}
Here, $\sigma_{rr^\prime}(\tau) = \pm 1$ are a set
of Ising fields living on the spatial links of the space-time lattice.
Within this restricted manifold the 
theory consists of chargons and spinons
hopping on a space-time lattice, minimally coupled
to an $Z_2$ gauge field.  

Hereafter we work under this fixed-magnitude approximation.
Within this approximation the full partition function
can be expressed as a functional integral,
\begin{equation}
\tilde{Z} = \int \prod_{i\alpha} d\bar{f}_{i\alpha} df_{i\alpha}
d\phi_{i} \sum_{N_i = - \infty}^\infty \prod_{\langle ij \rangle}
\sum_{\sigma_{ij} = \pm 1} e^{-S}    ,
\end{equation}
with $Z_2$ gauge fields $\sigma_{ij}$ living on the near neighbor links
of the space-time lattice, and 
\begin{equation}
S = S^f_\tau + S^\phi_\tau + S_0 + S_u + S_\Delta   ,
\end{equation}
with
\bea
 S^f_\tau &=& \sum_{i, j=i+\hat{\tau}} [\bar{f}_{i\alpha}(\sigma_{ij} f_{j\alpha}
-f_i  ) ]   , \\
S^\phi_\tau &=& -i \sum_{i, j=i-\hat{\tau}} N_i[\phi_i -\phi_j + {\pi \over 2}
(1-\sigma_{ij})    ]   , \\
 S_u &=& \epsilon u \sum_i (N_i - N_0)^2   , \\
 S_\Delta &=& \epsilon \Delta \sum_i (f_{i \uparrow} f_{i \downarrow} + \bar{f}_{i \downarrow} \bar{f}_{i \uparrow} )   , \\
 S_0 &=& - \epsilon \sum_{i,j=i + \hat{x}}  \sigma_{ij} (t_0 b^*_i b_j 
+ J_0 \bar{f}_{i\alpha} f_{j\alpha} + c.c.)  ,
\eea
 where we have defined $t_0 = 2t\chi_0$ and $J_0 = J \chi_0$.

Notice that the full action is local
in the integers $N_i$, so the summation can be performed
independently at each space-time point.
A straightforward Poisson resummation gives
\be
\sum_{N_i} e^{- \left(S_u + S^{\phi}_{\tau}\right)} = \exp[\sum_{i,j = i-\hat{\tau}}V(\Phi_{ij})],
\ee
where $\Phi_{ij} = \phi_i - \phi_j + \frac{\pi}{2}(1 - \sigma_{ij})$ is the gauge invariant
phase difference along a temporal link. Here, the periodic potential $V(\Phi)$ is given
by
\be
e^{V(\Phi)} = \sum_{l = -\infty}^{\infty} e^{-\frac{1}{4\epsilon u}[\Phi - 2\pi l]^2 + iN_0 (2\pi l - \Phi)}  ,
\ee
and we have dropped an overall multiplicative constant. 
In the limit of small $\epsilon u$, the sum over $l$ will be dominated by 
precisely one term which minimizes
$|\Phi - 2\pi l|$.
This occurs for integer $l$ satisfying $|\Phi - 2\pi l|< \pi$,
or equvalently,
\be
\label{l}
l = Int[\frac{\Phi}{2\pi} + \frac{1}{2}] .
\ee
Moreover, for small $\epsilon u$ we may approximate
\bea
e^{-\frac{1}{4\epsilon u}(\Phi - 2\pi l)^2}  & \sim & e^{\frac{1}{2\epsilon u}[ 1 - cos(\Phi - 2\pi l)]}  ,  \\
& = & e^{\frac{1}{2\epsilon u} [1 - cos(\Phi)]}  .
\eea
Within this approximation the sum over $l$ becomes simply, 
\be
e^{V(\Phi)} \approx e^{+\frac{1}{2\epsilon u}cos(\Phi) + iN_0(2\pi l - \Phi)} ,
\ee
with $l$ given by Eqn. \ref{l}.
We have again dropped an overall multiplicative constant.

The full $N$ sum in the action then leads to 
\be
\sum_{N_i}e^{-\left(S_u + S^{\phi}_{\tau}\right)}  =  e^{\sum_{i, j = i - \hat{\tau}}\frac{1}{2\epsilon u}
\sigma_{ij}cos(\phi_i - \phi_j) -S_B} ,
\ee
with the ``Berry phase'' term $S_B$ given by 
\bea
\label{SB}
S_B & = & -iN_0 \sum_{i, j = i - \hat{\tau}} (2\pi l_{ij} - \Phi_{ij})  ,\\
& = & -iN_0 \sum_{i, j = i - \hat{\tau}} [2\pi l_{ij} - \frac{\pi}{2}(1-\sigma_{ij})] .
\eea
In obtaining the last line, we have re-expressed $\Phi_{ij}$
in terms of $\phi$ and $\sigma$, and used the $\beta-$periodic boundary conditions on 
$\phi$ to drop the term involving $\phi_i - \phi_j$. 
The ``Berry phase'' term is the {\it only} 
term in the action which depends on the
(average)
occupation number per unit cell, $N_0$. It simplifies considerably for integer $N_0$. For {\em even}
integer $N_0$, we simply have $e^{-S_B} = 1$, while for odd integer $N_0$, 
\begin{equation}
\label{Sodd}
e^{-S_B} = \prod_{i,j=i-\hat{\tau}} \sigma_{ij}   ,\hskip0.4cm  N_0 \hskip0.1cm odd  .
\end{equation}
As we shall see,
the Berry's phase term will lead to subtle
yet important differences between Mott insulators
with odd integer $N_0$ and band insulators with even $N_0$.

The Euclidian path integral
is only identical to the Hamiltonian formulation
in the strict $\epsilon \rightarrow 0$ limit.  
But since the original lattice Hamiltonian is already
an effective low energy theory, the
time continuum limit which involves
arbitrarily high energies is not actually of interest.
For these reasons, hereafter we keep $\epsilon$
{\it finite}, viewing it as an inverse ``high energy" cutoff in the theory.
Since the kinetic ($t$) and interaction ($u$) energy scales
are the largest in the theory, it is convenient to
choose the value of $\epsilon$ so that the charge sector
of the theory is isotropic on the $2+1$-dimensional space-time
lattice.   To this end,  we require that the spatial chargon hopping strength
equals the temporal one:  ${1 \over {2\epsilon u}} =
2\epsilon t_0$, which implies
\begin{equation}
\label{epsilon}
{1 \over \epsilon} = 2 \sqrt{t_0 u}   .
\end{equation}

With this choice of $\epsilon$  
the full Euclidian action reduces to a much simpler
and more compact form:
\be
\label{IGAs}
S = S_c + S_s + S_B 
\ee
with
\bea
S_c &=& - t_c \sum_{\langle ij\rangle } \sigma_{ij}  (b^*_i b_j + h.c) , \\
S_{s} &=& \sum_{\langle ij \rangle} -(t^s_{ij} \sigma_{ij}  \bar{f}_i f_j + c.c.) + \delta_{ij} (t^{\Delta} f_{i\uparrow}
f_{i \downarrow} + c.c. - \bar{f}_i f_i 
) ,
\eea
and $S_B $ as defined above.  Here,
the dimensionless chargon coupling strength is given in terms of the microscopic parameters $t, u$ and $\chi_0$ to be 
\be 
\label{tc}
t_c = \epsilon t_0 ~~ = \sqrt{\frac{t \chi_0}{2u}} .
\ee
The dimensionless spinon coupling along the nearest neighbor spatial
links is
\be
\label{ts}
t^{s}_{ij} = \epsilon J_0 ~~ = J\sqrt{\frac{\chi_0}{8tu}},
\ee
whereas $t^s_{ij}=-1$ along the neighboring
temporal links. Similarly, the coupling constant for the spinon pairing
is
\be
\label{tDelta}
t^{\Delta} = \frac{\Delta}{\sqrt{8t\chi_o u}}  .
\ee
As will be shown in Section \ref{fracdwv}, for the physically interesting case 
of $d$-wave pairing near half-filling, the parameter $\chi_0$
may be roughly estimated to be
\be
\chi_0 \sim \left(\frac{tu}{J^2}\right)^{\frac{1}{3}}.
\ee
This can be used to obtain rough estimates of the three
dimensionless coupling constants, $t_c,t^s$ and $t^\Delta$. 
For the most part, 
however, we will treat these couplings as phenomenological parameters.  
 
The partition function involves an integration over the 
on-site chargon phase ($\phi_i$) and spinon Grassman fields
($\bar{f}_i,f_i$), as well as a summation over
the $Z_2$ gauge fields ($\sigma_{ij} = \pm 1$) which live
on the nearest neighbor links of the Euclidian space time lattice.
This ``final" form for the theory is exceedingly simple,
consisting of chargons and spinons hopping around,
minimally coupled to a dynamical $Z_2$ gauge field.
This form could have
essentially been guessed just using a knowledge
of the field content (chargons and spinons) and the required symmetries;
$U(1)$ charge conservation, $SU(2)$ spin conservation and the local
$Z_2$ gauge symmetry.  Perhaps the only subtlety is the
presence of the term $S_B$ in the action when the
filling factor $N_0$ is  not an even integer. 
Among the additional terms which are allowed by these symmetries,
is a field strength term for the $Z_2$ gauge field:
\begin{equation}
S_{\sigma} = -K \sum_{\Box} [ \prod_{\Box} \sigma_{ij} ]   .
\end{equation}
Here, the product denotes the gauge invariant
product of the Ising fields around
an elementary plaquette.  This Ising field strength
is then summed over all space-time plaquettes.
Clearly, $S_{\sigma}$ is the direct Ising analog
of the $F_{\mu \nu}^2$ term which enters 
the Lagrangian of ordinary $U(1)$ electromagnetism.
Even though not present in the derivation
presented here, this field strength term will be
generated upon integrating out the chargon or spinon 
matter fields,
as discussed below.
 
In Appendix \ref{dwaveigt} we show how the above analysis can be generalized to the
case in which local d-wave pairing correlations are incorporated
from the outset as in the Hamiltonian Eqn. \ref{dHam}, rather than s-wave as assumed above.  
The derivation
of the effective $Z_2$ gauge theory proceeds in much the same fashion,
and one arrives at the same model except with the
spinon action given instead by
\be 
S_s = -\sum_{\langle ij\rangle} \sigma_{ij}
(t^s_{ij} \bar{f}_{i\alpha} f_{j\alpha} + t^{\Delta}_{ij} f_{i\ua}f_{j\da} + c.c ) - 
\sum_i \bar{f}_{i\alpha} f_{i\alpha}   .
\ee 
 Here, $t^{\Delta}_{ij}$ denotes a d-wave pairing amplitude
living on the nearest neighbor spatial bonds,
with amplitude $+t^{\Delta}$ on the x-axis bonds and $-t^{\Delta}$
along the y-axis bonds.  Notice that the
$Z_2$ gauge field $\sigma_{ij}$ enters here,
because the d-wave pair field lives
on the {\it links}.  This form exhibits the
required Ising $Z_2$ gauge symmetry, being invariant
under the transformation in Eqn. \ref{igbf}. As shown in Section \ref{fracdwv},
a rough estimate of the various coupling constants in this case is
\be
t_c \sim \left(\frac{\sqrt{tu}}{J}\right)^{\frac{1}{3}} \sqrt{\frac{t}{u}}; ~~
t^s \sim \left(\frac{J}{t}\right) t_c; ~~t^{\Delta} \sim \frac{\Delta}{t}t_c. 
\ee
Here $t^s$ and $t^{\Delta}$ refer only to the spatial couplings. 
But we will once again regard these as phenomenological parameters.

\section{Fractionalization and Confinement}
\label{fracswv}

In this section we will analyze some of the phases which are described by the
$Z_2$ gauge theory 
model derived in Section II.
While the $Z_2$ gauge formulation is valid
in general dimension, for concreteness and simplicity
we specialize to two dimensions,
generalizing briefly to other dimensions
in Section \ref{Gsd}.  Moreover, for illustrative purposes
we focus first on the simplest case
with 
an even number of electrons
per site (unit cell), and presume
the presence of local s-wave pairing correlations.
As we shall see, in this case the model can 
exhibit a conventional band insulator.
In Section \ref{fracdwv} we will
turn to the more physically interesting situation
with an {\it odd} number of electrons per site.
At that stage we will focus on local d-wave
pairing correlations, which are more tenable
in the presence of a large positive on-site Hubbard $u$
as well as being of direct relevance
to the cuprates. Doping away from half-filling will be discussed in Section \ref{Dop}.

With even integer $N_0$ and local s-wave
pairing correlations the full action consists
of two contributions,  $S = S_c + S_s$,
corresponding to the charge and spin sectors, respectively:
\bea
\label{Ss-even}
S_c &=& - t_c \sum_{\langle ij\rangle } \sigma_{ij} (b^*_i b_j + c.c.)  , \\
S_{s} &=&   -\sum_{\langle ij \rangle} t^s_{ij} \sigma_{ij}  (\bar{f}_i f_j + c.c.) - \sum_i \bar{f}_i f_i \\
&+& t^{\Delta} \sum_i (f_{i\uparrow}
f_{i \downarrow} + c.c. ) .
\eea
The first term, which describes the dynamics of
the chargons, $b^* = e^{i\phi}$,
minimally coupled to an $Z_2$ gauge field, exhibits
the global $U(1)$ charge conservation symmetry.  The spinons also carry the $Z_2$ Ising ``charge".   
Due to the s-wave form of the
anomalous ``pairing" term the spinons,
which  are paired into singlets, should be gapped out.

\subsubsection{Correlated ``Band" Insulators}

We first consider electrically insulating states.
When the dimensionless
chargon coupling $t_c$ is much smaller than unity,
the chargons cannot propagate at low energies and a charge
gap results.  In this case, with both spinons and chargons gapped out,
it is possible to integrate them out from the theory,
leaving the $Z_2$ gauge field $\sigma$ as the only remaining
field.  This integration will generate
additional terms in the Lagrangian,
depending on $\sigma$, which will be local in space-time
and must also be gauge invariant.
The most important such term\cite{note1} is simply,
\be
S_\sigma =  -K \sum_{\Box} [ \prod_{\Box} \sigma_{ij} ]   ,
\label{Fmunu}
\ee
which describes a pure $Z_2$ gauge theory.

Remarkably, this simple gauge
theory exhibits a phase transition as the coupling $K$ is varied.
Indeed, as shown originally by Wegner\cite{Wegner,Kogut}, the pure $Z_2$ gauge
theory in $3D$ is {\it dual} to the familiar
three-dimensional Ising model:
\be
S_{dual} = - K_d \sum_{\langle ij \rangle} v_i v_j   ,
\ee
with Ising spins, $v_i = \pm 1$, living on the sites
of the dual lattice.
The dimensionless Ising model coupling, $K_d$, 
is simply related to $K$:  $tanh(K_d) = e^{-2K}$.
This form shows that the high and low ``temperature"
phases are exchanged under the duality transformation.
The details of this duality transformation are given in Appendix \ref{ISD}.

As emphasized originally
by Wilson\cite{Wilson}, a direct characterization of the
two phases of the pure gauge theory is given in terms of the correlator,
\be
\label{Wil} 
{\cal G}_{{\cal C}} = \langle \prod_{{\cal C}} \sigma_{ij} \rangle  ,
\ee
where the average
is for the pure gauge theory and the product is taken around a closed loop
in space-time, denoted ${\cal C}$.
For $K < K_c$ the Wilson-loop satisfies an ``area law",
with ${\cal G}_{{\cal C}} \sim exp(-c{\cal A})$, with loop area ${\cal A}$,
and $c$ a $K-$dependent constant.  
When $K > K_c$, ${\cal G}_{{\cal C}}$ decays more slowly,
only exponentially with the {\it perimeter} of the loop.

What do these two phases correspond to in physical terms?
Consider first the large $K$ limit,
which is the high temperature phase of
the dual Ising model.  As $K \rightarrow \infty$
all of the gauge field plaquette sums will be equal to plus one.
In this case it is possible to choose
a gauge in which all of the Ising link variables
are also unity, $\sigma_{ij}=1$.  In this phase the chargons
and spinons can {\it propagate} at energies above their respective gaps.
Apparently, the Hamiltonian
contains gapped excitations which carry the quantum numbers 
of spinons and chargons.  The electron has effectively been fractionalized!
We denote this exotic insulating state
with deconfined chargons and spinons as ${\cal I}^*$.
It is exceedingly important to emphasize that the splintering
of the electron into spin and charge carrying constituents
is conceptually unrelated to the presence or absence of spin order.
Indeed, electron fractionalization can occur
even in the presence of strong
spin-orbit interactions which destroys spin-rotational invariance - 
in that case the states of the fermionic $f$-particles
cannot be labelled by spin.  

As the coupling $K$ is reduced, so long as the gauge theory is
in its perimeter phase, the energy to separate
particles carrying the $Z_2$ charge remains finite, even
for infinite separation.  The chargons and spinons are deconfined.
Further, with $K < \infty$, 
configurations of the $Z_2$ gauge
theory with plaquette
products equal to minus one will become possible.  One can think of such 
plaquettes as being ``pierced" by non-zero ``$Z_2$ - flux"
or $Z_2$ vorticity.  Because the number of such plaquettes
on any given elementary space-time cube is even, the fluxes
form ``tubes" - analogous
to Abrikosov vortices in a Type II superconductor -  which propagate in space-time 
as particles .  These particles can scatter and can anihillate in pairs,
but since their number is conserved modulo 2 they
carry a conserved $Z_2$ ``charge".  
We will refer to these particle-like
$Z_2$ vortices as ``visons".  One can define
a vison ``3-current", $j_v$ - a field
which lives on the links of the dual lattice
and takes one of two values, zero or one - 
which satisfies,
\begin{equation}
(-1)^{j_v} = \prod_{\Box} \sigma_{ij}   ,
\end{equation}
with the plaquette pierced by the dual link.
In the deconfined phase, ${\cal I}^*$,
these vison particles exist as gapped excitations, in addition
to the spinons and chargons.  In terms of the dual Ising
model, $S_{dual}$, the Ising spins $v_i$
are essentially vison creation operators.
With the Ising model being disordered
for large $K$, the visons (Ising spins) are gapped. Thus, the distinct
gapped excitations in ${\cal I}^*$ are (i) the chargons (ii) the spinons
and (iii) the visons. An important property of these excitations
is the existence of long-ranged ``statistical" interactions
between them.
Specifically, when a chargon (or a spinon) is adiabatically
transported around a vison, it acquires a geometrical phase factor
of $\pi$ (because the chargon is
minimally coupled to the $Z_2$ gauge field).
Similarly, a vison picks up a $\pi$ phase factor upon encircling
either a chargon or a spinon.
Evidently, visons and chargons (or spinons) are ``relative semions".

As $K$ is reduced further the gauge theory undergoes a phase transition
at $K_c$ into its``area-law'' phase. 
This implies
that the energy to separate two spinons or chargons,
inserted as ``test" charges at spatial separation, $R$,
grows linearly with $R$.  
In this ``confined" insulating phase, denoted ${\cal I}$,
free chargons and spinons do not exist
in the spectrum.  The only allowed particle excitations
are those that are ``charge neutral" - that is,
invariant under the $Z_2$ gauge transformation.  Any bound state
with an even number of chargons plus spinons is ``neutral".
In addition to the electron, this includes any composite
built from electrons, such as a Cooper pair or a magnon.
In the phase ${\cal I}$ these electron-like
excitations will be gapped.  This phase is the familiar ``band insulator"
with an even number of electrons per unit cell.
 
Note that with $K <K_c$, the dual Ising model
orders, $\langle v_i \rangle \ne 0$.
This corresponds to a ``condensation" of the visons. 
Remarkably,  $Z_2$ vortex condensation leads directly to
a ``confinement" for the chargons and spinons.  To understand
confinement directly in terms of the dual
Ising model, consider the effect of inserting two static ``test"
chargons, separated by a distance $R$.  Each chargon
lives on a (spatial) plaquette
of the dual Ising model.  Due to the geometrical phase factor
between visons and chargons, the presence
of a chargon corresponds to a ``frustrated plaquette"
in the dual Ising model - that is, a plaquette
with an odd number of negative Ising couplings.  To frustrate
{\it two} plaquettes, it suffices to introduce an interconnecting string
of negative Ising bonds.
In the ordered phase of the dual Ising model,
the energy of this string will clearly be linear
in its length, thereby confining the two chargons.

It is worth drawing a very important distiction between the Ising
gauge theory considered here, and the gauge theories introduced by Baskaran and Anderson\cite{BA}
and generalized and extensively studied by several authors\cite{oldgauge}.  In the simplest
version of these theories, the spin itself is effectively
fractionalized, decomposed into a bi-linear
of spinful (complex) fermion operators, rather than splitting the Cooper pair
into two chargons as discussed above.  These spinful fermion operators - 
the spinons - are minimally coupled to a compact $U(1)$
gauge field.  But in contrast to the $Z_2$ gauge field
which exhibits both a confined and deconfined phase, the $U(1)$
theory has only a {\it single} phase\cite{Polyakov}.  In this phase,
point like monopole excitations in $2+1$-dimensional space-time
always proliferate, and drive spinon confinement\cite{RSSuN}.
The electron is, then, ultimately {\it not} expected to be fractionalized
in these theories.   

\subsubsection{Superconducting phases}

We now turn to a description of superconductivity within the
$Z_2$ gauge theory.  
Since the spinons will be gapped
into singlets within the superconducting phase, it is
legitimate to integrate them out, generating once again
a field strength term for the gauge field as
in Eqn.~\ref{Fmunu}.  
When the dimensonless chargon ``hopping" amplitude,
$t_c$, increases and becomes much larger than unity,
one expects the chargons to condense,
$\langle e^{i\phi} \rangle \ne 0$.
For large $K$ so that the gauge field
is effectively frozen, this chargon condensation transition
is simply a $3D$ classical $XY$ transition.
Since the chargon
carries electric charge $e$, in this phase
the charge $U(1)$ summetry is broken, and a Meissner effect
results.  But the chargon also
carries $Z_2-$ charge, so that the $Z_2$ gauge symmetry is also
spontaneously broken.
Within a conventional BCS description of superconductivity,
the order parameter (the Cooper pair) carries charge $2e$,
so one might be tempted to conclude that this ``chargon condensate"
is perhaps some sort of exotic unconventional
superconducting phase.  In particular,
it is not a priori clear that the chargon condensate
can support a conventional $hc/2e$ BCS vortex.  

To highlight the confusion, it's instructive to focus on the regime
with large $K$, where a good description of the ground state
can be obtained by setting $\sigma_{ij} = 1$ on every link,
and taking the chargon phase $\phi_i$ a space-time independent constant.
Consider placing an $hc/2e$ vortex at the (spatial) origin.
Upon encircling this $U(1)$ vortex at a large distance, the phase
of the chargon wavefunction must wind by $\pi$.  This is of
course not possible
with a smoothly varying phase field, but requires the introduction of a
``cut" running from the vortex to spatial infinity
across which the phase jumps by $\pi$.
The energy of this cut is, however, linear in its length
with a line tension proportional to $t_c |\langle e^{i\phi} \rangle |^2$.
It thus appears that $hc/2e$ vortices are themselves confined,
and not allowed in the superconducting chargon condensate.
But imagine changing the sign of
all the $Z_2$ gauge fields, $\sigma_{ij}$,
which ``cross" the cut. 
This corresponds to placing a $Z_2$ vortex at the origin.
These sign changes
``unfrustrate" the $XY$ couplings across the cut,
so that the line tension vanishes.  It is thus apparent
that a bound state of a $Z_2$ vortex and the $hc/2e$ $U(1)$ vortex (in the
phase of the chargon) can exist within the
chargon condensate.  It is this bound state which
corresponds to the elementary BCS vortex in the conventional
description of a superconductor.

It is worth emphasizing that both the ``naked" $hc/2e$ $U(1)$ vortex
and the $Z_2$ vortex  - the vison - are confined in the superconducting phase.
For example, the energy cost to pull apart {\it two} $Z_2$ vortices
also grows linearly with separation.  To see this,
introduce two visons by 
changing the sign of the $Z_2$ gauge field
along an interconnecting ``line".
Due to the chargon condensate which breaks the $Z_2$ gauge symmetry
making the gauge field ``massive",
{\it each} negative bond costs
an energy $4t_c$, implying linear confinement.

\begin{figure}
\epsfxsize=3.5in
\centerline{\epsffile{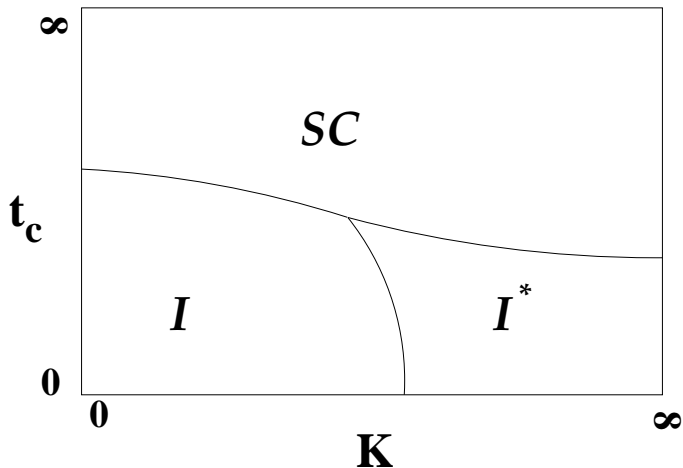}}
\vspace{0.15in}
\caption{Schematic zero temperature phase diagram in the $K-t_c$ plane for 
local $s$-wave pairing with an even number of electrons per unit cell.}
\vspace{0.15in}
\label{fracswvf}
\end{figure}

Thus the distinct massive excitations (apart from the Anderson-Higgs plasma mode
necessitated by the $U(1)$ symmetry breaking) in the chargon condensate are the spinons
and the BCS $hc/2e$ vortices. This is exactly as required in a conventional superconducting phase.
Further, 
since the spinons are minimally coupled to the $Z_2$ gauge field,
there is a long range statistical interaction 
between the spinons and the BCS vortices.  In effect, 
a spinon ``sees" the $Z_2$ vortex  - which is bound to the $hc/2e$
vortex - as a source of ``Ising flux". This too is as required in a 
conventional superconductor. Thus, the chargon condensate does in fact describe
a conventional superconducting phase  - denoted hereafter
as ${\cal SC}$.   

A schematic phase diagram is shown in the $K-t_c$ plane in Fig. \ref{fracswvf}.
The transition from the
fractionalized insulator ${\cal I}^*$ into ${\cal SC}$ is essentially
a superconductor-insulator transition
for the charge $e$ chargons. These exist as finite energy
excitations in ${\cal I}^*$ - superconducting order is
obtained if they condense. 
On the other hand, the transition from
the conventional insulator ${\cal I}$ into ${\cal SC}$ can be
viewed as a superconductor-insulator transition for
charge $2e$ Cooper pairs.  This can be seen by considering
the $K=0$ limit, where it is possible to integrate
out the $Z_2$ gauge field and arrive at an effective
theory of Cooper pair hopping:  
\be 
\label{pairhop}
S_{pair} = -2t_2  \sum_{\langle ij \rangle } cos[2(\phi_i - \phi_j)] .
\ee

\section{Odd number of electrons per unit cell with d-wave pairing}
\label{fracdwv}
Having explored the physics
of electron fractionalization
which follows from the $Z_2$ gauge theory
in the simplest of cases with 
an even number of particles per site in the presence of s-wave pairing correlations, we turn now
to a much more interesting and challenging
situation:  Correlated Mott insulators
with one electron per unit cell
in the presence of local d-wave pairing correlations.
As we shall see, in this case 
the $Z_2$ gauge theory has two simple
limiting regimes - one describing a d-wave superconductor
and the other a conventional antiferromagnetic insulator.
But in the interesting crossover regime between these two limits, 
a number of other phases can be readily described within the $Z_2$ gauge theory.
Besides a spin-Peierls ordered phase, the theory
gives a simple description of the {\it nodal liquid} - an exotic
fractionalized insulator with gapless fermionic quasiparticles.
With one electron per unit cell, 
{\it confinement} transitions out of
the d-wave superconductor or nodal liquid are inextricably linked
to breaking of translational symmetry.  

The full theory of interest can be written as
\bea
\label{IGAodd}
S & = & S_{c} + S_{s}  + S_{B}, \\ 
S_c &=& - 2t_c \sum_{\langle ij\rangle } \sigma_{ij} cos(\phi_i - \phi_j) , \\
S_{s} &=& -\sum_{\langle ij \rangle} \sigma_{ij} (t^s_{ij}  \bar{f}_i f_j +  t^{\Delta}_{ij} f_{i\uparrow}
f_{j \downarrow} + c.c.) - \sum_i \bar{f}_i f_i  .
\eea
As shown in Eqns. \ref{SB} and \ref{Sodd}, with {\it odd} integer $N_0$ 
there is an extra Berry's phase term in the action,
\be
S_B =  -i{\pi \over 2} \sum_{i,j=i-\hat{\tau}} (1-\sigma_{ij}) .
\ee

It is instructive to consider various limiting cases described by the above action.
First consider the limit $t_c = 0$. Then  $S_{c} = 0$, and the 
$\phi$ fields may be trivially integrated out.
Surprisingly, 
the partition function for the remaining spin sector
of the theory is formally equivalent to the
Heisenberg antiferromagnetic spin model.  To demonstrate this we first
trace over the two allowed values of the $Z_2$ gauge field
$\sigma_{ij}$ on each link.
Consider first the {\it spatial} links, which enter the
action in the form,
\bea
S^r_s & = & \sum_{<rr'>}\sum_{\tau}\sigma_{rr'}{\cal A}^{\tau}_{rr'} ,\\
{\cal A}^{\tau}_{rr'} & = &  - t^s_{rr'}( \bar{f}_{r} f_{r'} + c.c.) -  t^{\Delta}_{rr'} (f_{r\uparrow}
f_{r' \downarrow} - (\ua \ra \da) + c.c.) .
\eea
For notational simplicity we have suppressed the $\tau$ index on the fermion fields.   
Tracing over the $\sigma_{rr'}$ fields
for each (independent) spatial link and exponentiating the result
generates a term in the action of the form,
\be
S_r = -\sum_{<rr'>}\sum_{\tau}\ln cosh \left({\cal A}^{\tau}_{rr'} \right) .
\ee
Since ${\cal A}$ is bi-linear in the
fermion fields, upon expanding
in powers of ${\cal A}$ one generates
a series of terms that involve
multiples of four spinons. 

Now consider the trace of $\sigma_{ij}$ along the temporal links. 
Recall that the effect of the
gauge field $\sigma_{i, j = i- \hat{\tau}}$ along the temporal
links is precisely to impose the constraint Eqn. \ref{constr} on the Hilbert space
in a Hamiltonian formulation. With the 
$\phi$ fields integrated out, at $t_c = 0$,  this constraint reduces to requiring
\be
(-1)^{n_{f}} = -1 ,
\ee
at each site of the spatial lattice. 
Due to Pauli exclusion this is
equivalent to the constraint that
$n_{f} = 1$ at each site. Thus, after tracing out the $\sigma$ field, 
the Hamiltonian obtained from $S_r$ is constrained to operate on a Hilbert space with exactly one spinon per site.  This Hamiltonian consists
of a sum of terms for each nearest neighbour spatial link. With the 
additional requirement of spin rotation symmetry, the Hamiltonian must take the form of the 
Heisenberg spin Hamiltonian,
\be
H = J\sum_{<rr'>} \bbox{S}_r \cdot \bbox{S}_{r'} .
\ee
This can be verified directly from $S_r$ by expanding out the
$ln cosh$ term, and re-expressing the spinon operators
in terms of the spin operators, $\bbox{S}_r = f^\dagger_r \bbox{\sigma}
f_r$.  This leads to an explicit expression for the exchange interaction:
\be 
\label{Jval}
J = \frac{1}{\epsilon}\left((t^s)^2 + \frac{(t^{\Delta})^2}{4}\right) ,
\ee 
where $\epsilon$ is the discrete time slice defined in Eqn.~\ref{epsilon}.

Recovering the Heisenberg antiferromagnet in the limit $t_c \ra 0$ provides  
a way to obtain a rough estimate for the 
saddle-point parameter $\chi_0$. First, we note that $t^s$ and $t^{\Delta}$ can be re-expressed
in terms of the parameters $t, u, J, \Delta$ and $\chi_0$ using Eqns. \ref{ts}
and \ref{tDelta}. Though these relations are
strictly valid for $s$-wave pairing, they suffice to give rough estimates even for the $d$-wave case. It is, however,
necessary to modify the equation for $t^{\Delta}$ due to the 
slightly different decoupling in the $d$-wave case (See Appendix \ref{dwaveigt}). Assuming that
the saddle point value $\eta_0 \sim \chi_0$, we get
\be
\label{tDelta-d}
t^{\Delta} \sim \frac{\Delta}{J}t^s  .
\ee
Combining Eqns. \ref{ts} and \ref{tDelta-d} with Eqn. \ref{Jval}
and assuming $\Delta << J$,
leads to an estimate for $\chi_0$,
\be
\chi_0 \sim \left(\frac{tu}{J^2}\right)^{\frac{1}{3}},
\ee
which is appropriate in the limit of large $u/t$.
Having estimated $\chi_0$, one can use Eqns. \ref{tc}, \ref{ts} and \ref{tDelta-d} to obtain estimates for the
three dimensionless coupling constants, 
$t_c$, $t^s$ and $t^\Delta$, respectively.  The resulting 
estimates are given in Eqn.
\ref{couplings}.

Having established the equivalence of the
action in Eqn. \ref{IGAodd} to the Heisenberg antiferromagnet
in the limit $t_c \rightarrow 0$, we briefly consider the opposite
large $t_c$ limit.  With sufficiently large $t_c$
the chargons will condense, and as 
argued in the previous section this describes
a conventional superconducting phase.
But due to the assumed form of the pairing correlations,
the pairing symmetry here will be 
$d_{x^2 - y^2}$.
Thus, the $Z_2$ gauge theory in Eqn. \ref{IGAodd} has
the remarkable property that it describes a conventional antiferromagnet
for small chargon coupling, 
and a conventional $d_{x^2 - y^2}$ superconductor in the opposite extreme.
We now turn our attention to the
exceedingly interesting regime between these two limits.

\subsubsection{Correlated Mott insulators}

When the chargon coupling strength $t_c$ is small, 
the chargons will be gapped out, and the system 
in an insulating phase.  In this case,
it is appropriate to integrate 
out the chargon fields to obtain an effective action for the 
spinons and the gauge field $\sigma$. The main
result  of this integration will be to generate a plaquette product
term of the form,
\be
S_\sigma =  -K \sum_{\Box} [ \prod_{\Box} \sigma_{ij} ]   .
\ee
The full remaining action which
is valid within the insulating phases is then simply,
\be
\label{sfsigma}
S = S_s + S_{\sigma} + S_{B} .
\ee 
The parameter $K$ depends on the coupling $t_c$, 
vanishing at $t_c=0$ and increasing monotonically
with $t_c$.
The transition to superconductivity
will occur when $t_c \sim 1$. Near this limit, but on the 
insulating side, the 
value of $K$ will also be of order one. 
Keeping this in mind, we first find it convenient to analyze the phase diagram of the above action
for {\it arbitrary} $K$, 
incorporating later the superconducting phase.

The action in Eqn. \ref{sfsigma} has three dimensionless
coupling constants, $t^s,t^\Delta$ and $K$.
Considerable progress can be made in determining the phase diagram by focussing on three different limits.
The first, considered above, is 
$K =0$ 
where the model reduces to the Heisenberg spin model. 
The second tractable limit is large $K$.
If $K = \infty$ the 
gauge field is frozen out and it is possible to choose a gauge with 
$\sigma_{ij} = 1$ on every link. Then, the only remaining piece of the action
describes non-interacting spinons with a
gapless ``$d$-wave'' dispersion at four points in the Brillouin zone.
This is the ``nodal liquid" phase - obtained in earlier work\cite{NLI,NLII}
by vortex-pairing within a dual vortex formulation.
The nodal liquid is
a fractionalized insulator
with deconfined, gapless spinons and gapped chargons. 
For large
but finite $K$ and in the absence of $S_{B}$, the $Z_2$ gauge theory is in
its perimeter law phase.  
As we show below, this continues to hold even in the presence of $S_B$ - in fact, the 
region of stability of the perimeter phase is {\em enhanced} by the $S_B$ term. 
Thus, the chargons and spinons remain
deconfined and the nodal liquid phase survives for large but finite $K$.

As with the fractionalized insulator discussed in
Section \ref{fracswv}, apart from the chargons and the
spinons there is another distinct excitation in the nodal liquid phase - the $Z_2$
vortex configuration in the $\sigma$ field, 
dubbed the ``vison". The vison is a gapped 
excitation in the nodal liquid.  As before, due to the minimal coupling of the chargons
and the spinons to the $Z_2$ gauge field $\sigma$, they each acquire a phase of $\pi$ upon encircling a
vison. There is thus a long ranged statistical interaction between a chargon (or a spinon) and a vison.  
  
The third tractable limit of the action Eqn. \ref{sfsigma} is 
small $t^s$ and $t^{\Delta}$.  (Estimates
appropriate to the cuprates obtained from Eqn. \ref{couplings},
suggest that these couplings will most likely be much smaller than one.)
In the extreme limit of $t^s = t^{\Delta} = 0$, we are  
left with a pure $Z_2$ gauge theory described by $S_{eff} = 
 S_\sigma + S_{B}$.  
To explore the effects of the Berry's phase term $S_{B}$
on the gauge theory, it is useful
to pass to the dual representation.  Recall that for $S_B=0$ the dual
theory is simply the $2+1-$dimensional Ising model, with the Ising
spin operators  
($v_i = \pm 1$)
creating the ``vison" excitations.  
To implement the duality transformation with the Berry's phase term present, it is convenient to first rewrite it as,
\be
\label{Berry-2}
S_{B} = i{\pi \over 4} \sum_{\langle ij \rangle} (1-\sigma_{ij})
(1- \prod_{\Box} \mu^{ext}_{ij} )  .
\end{equation}
Here $\mu^{ext}_{ij}$ can be viewed as an
``external" $Z_2$ gauge field living on the links
of the dual lattice,
which satisfies $\prod_{\Box} \mu^{ext}_{ij} = -1$ 
through every {\it spatial} plaquette.  In this form one can readily
generalize the duality transformation in Appendix \ref{ISD} to give,
\be
S_{dual} = - K_d \sum_{\langle ij \rangle} v_i \mu^{ext}_{ij} v_j  ,
\ee
with dual coupling satisfying; $tanh(K_d) = e^{-2K}$.
Due to the Berry's phase term, every spatial plaquette
(with normals along the time direction) in the dual Ising model
is {\it frustrated}.  In the time continuum limit this becomes
a 2d quantum transverse-field Ising model
which is
{\it fully frustrated}.  

The quantum Ising model on a fully frustrated square lattice
has been studied extensively by several authors\cite{FTFIM1,FTFIM2}. 
In particular, Jalabert and Sachdev\cite{FTFIM2} studied the model numerically (not coincidentally)
in the context of frustrated quantum Heisenberg spin models.
For small $K_d$ the Ising model
exhibits the usual paramagnetic phase, in which
the visons are gapped (uncondensed) with $\langle v_i \rangle =0$.  
This corresponds to the ``low temperature" phase
of the gauge theory.  Deep within this phase one can
set $\sigma_{ij}=1$ on all the links, which implies (for $t^s, t^{\Delta} \neq 0$)
that the chargons and spinons are {\it deconfined}.  This is 
the nodal liquid phase discussed earlier. 
It is noteworthy that the frustration in the Ising
model  - which is a direct consequence of 
being in a Mott insulator with one electron per site - 
{\it enhances} the stability of the fractionalized nodal liquid phase 
(the paramagnetic phase of the Ising model).

\begin{figure}
\epsfxsize=3.5in
\centerline{\epsffile{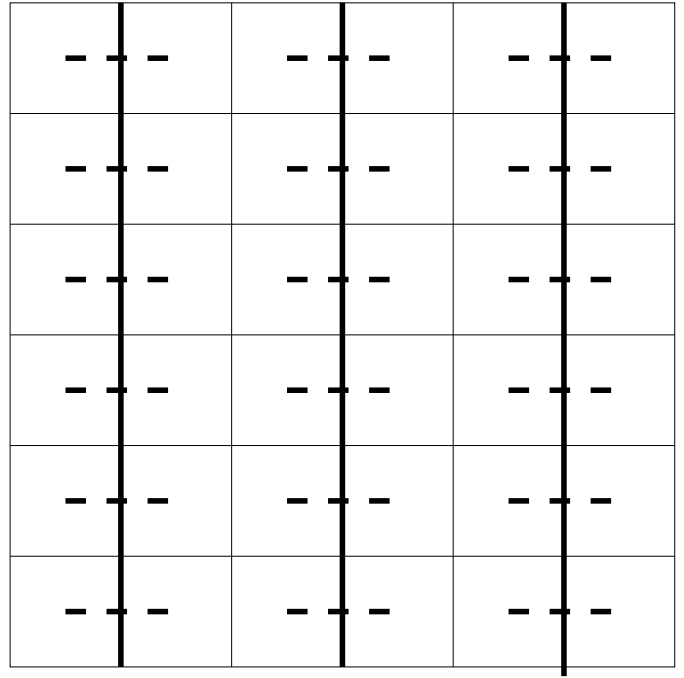}}
\vspace{0.15in}
\caption{One possible ordered low temperature phase of the fully frustrated transverse field Ising model
in two spatial dimensions. The thick lines represent the frustrated bonds. The dashed lines 
denote the links of the dual lattice where the corresponding ``singlet bonds'' live.}
\vspace{0.15in}
\label{colsp}
\end{figure}

As $K_d$ is increased, it has been found\cite{FTFIM2} that the Ising model
orders - breaking the global $Z_2$ spin flip symmetry.
But due to the frustration, this ordering is accompanied by a spontaneous
breaking of translational symmetry.  It is convenient to 
characterize this symmetry breaking in terms of 
the gauge-invariant energy densities of the
near-neighbor bonds:  ${\cal E}_{ij} = -\langle v_i \mu^{ext}_{ij}
v_j \rangle$.  It is found that some of the bonds are ``frustrated"
with positive ${\cal E}_{ij}$, while the remaining are ``happy" with
negative bond energies.  In the spatially broken ordered phases, it is found
that these frustrated bonds form
lines (see Fig. \ref{colsp}), which run along the principle axis
of the square lattice (columns or rows).  There are four
favored configurations,
corresponding to frustrated bonds along
every other column, or along every other row. 
Within each of these phases, a particular gauge choice
can be made with  
$\mu^{ext}_{ij} = -1$ on each ``frustrated" bond.
With this choice of gauge, the Ising spins, $v_i$, exhibit
a global ferromagnetic ordering.  Altogether,
the ground state is thus eight-fold degenerate and breaks the $Z_2$ spin flip, translational and 
rotational symmetries. 

In general, several other ordered phases of the fully frustrated 
Ising model are possible - some of these 
are explored in the Landau theory of the first reference in Ref. \cite{FTFIM1}. These phases may 
perhaps be stabilized by very large $K_d$, and/or longer ranged interactions 
beyond the simplest nearest neighbour model studied in Ref. \cite{FTFIM2}. 
We will not consider these other possibile phases in the present paper.

What are the effects of a small non-zero $t^s$ and $t^{\Delta}$ which
couple the spinons
to the $Z_2$ gauge field?
In the context of quantum antiferromagnets, Sachdev\cite{FTFIM2,subir_pc}
has suggested that the {\it spatial} ordering of the Ising model corresponds to
a spin-Peierls ordering.  This interpretation appears to be consistent
within our present framework.
Specifically, associated with each frustrated bond in
the Ising model, is a corresponding frustrated plaquette
on the dual lattice ``pierced" by that bond.  The expectation
value of the plaquette product in the gauge theory
will therefore be modulated
in these ordered phases, with $\langle \prod_{\Box} \sigma_{ij} \rangle \approx -{\cal E}_{ij}$.
Upon including the coupling to the spinons, this modulation of the energy density will, in general, 
induce a modulation in various other physical quantities. In particular, the quantum expectation 
value $<{\bf S}_r \cdot {\bf S}_{r'}>$ evaluated for each bond will be spatially modulated - bonds which 
``cross'' the frustrated lines of the dual lattice will have a different value for this expectation value from 
other bonds. Presuming the spin rotation invariance remains unbroken, this state corresponds to a 
spin-Peierls phase - which we denote as ${\cal SP}$. The ``singlet bonds'' in this phase are 
arranged in a columnar fashion - running perpendicular to the lines of frustrated bonds in the dual 
Ising model as depicted in Fig. \ref{colsp}.

Since the Ising spins in the fully frustrated Ising model
order ferromagnetically in these modulated phases
(with an appropriate gauge choice for $\mu^{ext}_{ij}$)
implying a vison condensation, $\langle v_i \rangle \ne 0$,
confinement is expected.  To see this, consider
evaluating the Wilson loop correlator
defined in Eqn. \ref{Wil}.  In the dual frustrated Ising model,
this corresponds to changing the sign of all the
Ising couplings on bonds which ``pierce" throught the loop.
Being ferromagnetically ordered, this will cost
an energy (action) proportional to the area
of the loop  - the signature of confinement.
Thus, as expected, the spin-Peierls state 
is a conventional insulator, with confined
spinons and chargons.  The gapped spin-one excitations
made by breaking the singlet bonds can then be thought
of as a (confined) pair of spinons.

The three limiting cases discussed above suggest the phase diagram shown in Fig. \ref{fracdwvf}
for the action in Eqn. \ref{sfsigma}.  Consider first the regime with small 
$t^s$ and $t^{\Delta}$.
At very small $K$ a conventional
antiferromagnetic insulator is expected. With increasing $K$ there is 
presumably a phase transition 
into a conventional spin-Peierls insulator with confined chargons and spinons. 
Upon further increasing $K$,
the spin Peierls phase undergoes a {\it deconfinement}
transition into the fractionalized nodal liquid phase.
For large $t^s$ and $t^\Delta$, the antiferromagnet and nodal liquid phases 
will still be present in the limit of very small and large $K$, respectively.
But it is not clear which phases will
be present when all three
of the coupling constants are of order one.
In particular, it is unclear whether
it is possible to have a direct second order phase
transition from the antiferromagnet into the nodal liquid ,
or whether there will always
be an intervening (spin-Peierls) phase.

\begin{figure}
\epsfxsize=3.5in
\centerline{\epsffile{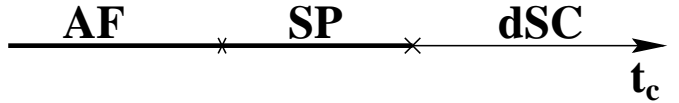}}
\vspace{0.15in}
\caption{Schematic zero temperature phase diagram showing one possible scenario for the evolution
from the antiferromagnet ($AF$) to the $d$-wave superconductor $dSC$. In this scenario, all the insulating phases are conventional.
The thick lines indicate confinement of the chargons and spinons. For concreteness, we have chosen to display a 
particular sequence of confined phases, namely, a transition from $AF$ to a spin-Peierls ($SP$) insulator, 
and a further transition to $dSC$.}
\vspace{0.15in}
\label{afscconf}
\end{figure}

\begin{figure}
\epsfxsize=3.5in
\centerline{\epsffile{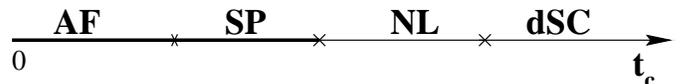}}
\vspace{0.15in}
\caption{The other qualitatively different scenario for the evolution from the antiferromagnet to the 
$d$-wave superconductor. In this case, on increasing $t_c$, a transition to the 
fractionalized nodal liquid ($NL$) phase occurs before the onset of superconductivity.}
\vspace{0.15in}
\label{afscdeconf}
\end{figure}

We now discuss the implications of these results for the phase diagram of the full $Z_2$ gauge theory in which
the charge degrees of freedom are present
and superconductivity is possible.
Of primary interest is the evolution from the antiferromagnet
to the $d-$wave superconductor upon increasing the chargon
coupling, $t_c$.  A transition into the superconductor is expected to occur at some critical
chargon coupling, $t_c^*$, of order one.  For smaller $t_c$
in the insulating regime, the dimensionless coupling $K$ will
at most be of order one.  
One can imagine two qualitatively distinct possibilities 
upon tuning towards the superconductor
from the insulating phases.  First, it may be that even when $t_c$ increases to $t_c^*$, the value of $K$ will remain {\it smaller} than the 
critical value needed for deconfinement, $K_c$. In this case, all the intermediate phases between the 
antiferromagnet and the superconductor will be conventional confined phases. This is illustrated 
in Fig. \ref{afscconf}. Alternately, it 
may be that $K$ exceeds $K_c$ {\it before}
the onset of superconductivity.  This would
imply the existence of the deconfined nodal liquid phase 
intervening between the $d-$wave superconductor
and a conventional insulator. This is illustrated in Fig. \ref{afscdeconf}.

Which one of these two possibilities is realized will 
presumably depend sensitively
on microscopic details. 
Indeed,  
since $K$ is of order one when $t_c$ approaches $t_c^*$,
it seems likely that the onset of superconductivity 
will occur close to the boundary between the confined and deconfined 
insulating phases.  
But in any event,
our analysis has firmly established
the {\it possibility} of the deconfined nodal liquid phase.
It remains as a challenge to determine whether this
exotic fractionalized insulator is realized in the cuprates.

In the next section, we will describe much of the physics discussed here
in a dual formalism in terms of vortices rather than the chargons. This will provide
considerable further insight, and make connections with earlier approaches.

\section{Dual vortex representation}
\label{DVT}
For a system of interacting bosons in two spatial dimensions, it is well-known that
the insulating phases can be described as a condensate of vortices in the 
many particle wave-function. More formally, it is possible to set up a dual 
description where the vortices, rather than the particles, 
are the fundamental degrees of freedom such that the insulating phase is a 
vortex condensate while the superfluid phase is the vortex vacuum.
For the electronic systems considered in this paper, 
it is natural to attempt to 
do the same, and work with a dual description in terms of vortices in the Cooper
pair phase $\varphi$, and the spinons. 
Since the Cooper pair
has charge $2e$, these are the $\frac{hc}{2e}$
vortices which occur in a conventional
superconductor.
Besides providing additional insight into the mechanism
and nature of electron fractionalization, passing to a dual vortex description 
enables us to make contact with earlier work which 
describes fractionalization in terms of vortex pairing. 
 
We will start with the full
chargon-spinon action $S = S_c + S_s + S_B$
discussed in the last section, and  
perform a duality transformation to trade the chargon fields for the 
$\frac{hc}{2e}$ vortices. This differs somewhat from the conventional duality 
transformation\cite{xydual} from bosons to vortices due to the coupling of the chargons to the 
$Z_2$ gauge field.

To understand how to deal with the chargon coupling to the $\sigma$ field, 
it is useful to first review the well-known self-duality of the $Z_2$ gauge 
theory with Ising matter fields 
in $2+1$ dimensions. This is done in detail in 
Appendix \ref{ISD}. The duality proceeds by first rewriting the partition function
in terms of a $Z_2$ {\it current} for the Ising matter fields and the $Z_2$ gauge field,
$\sigma_{ij}$. The $Z_2$ current lives on the links
of the lattice and can take one of 
two values $0,1$. 
It is
conserved modulo $2$ at each site of the lattice. This conservation law can be implemented
by writing the $Z_2$ current as the flux of a dual $Z_2$ gauge field,
denoted as $\mu_{ij}$. (This is completely
analogous to the duality of the three dimensional classical $XY$ model). Eliminating the 
$Z_2$ current in favor of the dual gauge field gives an action written entirely in terms of 
two $Z_2$ gauge fields ($\sigma_{ij}$ and $\mu_{ij}$) which are duals of each other. The original $Z_2$ gauge field, $\sigma_{ij}$
may be eliminated by expressing its flux as the current of a dual Ising matter field, the vison $v_i$.
The resulting theory has the same form as the original $Z_2$ gauge theory with matter fields,
but is dual to it. 

To obtain a dual representation of the system of chargons and spinons 
coupled to 
the $Z_2$ gauge field $\sigma_{ij}$, we need to combine the dual representation of the $Z_2$ gauge theory 
with the standard duality transformation of the $XY$ model. As shown in detail in 
Appendix \ref{u1z2d},
this is readily done. For the time being, we will only consider the situation with local $d$-wave
pairing and an odd number of electrons per unit cell.
The result is a lattice action in terms of 
$\frac{hc}{2e}$ vortices,
which are minimally coupled to a fluctuating $U(1)$ gauge field
$a$ whose circulation is the total electrical current. In addition, the $hc/2e$
vortices are minimally coupled to
a $Z_2$ gauge field $\mu_{ij}$. The full action is given by 
\bea
\label{sdual}
S & = & S_{v} + S_a +S_s + S_{CS} + S_B, \\
S_{v} & = & -t_{v}\sum_{<ij>}  \mu_{ij}  cos(\theta_i - \theta_j + \frac{a_{ij}}{2}), \\
S_a & = & \frac{\kappa}{8\pi^2} \sum_{\Box} (\Delta \times a_{ij})^2 ,\\
S_s &=& -\sum_{\langle ij \rangle} \sigma_{ij}[t^s_{ij}   
\bar{f}_i f_j +  t^{\Delta}_{ij}  f_{i\uparrow}
f_{j \downarrow} ] - \sum_i\bar{f}_i f_i ,\\
S_{CS} & = & \sum i\frac{\pi}{4}(1- \prod_{\Box}\sigma)(1 - \mu_{ij}) .
\eea 
Here $e^{i\theta_i}$ creates the $\frac{hc}{2e}$ vortex, and $f_i$ is the spinon
as before. The first term represents single vortex hopping,
while the second is a kinetic term for the 
$U(1)$ gauge field $a_{ij}$.  
The flux of $a$ is the total electrical current - in particular
a flux of $2\pi$ through a spatial plaquette adds an electric charge of one - 
a ${\em chargon}$. 
Together these two terms
comprise the usual dual vortex representation of a set of
charge $2e$ Cooper pairs, except that here the vortices
are minimally coupled to an {\it additional} $Z_2$ gauge field
$\mu_{ij}$.  This leads to a vortex-spinon coupling
mediated by $S_{CS}$.       
This term has a structure very similar to 
a Chern-Simons term (although it is for the group $Z_2$), and as discussed below, plays a similar
role. 
The Berry's phase term $S_B$ is the same as before.

The full dual action is invariant under a local
$U(1)$ gauge transformation
\bea
\theta_i & \ra & \theta_i + \Lambda_i, \\
a_{ij} & \ra & a_{ij} - \frac{\Lambda_i - \Lambda_j}{2} 
\eea
This is standard in the dual vortex description of $XY$ models in three dimensions.
The corresponding conserved charge is the vorticity. 
The action has an additional $Z_2$ gauge symmetry under which
\be
e^{i\theta_i}  \ra  \epsilon_ie^{i\theta_i} ; \hskip0.4cm
\mu_{ij}  \ra  \epsilon_i  \mu_{ij} \epsilon_j.
\ee
with $\epsilon_i = \pm 1$. We emphasize that this gauge symmetry is 
distinct from the local $Z_2$ gauge symmetry of the spinon-chargon action,
but in fact is dual to it.

To get some intuition for the term $S_{CS}$, it is instructive
to replace the vortex hopping term in the action by a Villain potential,
\be
e^{t_v cos\Theta_{ij}}  \rightarrow  \sum_{J_v = - \infty}^\infty
e^{-J_v^2/2t_v} e^{iJ_v \Theta_{ij}}  ,
\ee
where $\Theta_{ij} = \theta_i - \theta_j + {a_{ij} \over 2} + {\pi \over 2} (1 -
\mu_{ij})$ is the gauge invariant phase difference.
Here the integer field $J_v$ that lives on the links of the lattice
represents the 3-current of the $hc/2e$ vortices.
After this replacement it is possible to explicitly perform the summation
over the gauge field $\mu_{ij}$.
For each link of the lattice this contributes
a term to the partition function of the form,
$1+ (-1)^{J_v} \prod_{\Box} \sigma$, which vanishes unless
\be
(-1)^{J_v} = \prod_{\Box}  \sigma  .
\ee
Thus, the Chern-Simons term has effectively attached
a $Z_2$ flux of the gauge field $\sigma$  - a ``vison" - to each
$hc/2e$ vortex.  As discussed in Section \ref{fracswv},
this composite comprised of an $hc/2e$ vortex
bound to the $Z_2$ vison is nothing but the familiar
BCS vortex.  Due to the attached vison, when
a spinon is taken around the BCS vortex
it acquires the expected $\pi$ phase factor.

Alternatively, it is possible to perform an ``integration by parts"
on $S_{CS}$ which effectively exchanges the role of $\sigma$ and $\mu$,
and then perform a summation over $\sigma$.  This leads to the
additional constraint,
\be
(-1)^{J_f} = \prod_{\Box} \mu  ,
\ee
with $J_f$ the spinon 3-current.  A
$Z_2$ flux in the gauge field $\mu$ has thereby been attached
to each spinon. 
More precisely,
since the spinon number is only conserved
modulo $2$ due to the anomalous pairing term, the $Z_2$ flux is attached 
whenever an {\it odd} number of spinons propagates.
The net effect of this $Z_2$ Chern-Simons term is to implement mathematically
the long-ranged statistical interaction between BCS vortices
and spinons.  
This kind of flux attachment may be familiar to many readers for the $U(1)$ 
group from theories of the quantum Hall effect.   
But since
the spinon number itself is not conserved,
implementing this statistical interaction with a $U(1)$ Chern-Simons
term is problematic.  It is a remarkable aspect of the duality
transformation in Appendix \ref{u1z2d}, that this 
Ising-like Chern-Simons terms emerges so naturally.

\subsection{Phases}

We now analyze the phases in this dual
vortex description, focussing on the most interesting
case of an odd number of electrons per site with local
d-wave pairing correlations.
In the vortex description the superconducting phase
corresponds to a vortex vacuum, and
the insulating phases are vortex condensates.
We consider first two simple limiting cases,
firstly the superconductor with vanishingly small vortex hopping
$t_v \rightarrow 0$, and then the insulator with
$t_v \rightarrow \infty$.

When $t_v$ is zero the summation over the gauge field
$\mu$ can be performed, giving the constraint
$\prod_{\Box} \sigma =1$.  It is then possible
to pick a gauge with $\sigma_{ij}=1$ on every link.
The resulting action has two pieces, $S_a$ which
describes the gapless sound mode of the superconductor
(gapped when long-ranged Coulomb interactions
are included) and the spinon piece $S_s$.
With $\sigma_{ij}=1$ the spinons can freely propagate
and describe the gapless nodal quasiparticles.
A correct description of a conventional d-wave
superconductor is thereby recovered.

Consider next the opposite limit with $t_v \rightarrow \infty$.
In this regime the $hc/2e$ vortices will
condense, $\langle e^{i\theta_i} \rangle \ne 0$.
The dual ``Anderson-Higgs" mechanism
leads to a mass term for the gauge field $a_{ij}$,
indicative of a charge gap.  
With one electron per unit cell
the resulting phase is thus a Mott insulator.
In the absence of any gapped charge excitations ($\Delta \times a=0$),
it is possible to choose a gauge with $a_{ij}=0$ on every link.
The vortex hopping term becomes, $S_v = -h\sum_{ij} \mu_{ij}$,
with a non-zero ``field":
$h = t_v |\langle e^{i\theta_i} \rangle|^2$.
When this ``field" is large one can set $\mu_{ij}=1$ on each link,
so that the Chern-Simons terms vanishes.  The full action then
reduces to $S_{eff} = S_s + S_B$.  
At this stage the summation over the $\sigma$ gauge field
can be performed explicitly.  As detailed in the previous section,
the resulting model reduces to a simple
2d near-neighbor Heisenberg antiferromagnet.  Thus, we readily
recover the simple antiferromagnet from the dual representation
by condensing
$hc/2e$ vortices. 

Finally, in this section we wish to recover
a dual description of the fractionalized
``nodal liquid".  Since the nodal liquid is electrically insulating
it requires vortex condensation.  But as established in the previous section, 
the nodal liquid supports gapped $Z_2$ vortices - the``vison" excitations.
Since the Chern Simons term attaches a vison to each
$hc/2e$ vortex, it is clear that to obtain the nodal liquid
the $hc/2e$ vortices cannot be condensed.
But since the {\it square} of the vison operator 
is unity ($v_i^2 =1$), a {\it pair} of $hc/2e$ BCS vortices
does not carry a vison with it.  As we now show,
the nodal liquid can be obtained
from the d-wave superconductor by {\it pairing}
BCS vortices, and then condensing the $hc/e$ vortex composite.

To this end, we add an extra vortex pair hopping term to the
action,
\be
S_{2v} = -t_{2v} \sum_{\langle ij \rangle}
cos(\theta_{2i} - \theta_{2j} + a_{ij})   .
\ee
Here, $e^{i\theta_{2i}} = (e^{i\theta_i})^2$,
thus creating a pair of BCS vortices.  Notice that 
the $hc/e$ vortex is also minimally coupled to the $U(1)$
gauge field - as required by the dual $U(1)$ symmetry
of the action - but is {\it not} coupled to the $Z_2$ gauge field,
$\mu_{ij}$, because it carries no vison charge.
We now consider taking $t_{2v}$ large
and condensing the $hc/e$ vortex, $\langle e^{i\theta_{2i}} \rangle \ne 0$,
keeping the $hc/2e$ vortex uncondensed.
Before doing this it is convenient to
re-express the $hc/2e$ vortex as 
\be
e^{i\theta_i} = v_i e^{i\theta_{2i}/2}   ,
\ee
with $v_i= \pm 1$ the vison operator.
Notice that with this identification the field
$\theta_2$ can be treated as an angular variable,
since the right side is invariant under the
combined transformation, $\theta_2 \rightarrow \theta_2 + 2\pi$
and $v_i \rightarrow - v_i$.
We finally find it convenient
to absorb the field $\theta_{2i}$ into the gauge field $a_{ij}$
by the gauge transformation,
\be
a_{ij} \ra a_{ij} + \theta_{2i} - \theta_{2j} .
\ee
In this gauge, the vortex hopping terms become
\bea
\label{vishop}
S_{v} & = & -t_{v}\sum_{ij} v_i \mu_{ij} v_j 
cos\left(\frac{a_{ij}}{2}\right) , \\ 
S_{2v} & = & -t_{2v} \sum_{ij} cos(a_{ij}) .
\eea 

In the insulating phase with large $t_{2v}$
there will again be a charge gap
due to the dual Anderson-Higgs mechanism,
coming from the $hc/e$ vortex condensate.
Above the gap will be charge $e$ chargons, corresponding
to a 
$2\pi$ flux tube in $a_{ij}$.  In the absence of
any charged excitations one can set $a_{ij} =0$,
and the single vortex hopping term becomes,
\be
S_{v} = -t_v \sum_{\langle ij \rangle} v_i \mu_{ij} v_j .
\ee
The full effective action is
$S_{eff} = S_v + S_s + S_{CS} + S_B$.
When $t_v$ is small the visons will be uncondensed
$\langle v_i \rangle =0$.  In this limit the summation
over the $\mu$ gauge field can be performed, and
due to the Chern Simons term leads to the
constraint, $\prod_{\Box} \sigma =1$.
One can then choose a gauge with $\sigma_{ij}=1$ on each link,
which sets $S_B =0$.  The only remaining term
in $S_{eff}$ 
describes free propagating spinons.  These are the gapless
nodons in the insulating nodal liquid.

We thereby recover a decription of the nodal liquid
from the dual vortex formulation.  In addition to the
gapless nodons, the nodal liquid supports two
gapped excitations; the chargon and the vison.
As clear from the above analysis,
the vison is simply a remnant of the $hc/2e$ BCS vortex
which survives into
the nodal liquid upon condensation of the $hc/e$
vortex pair.  Physically, since the vorticity is only
conserved modulo 2 (in units of $hc/2e$) once the field $e^{i\theta_{2i}}$ has condensed, only a conserved $Z_2$ remains
from the $hc/2e$ BCS vortex.
As before, the vison picks up a $\pi$ phase change
when it is transported around either a spinon or a chargon.
To see this, note that
a chargon corresponds
to a $\pi$ flux in $a_{ij}/2$ and the nodon (spinon)
a $\pi$ flux in $\mu_{ij}$.
As seen in Eqn. \ref{vishop}, the vison is minimally coupled to {\it both} of these
gauge fields, thus acquiring a sign change upon encircling
the spinon or chargon. 

It is worth emphasizing that a clear mechanism
for vortex pairing can be found from the analysis
in the previous section.  Since the chargons
and visons (or vortices) have a long-ranged statistical
interaction, motion of the charge is greatly
impeded by the presence of unpaired visons. 
On the other hand, once the $hc/2e$ vortices
are paired, the charge can move coherently. 
Thus, the presence of a large kinetic energy makes 
vortex pairing energetically favorable.

It is finally worth mentioning that in the limit
$S_s =0$, one readily recovers the fully frustrated
Ising model considered in the previous section.
To see this, note first
that $S_B$ can be re-written
in the form of a Chern-Simons terms with $\mu$ replaced by $\mu^{ext}$,
where $\prod_{\Box} \mu^{ext} = -1$ through all spatial plaquettes.
With $S_s=0$, one can then perform the summation over the $\sigma$ gauge field,
and this sets $\mu_{ij} = \mu^{ext}_{ij}$.
The remaining term in $S_{eff}$ is the fully frusrtated
Ising model:
\be
S_v = - t_v \sum_{\langle ij \rangle} \mu^{ext}_{ij} v_i v_j   .
\ee

\section{Doping}
\label{Dop}
Our analysis has so far focussed only on situations with an integer number, $N_0$,
of electrons per unit cell. Finite doping leading to non-integer
$N_0$ does not crucially modify our discussion of fractionalization issues. 
Indeed, both confined and fractionalized insulating phases can exist for non-zero doping.
At a qualitative level, in both kinds of insulating phases, 
the main effect of non-integer $N_0$ will be to induce charge order,
accompanied by translational symmetry breaking. The precise nature of this charge order presumably
depends on the details of the system, and may be sensitive to the presence of long-ranged
Coulomb interactions. 

Formally, non-integer values of $N_0$ can be incorporated into either
the particle or vortex representations as follows. In the particle representation,
as discussed in Section \ref{Mod}, the main effect of non-integer $N_0$ is to 
modify the Berry phase term to 
\be 
S_{B} = -i\sum_{i, j = i - \hat{\tau}}N_0 ( 2\pi l_{ij}- \frac{\pi}{2} (1 - \sigma_{ij})) .
\ee  
Here, $l_{ij}$ is an integer defined on each temporal link given by
\be
l_{ij} = Int\left[\frac{\Phi_{ij}}{2\pi} + \frac{1}{2}\right]  ,
\ee
where $\Phi_{ij} = \phi_{i} - \phi_{j} + \frac{\pi}{2}(1-\sigma_{ij})$ is the 
gauge-invariant phase difference between two sites. 
When $N_0$ is not an integer, this Berry phase
term leads to {\em complex} Boltzmann weights in the partition function
sum. This is 
not too surprising- even in the absence of any gauge field coupling, the partition function 
for simple Boson Hubbard models
at arbitrary chemical potential involves complex weights.
 
The presence of such complex weights does not pose a problem for the existence 
of the fractionalized insulator.  We recall that the fractionalized phase is obtained
when the gauge field $\sigma_{ij}$ is in its perimeter phase. Deep in this
phase, we may set $\sigma_{ij} \approx 1$ on each space-time link so that
the Berry phase term $S_B$ becomes independent of $\sigma_{ij}$. The resulting action then 
describes a lattice model of bosonic chargons at filling $N_0$ and the 
fermionic spinons, decoupled
from one another. Thus the chargons and spinons will still be deconfined.
However, 
the ground state will generally exhibit charge ordering accompanied by broken 
translational invariance.  Confined conventional insulating phases at non-integer $N_0$ clearly 
also exist. 

Numerical simulations of the $Z_2$ gauge theory at arbitrary $N_0$ to determine the 
precise nature of the charge ordering in these insulating phases will be seriously
hampered by the presence of these complex weights in the partition function. 
Fortunately, in the dual vortex representation, non-integer $N_0$ enters in a 
more innocuous manner.  To generalize the duality transformation
to arbitrary $N_0$ is straightforward, because the Villain
representation of the chargon hopping term in 
Eqn. \ref{svill} 
is simply modified to read,
\be
\frac{\kappa}{2}\sum_{<ij>} (J_{ij} - 2\pi N_{ij})^2 .
\ee
Here, $N_{ij} =  N_0$ for temporal links, and is zero otherwise. 
Proceeding with the duality transformation gives the action,
\be
\label{sduald}
S  =  S_{v}  +S_s + S_{CS} + \tilde{S}_a,
\ee
where the first three terms are the same as before in Eqn.~\ref{sdual}.
The last term, which was equal to $S_a + S_B$ for integer
$N_0$, becomes instead,
\be
\tilde{S}_a  =  \frac{\kappa}{8\pi^2}\sum_{\Box} (\Delta \times a_{ij} - 2\pi N_{ij} )^2 .
\ee
Notice that in this dual
representation, $N_0$ acts like an external ``magnetic field"
piercing each spatial plaquette.

For the particular case of odd integer $N_0$, it is instructive to see how the 
term $S_B$ may be recovered. To that end, 
we define a new ``external'' gauge field $a^{ext}$ on the links of the dual lattice
such that, 
\be
\Delta \times a_{ij}^{ext} = 2\pi N_{ij} .
\ee
We now absorb $a^{ext}$ into $a$ by the shift $ a \ra a - a^{ext}$. This 
eliminates 
$a^{ext}$ so that $\tilde{S}_a \rightarrow S_a$, but modifies 
the vortex hopping term which becomes,
\be
S_v = -t_{v}\sum_{<ij>}  \mu_{ij}  cos\left(\theta_i - \theta_j + \frac{a_{ij} + a^{ext}_{ij}}{2}\right).
\ee
For odd integer $N_0$ (say $N_0=1$)
one may choose,
\be
a^{ext}_{ij} = 2\pi n_{ij} ,
\ee
with {\it integer} $n_{ij}$,
which satisfies $\Delta \times n_{ij} = N_0=1$ for every {\em spatial} plaquette
and is zero for all other plaquettes. 
With this choice we may write,
\be
S_v = -t_{v}\sum_{<ij>}  \mu_{ij}\mu^{ext}_{ij}  cos(\theta_i - \theta_j + \frac{a_{ij}}{2}),
\ee
where $\mu^{ext}_{ij} = (-1)^{n_{ij}}$. Notice that the flux $\prod_{\Box} \mu$ is $-1$
for every spatial plaquette, and zero for other plaquettes. 
If we now perform the shift, $\mu \ra \mu \mu^{ext}$, the
field $\mu^{ext}$ is eliminated from $S_v$
but reappears in $S_{CS}(\mu \mu^{ext})$.  But upon
noting the form of the Berry's phase term in Eqn.~\ref{Berry-2},
one can easily demonstrate that $S_{CS}(\mu \mu^{ext}) = S_{CS}(\mu) + S_B$.
We thereby recover the earlier Berry's phase form for
the case with odd integer $N_0$.

The dual representation for arbitrary $N_0$ is simpler looking than the one in the particle formulation, 
and is probably better suited to discuss
issues such as the nature of charge ordering at finite doping. In particular, 
if we ignore the coupling to the spinons and set $\prod_{\Box} \mu =1$,
the remaining partition function sum involves only {\it real} weights,
and can presumably be evaluated numerically.

\section{Other exotic fractionalized phases}
\label{SCstar}
In this section we will briefly explore the possibility of obtaining 
other fractionalized phases different than the ones discussed so far.
The most interesting phase that emerges is a novel 
fractionalized {\it superconductor} - we will describe its properties in both
the particle and vortex formulations.

\subsection{Particle description}
\label{SC*p}
In earlier sections we argued that when the charge $e$
chargons condense, the resulting phase is
a conventional superconductor. This is perhaps surprising, since
in a conventional BCS description the order parameter carries
charge $2e$. 
One might ask whether it is possible to have a superconducting phase
in which the chargon {\em pairs} ({\em i.e} the Cooper pairs) have condensed, 
while single chargons have not.
As we now demonstrate,
such a superconducting phase - which
we denote as ${\cal SC}^*$ - can exist and has a surprisingly simple
description in terms of our $Z_2$ gauge theory.
For simplicity, we will initially present the discussion for $s$-wave pairing with an
even number of electrons per unit cell. 

The appropriate action from Eqn`s. \ref{Ss-even} and \ref{Fmunu} in Section III,
takes the form $S=S_c + S_s + S_K$.
As discussed there, the kinetic term for
the gauge field $S_{\sigma}$, although 
not present in the original action, will in any case be generated
upon integrating out high-energy modes.
To access the chargon pair condensate phase,
it is extremely convenient to add 
an explicit pair hopping term
to the action, $S_{pair}$ from Eqn. \ref{pairhop}.
For large pair-hopping amplitude, $t_2$,  
the chargon pairs will condense,
leaving the single chargons uncondensed:
\be 
\langle e^{2i\phi} \rangle \ne 0  ; \hskip0.4cm
\langle e^{i\phi} \rangle = 0    .
\ee
This still breaks the global $U(1)$ charge symmetry,
and so describes a superconductor, but one with
rather exotic properties.  
To examine this phase
it suffices to take $t_2 \rightarrow \infty$
which allows one to set $2 \phi_i$ equal to $2\pi$ times an integer,
or equivalently,
\be
\phi_i = {\pi \over 2} (1-s_i)   ,
\ee
with the value of the Ising spins, $s_i =\pm 1$, arbitrary.
In this limit, the chargon creation operator
equals the Ising spin: $e^{i\phi_i} = s_i$.
After integrating out the massive spinons, this leaves
an effective theory of the form:
\be
\label{igm} 
S_{I-gauge} = -2t_c \sum_{\langle ij \rangle} s_i \sigma_{ij} s_j
-K \sum_{\Box} [\prod_{\Box} \sigma_{ij} ] ,
\ee
with $t_c$ the chargon ``hopping" strength.

\begin{figure}
\epsfxsize=3.5in
\centerline{\epsffile{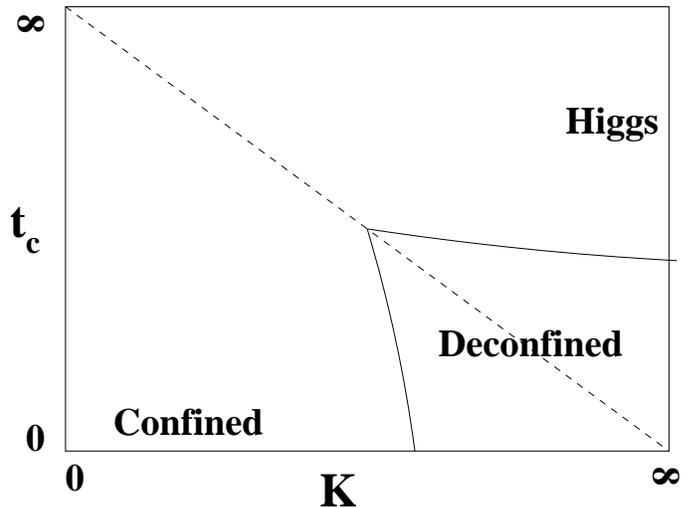}}
\vspace{0.15in}
\caption{Schematic zero temperature phase diagram for the $Z_2$ gauge theory coupled to 
matter fields described by the action Eqn. \ref{igm}.}
\vspace{0.15in}
\label{z2gpm}
\end{figure}

This theory, which describes Ising spins ``minimally coupled"
to a $Z_2$
gauge field, has been extensively studied by Fradkin and Shenker\cite{FrSh}
as a toy model of confinement.  The phase diagram in the $t_c-K$ plane
is shown in Fig. \ref{z2gpm}.  In the $K \rightarrow \infty$ limit
the model reduces to a global Ising model for the spins.
With increasing $t_c$ there is an Ising transition
into a phase with $\langle s_i \rangle \ne 0$ (the ``Higgs" phase),
which corresponds to the chargon-condensed ${\cal SC}$ phase.
Along the $t_c =0$ axis the pure $Z_2$ gauge field exhibits
a confinement transition with decreasing $K$.
Fradkin and Shenker argued that the ``Higgs" and confined phases
could be continuously connected, by noting the absence of a
phase transition along the $t_c = \infty$ and $K=0$ lines.
Moreover, as detailed in Appendix \ref{ISD}, this model is in fact self-dual,
and maps into an equivalent model with new parameters
reflected across the dashed line. 

The phase with large $K$ but small $t_c$ corresponds to the
exotic new superconducting phase, ${\cal SC}^*$.  In this phase
there are four deconfined massive excitations: (i) the spinon,
(ii) an $hc/2e$ $U(1)$ vortex,
(iii) the Ising spin $s_i$ and 
(iv) the $Z_2$ vortex in the gauge field $\sigma$ - the ``vison''. 
In striking constrast to a conventional superconducting
phase, in ${\cal SC}^*$ the $U(1)$ and $Z_2$ vortices can
exist as {\it separate} excitations, and are {\it not}
confined to one another. In order to distinguish this $hc/2e$ vortex 
from the 
BCS vortex, we will refer to it as an $hc/2e$ {\em vorton}.  
The Ising spin excitation $s$
is a remnant of the chargon.  In the paired-chargon condensate
${\cal SC}^*$ phase, the global $U(1)$ charge symmetry is not fully
broken - there is an unbroken $Z_2$  ``charge" symmetry ($s_i \rightarrow -s_i$)
corresponding
to an invariance under a sign change of the chargon operator.
Although the electrical $U(1)$ charge of the chargon
is not conserved, the chargon number is conserved modulo 2,
a reflection of this unbroken Ising symmetry.
Indeed, one can define a conserved Ising ``charge" as,
$Q_2 = (-1)^{N} = \pm 1$, where $N$ is the 
chargon number operator.  
Since the Ising spin operator changes the sign of $Q_2$,
the massive spin excitation carries the conserved
$Z_2$ electrical charge of the chargon.
We refer to this excitation as an ``ison".

To gain some physical insight into this strange 
ison particle, consider what happens when an
electron is added to a superconductor.
The electron creation operator can be decomposed into the
product of a spinon and a chargon,
\be
c^\dagger_{i \alpha} = b^\dagger_i f^\dagger_{i \alpha} \approx s_i f^\dagger_{i \alpha}   .
\ee
The second equality is valid within the two superconducting
phases.  
In the conventional superconductor ${\cal SC}$,
the ison is also condensed, $\langle s_i \rangle \ne 0$,
so that the electron is essentially equal to the spinon.
Thus the spin of the added electron is carried away
by the spinon - the conventional BCS quasiparticle - whereas
the electrical charge is carried by the condensate.
On the other hand, in ${\cal SC}^*$ adding an electron not only
increases the conserved spin by $1/2$, but changes
the conserved $Z_2$ ``electrical charge".
The spin and $Z_2$ charge
are carried away by two {\it separate} massive excitations - 
the spinon and ison.  Thus, the ${\cal SC}^*$ phase exhibits
an exotic form of spin-charge separation.

It is again important to ask about geometric phase factors acquired when any of the four massive
excitations in ${\cal SC}^*$ encircle another. First, note that both the ison
and the spinon are minimally coupled to the gauge field $\sigma$. 
Consequently, they both acquire a phase factor of $\pi$ on encircling the 
$Z_2$ vortex, namely the vison. The ison, being a remnant of a 
chargon, also acquires a 
phase of $\pi$ on encircling an $hc/2e$ vorton. Thus the pairs
- (spinon, vison), (ison, vison), and (ison, $hc/2e$ vorton)-
acquire phase factors of $\pi$ upon encircling one another.  Equivalently, there
are long-ranged statistical interactions between any two members of a pair.
All other pairs of excitations do not acquire any geometrical phase factors. 
Note in particular  
that {\em the $hc/2e$ vorton, being unbound from the $Z_2$ vison,
does not have a long range statistical interaction
with the spinon in ${\cal SC}^*$}.  This distinguishing feature will have several
important consequences in the dual vortex description developed in the next section.

\begin{figure}
\epsfxsize=3.5in
\centerline{\epsffile{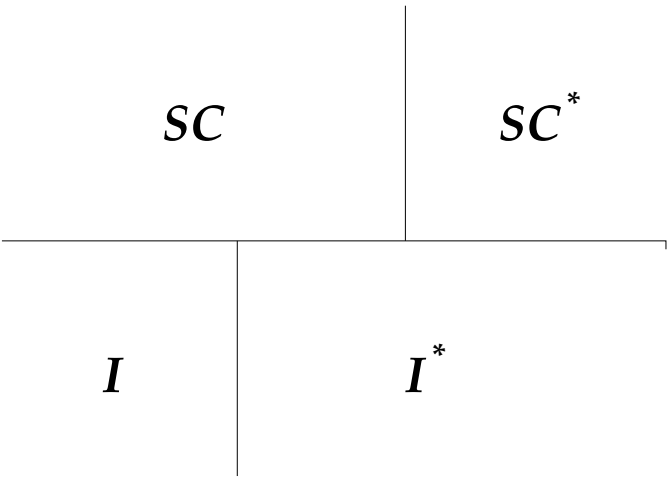}}
\vspace{0.15in}
\caption{Schematic zero temperature phase diagram displaying the four phases ${\cal SC}$,
${\cal SC}^*$, ${\cal I}$, and ${\cal I}^*$.}
\vspace{0.15in}
\label{scstar}
\end{figure}  

The transition from ${\cal SC}^*$ to ${\cal SC}$ occurs on condensing the ison
- so that single chargons are themselves condensed. Note that ison condensation 
leads to confinement of the excitations it has long-ranged statistical interactions with
- the $hc/2e$ vorton and the vison ({\em i.e} the $Z_2$ vortex). The result is the 
BCS $hc/2e$ vortex, as discussed earlier in  Section \ref{fracswv}.

The transition from ${\cal I}^*$ into ${\cal SC}^*$ 
upon increasing $t_2$, can be understood
as a superconductor-insulator transition
of charge $2e$ chargon (or Cooper) pairs. Note that a direct transition from the 
conventional insulator ${\cal I}$ to ${\cal SC}^*$ is not generically 
possible.

Figure \ref{scstar} is a schematic phase diagram
exhibiting the four phases - ${\cal SC}, {\cal SC}^*,{\cal I}$ and
${\cal I}^*$  - as well as the intervening transitions.
Of the four, it is only in the band insulator
${\cal I}$ that spinons are confined.  In the other three
phases the $Z_2$ vortex is gapped out and uncondensed.
These three phases exhibit excitations with 
``fractionalized" quantum numbers.
It is the condensation of the $Z_2$ vortex    
which leads to confinement, leaving only the electron
in the spectrum.

\subsubsection{Odd number of electrons per unit cell}

We now briefly consider the superconducting phases
with odd integer filling, but still presuming $s-$wave
pairing.  Since chargon pairs
are condensed in both ${\cal SC}$ and ${\cal SC}^*$,
it suffices again to consider very large pair hopping
amplitude, $t_2$.  Moreover,
with condensed chargon-pairs,
the chargon operator can be replaced by the Ising spin, $b^\dagger_i = s_i = \pm 1$ - the ``ison" - 
as discussed above.
After integrating out the gapped spinons,
the effective theory again reduces to the Ising matter-plus-gauge theory
as in Eqn. \ref{igm}, but with the addition of the Berry's phase term, $S_B$;
\be
S_{eff} = -2t_c \sum_{\langle ij \rangle} s_i \sigma_{ij} s_j
-K \sum_{\Box} [\prod_{\Box} \sigma_{ij} ] + S_B[\sigma_{ij}]   .
\ee
Note that the $SC^*$ phase is realized only for large $K$,
as discussed above.
In this limit, as we have emphasized several times, the effects of the Berry phase
term $S_B$ are expected to be innoccuous. Thus, $SC^*$ will continue to exist
even in the presence of $S_B$. To see this in more detail, 
it is once again illuminating to pass to a dual representation,
which exchanges the isons for the visons:
\be 
\label{dualZ2matt}
S_{dual} = -K_d \sum_{\langle ij \rangle} v_i \mu_{ij}
v_j - t_d \sum_{\Box} \prod_{\Box} [\mu^{ext}_{ij} \mu_{ij} ]   ,
\ee
with $tanh( t_d )= e^{-4t_c}$ and $tanh(K_d) = e^{-2K}$.
Here $\mu_{ij}$ is a dynamical $Z_2$ gauge field, and as before
$\mu_{ij}^{ext}$ is an ``applied" field with
$\prod_{\Box} \mu^{ext}_{ij} = -1$ through all spatial plaquettes.
This theory is a direct $Z_2$ analog of a $U(1)$ 
superconductor in the presence of an applied magnetic field.

\begin{figure}
\epsfxsize=3.5in
\centerline{\epsffile{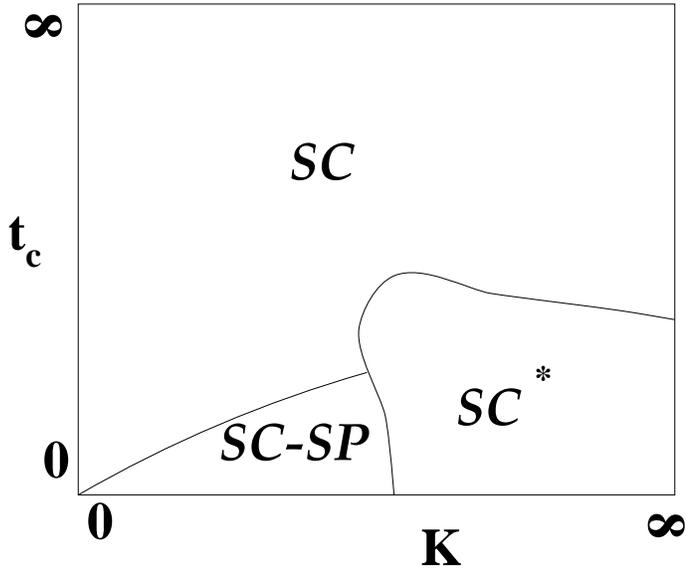}}
\vspace{0.15in}
\caption{Schematic zero temperature phase diagram for the superconducting phases with an odd number of
electrons per unit cell. The ${\cal SC-SP}$ phase is discussed in the text. The precise topology
of the phase diagram when the couplings $t_c$ and $K$ are both of order one is not firmly established.}
\vspace{0.15in}
\label{scsp}
\end{figure}

Consider briefly the phase diagram in the $t_c-K$ plane. A schematic phase diagram is
shown in Fig. \ref{scsp}.
Progress can be made in various limiting regimes. For $K_d=0$ the theory
reduces to a pure $Z_2$ gauge theory with
gauge field, $\tilde{\mu}_{ij} = \mu^{ext}_{ij} \mu_{ij}$.
Since $\mu^{ext}_{ij}$ plays no role in this limit,
the resulting phases are
identical to that with even integer
$N_0$ analyzed in the previous subsection. In particular, for large $t_c$,
we have a conventional superconductor ${\cal SC}$ with broken $Z_2$
gauge symmetry, while for small $t_c$, 
we get the exotic superconductor ${\cal SC}^*$. These phases survive 
for small $K_d$. 
It is easy to establish the absence of phase transitions
for $t_c= \infty$ and $K=0$.  For $t_d = \infty$,
on the other hand, one can set $\mu_{ij} = \mu^{ext}_{ij}$,
and the model reduces to the fully frustrated Ising model.
As discussed extensively in Section \ref{fracdwv}, the results of Ref. \cite{FTFIM2}
show the existence of an ordered phase for 
large $K_d$ where translationally symmetry is spontaneously broken.
In general, this is expected to lead to spin-Peierls order. In this case, though, the spin-Peierls
order co-exists with superconductivity. We will denote this phase as
${\cal SC-SP}$. Several other ordered phases are presumably also 
possible though we will not discuss these here.

In the ${\cal SC}-{\cal SP}$ phase the external gauge field
``penetrates" with $\mu_{ij} \approx \mu^{ext}_{ij}$,
and the Ising model is frustrated.  But as $t_d$ is reduced,
it eventually becomes favorable to ``screen" out this
external field, and enter a ``Meissner" phase
with $\langle \prod_{\Box} \mu_{ij} \rangle \approx 1$.
When this happens the broken translational
symmetry disappears - along with the frustration  - 
and one enters into ${\cal SC}$.

\subsubsection{$d$-wave pairing and doping}
The discussion above generalizes readily to the case of $d$-wave pairing. In particular,
a $d{\cal SC}^*$ phase where chargon pairs, but not single chargons, have condensed
is an allowed phase in the model. It's properties are the
same as for the $s$-wave case above, except that the spinons
have a 
gapless $d$-wave dispersion. Also possible 
is a $d{\cal SC}$ phase coexisting with spin-Peierls order, just as in the $s$-wave case. 

In the presence of finite doping with non-integer $N_0$, in either the $s$-wave or the $d$-wave
case, the ${\cal SC}^*$ phase is expected to survive, since the $S_B$ term is innoccuous 
in this phase. The conventional superconducting phases will be more sensitive to the value of $N_0$
- several additional superconducting phases with broken lattice symmetries are presumably possible.

\subsection{Vortex description}
In this subsection we show how the superconductor ${\cal SC}^*$ 
may be described in the dual vortex formulation. 
The discussion in Section \ref{DVT} was based on the action in Eqns. \ref{sdual}
for the spinons and $hc/2e$ vortices.
The symmetries of the action allow the addition of ``kinetic'' terms for {\it both}
$Z_2$ gauge fields $\sigma$ and $\mu$. Once again, though not present in the original 
action, these terms will be generated upon integrating out high energy modes:   
\bea
S_{\sigma} & = & -K_{\sigma} \sum_{\Box} \prod_{\Box} \sigma_{ij} , \\
S_{\mu} & = & -K_{\mu} \sum_{\Box} \prod_{\Box} \mu_{ij} .
\eea
It is of interest to explore the phase diagram for arbitrary positive values of
the couplings $K_{\sigma}$ and $K_{\mu}$. We will show that the superconductor 
${\cal SC}^*$ emerges quite naturally for large $K_{\sigma}$ and $K_{\mu}$. 
As shown below, an important physical consequence of the addition of these $K_{\sigma}$ 
and $K_{\mu}$ terms is that the Chern-Simons term $S_{CS}$ is no longer
effective in attaching flux to the vortices and the spinons. Note that, in the absence of
flux attachment, the field $e^{i\theta_i}$ creates a ``naked'' $hc/2e$ vortex, {\em i.e},
an $hc/2e$ {\it vorton}. Attaching flux of the field $\sigma$, {\em i.e} a vison, converts this into a 
regular $hc/2e$ BCS vortex.  
  
For ease of presentation, we specialize to the case of $s$-wave pairing and 
an even number of electrons per unit cell. In that case, the term $S_B$ may be dropped
from the action. Further, the spinons are gapped and can be integrated out. This will
lead to an innoccuous renormalization of the value of $K_{\sigma}$. 

In the vortex description, superconducting phases correspond to vortex vacuua.
To analyse these, it is then appropriate to imagine integrating out the vortices. This
will renormalize the value of $K_{\mu}$ (or generate
it if not present originally). The
resulting action has only the terms,
\be
S = S_a +
S_{\sigma} + S_{\mu} +S_{CS}  .
\ee
The term $S_a$ leads to a gapless linear dispersing excitation (in the absence
of long-ranged Coulomb interactions), and corresponds
physically to the sound modes of the superconductor. The remaining three 
terms only involve the two $Z_2$ gauge fields $\sigma$ and $\mu$. As shown in Appendix
\ref{ISD}, this action 
is equivalent to that of the $Z_2$ gauge theory with Ising matter fields. If we choose to 
integrate out the $\mu$, this is exactly the same 
as the Ising effective action derived in the previous subsection (Section \ref{SC*p}) to discuss the 
superconducting phases.  Alternatively,
we can integrate out the 
$\sigma$ field to obtain the dual theory as in Eqn.~\ref{dualZ2matt}:
\bea
S & = & S_{vis} + S_{\mu} , \\
S_{vis} & = & -K^d_{\sigma}\sum_{ij}v_i \mu_{ij} v_j  .
\eea
Here $tanh(K^d_{\sigma}) = e^{-2K_{\sigma}}$, so that $K^d_{\sigma}$ is the coupling dual
to $K_{\sigma}$.  Once again, $v_i$ creates
a vison, whose $Z_2$ current is equal to the flux 
of the $\sigma$ field.  On the other hand, the 
vortex configurations of the gauge field $\mu$ correspond to the 
{\em ison} excitations. 

As discussed earlier, the $Z_2$ gauge theory with matter fields has two phases
- a Higgs-confined phase and a deconfined phase. The Higgs-confined phase describes the 
conventional superconductor ${\cal SC}$,
and is perhaps easiest to understand in the
limit in which both $K_\mu$ and $K_\sigma^d$ are small.
With small $K_\mu$ the gauge field
is in its confining phase, so that
test charges coupling to the gauge field 
$\mu$ are confined. There are actually two different particles 
minimally coupled to $\mu$ - 
the $\frac{hc}{2e}$ vorton and the vison, with creation operator $e^{i\theta_i}$ and $v_i$, respectively.  As before, the confined vorton-vison bound state
is the conventional
$\frac{hc}{2e}$ BCS vortex.

The deconfined phase describes the exotic superconductor ${\cal SC}^*$. In this phase, test charges that 
couple to $\mu$ are deconfined. This implies
that the $\frac{hc}{2e}$ vorton and the vison are {\it not}
bound together, and can propagate as independent
gapped excitations, in agreement with the earlier discussion. 
In effect, within ${\cal SC}^*$ the Chern-Simons term has been 
rendered ineffective and does not
attach flux.
Also, configurations with $\pi$ flux in the gauge field $\mu$, 
corresponding to the ``ison",
exist as finite energy excitations. 
Thus, as before we conclude
that there are {\it four} gapped
excitations in ${\cal {\cal SC}}^*$ - the 
$hc/2e$ vorton, the spinon, the vison and the ison.

Note that a transition from ${\cal SC}^*$ to an insulator 
obtained by condensing the 
$hc/2e$ vortons (which are the fundamental $U(1)$ vortices in this phase) leads naturally
to the fractionalized insulator ${\cal I}^*$. This is because the vison is unbound from the 
$hc/2e$ vorton in ${\cal SC}^*$, so that condensation of the latter leaves the former 
uncondensed.  Indeed, the distinct
excitations in the resulting insulator
are the chargons, the spinons, and the visons - as appropriate
to ${\cal I}^*$.  {\em Thus, the exotic insulator ${\cal I}^*$ may either 
be reached from ${\cal SC}$ by condensing $\frac{hc}{e}$ vortices
or from ${\cal SC}^*$ by condensing $\frac{hc}{2e}$ vortons}.
In either case, the vison remains uncondensed.

This completes the dual description of ${\cal SC}^*$. Complications such as 
$d$-wave pairing or arbitrary filling $N_0$ can be handled
straightforwardly in this dual formulation as well, though we shall not do so here.

\section{Extension and generalizations}
\label{ext}

\subsection{General spatial dimension}
\label{Gsd}
The $Z_2$ gauge theory formulation (in the particle representation) is readily generalized to 
arbitrary spatial dimension. The cases of physical interest are 
three dimensions ($3d$) and one dimension ($1d$),
which we discuss in turn.
For simplicity, we will restrict our attention to 
situations with integer filling per unit cell.
The most important effect of spatial dimensionality enters
in the properties of the pure $Z_2$ gauge theory with action,
\be
S = S_{\sigma} + S_B ,
\ee
with $S_B$ included when there are an odd number of electrons per unit
cell. 

\subsubsection{$d=3$}
In $3d$ and in the absence of $S_B$, 
the $Z_2$ gauge theory again has two phases 
distinguished by the behaviour of the Wilson loop correlator (``area law'' versus ``perimeter law'').
As in $2d$, the presence of $S_B$ will enhance the stability
of the perimeter phase,
but the area law phase will still be present.
The presence of the perimeter law phase,
implies the existence of $3d$ insulators
with electron fractionalization.
But in constrast to $2d$,
the flux tubes in the $Z_2$ gauge field 
- the visons- are not
point-like excitations,
but become extended {\em string}-like excitations in $3d$.
The area law phase again describes various confined insulating phases. 
Whether the presence of $S_B$ leads to broken translational symmetry as in $2d$ is 
an interesting open question. Note, however, that in $3d$ it is not possible to
pass to a dual global Ising model.
In fact, 
the pure $Z_2$ gauge theory (in the absence of $S_B$) is in fact self-dual\cite{Wegner} in three spatial dimensions. 

To discuss the superconducting phases ${\cal SC}$ and ${\cal SC}^*$, it
is necessary to 
understand the properties of the $Z_2$ gauge theory coupled to Ising matter fields. 
In the absence of $S_B$, it is known\cite{FrSh} that in 
three spatial dimensions, there are
again two phases - the Higgs-confined phase, and the deconfined phase. These correspond to 
${\cal SC}$ and ${\cal SC}^*$, respectively.  
Their distinguishing properties will be qualitatively similar to the $2d$ case.    As in $2d$, we expect that the main effect of $S_B$ would only be to 
make possible the existence of an ${\cal SC}$ phase with broken translational symmetry.
 
In layered quasi-two dimensional systems, fractionalized 
insulating phases in which each layer is
decoupled from the others are possible, and exist as distinct phases from the
isotropic ones discussed above. Such phases are currently under further investigation.

Finally, it is worth emphasizing that while the extension to $3d$
is straightforward
in the particle representation, the dual vortex
representation
necessarily involves string-like vortex degrees of freedom.

\subsubsection{$d =1$}
In one spatial dimension ($1d$), the $Z_2$ gauge theory is always in its area law phase,  
with or without the $S_B$ term. Thus, our 
formulation is incapable of describing electron fractionalization in one dimension.  Evidently,
fractionalization in $d =1$ must have different 
physical origins than for $d >1$.
To highlight this point, note that $1d$ fractionalization can be
{\em continuous},
as exemplified by the spinless Luttinger liquid which
supports charge-carrying excitations 
with essentially arbitrary (even irrational) charge.
For $d >1$, on the other hand, fractionalization is
{\em discrete} - the fractionally charged excitations carry
a definite rational fraction of the electron charge. 
As in the fractional quantum Hall
effect, this discreteness
can be traced to the binding (and condensation) of a discrete number
of vortices.   This physics appears
to be qualitatively different than the ``solitonic"
mechanism responsible for fractionalization in $1d$,

\subsection{Finite temperature}
\label{FT}
In our formulation there is a sharp distinction between fractionalized and confined phases at
zero temperature, which is
independent of whether or not the phases in question
have any sort of conventional long-ranged order.  It is
extremely interesting to ask
whether this sharp distiction
survives at finite temperature.
Consider first the deconfined phases in $2d$.  In these phases,
the point-like vison excitations are gapped at
zero temperature.  However, since the energy cost to create a vison is finite,
at any non-zero temperature there will be a non-vanishing
density of thermally excited visons.  
In the absence of 
other kinds of order (eg. magnetic), this low temperature
regime will be
smoothly connected to the high temperature limit,  without an
intervening finite temperature transition.  Thus, in $2d$
the sharp distinction between fractionalized and confined
insulators does {\it not}
survive at finite temperature.

But in $3d$,
the vison excitations
in the deconfined phase are {\it string-like} extended objects,
with an energy cost proportional to their length.
Consequently, at low temperatures arbitarily 
large vison loops will not be thermally excited - 
the vison loops will be ``bound".
As the temperature increases, there
will be a transition at which the vison loops unbind
and proliferate.
Thus, {\em the fractionalized insulator 
in three spatial dimensions undergoes a finite temperature phase transition 
associated with the unbinding of vison loops}.  A defining
characteristic
of the low temperature phase is that vison loops
will cost a {\it free energy} linear in their length.
Equivalently,  $hc/2e$ (or $Z_2$) magnetic monopole ``test charges"
are confined even at finite temperature,
with an infinite free energy cost to separate them.
A confinement of monopoles is also
one of the characteristics of a $3d$ superconductor,
but quite remarkably the confinement here is occuring in a ``normal"
non-superconducting phase.  The conventional insulating
phases with confinement at zero temperature, on the other hand,
will not exhibit finite temperature transitions
(other than those associated
with the loss of conventional long-ranged order - eg. magnetic).

To understand the origin of these results, we briefly discuss
the properties of the pure $Z_2$ gauge theory (with no matter fields)
in $3+1$ space-time dimensions in more detail. 
At zero temperature the theory is self-dual\cite{Wegner} - the
duality transformation interchanges the ``electric'' and ``magnetic'' fields of the gauge theory. 
For $K > K_c$ when the gauge theory is 
in its deconfining phase,
the theory has string-like vison excitations
(which are $Z_2$ ``magnetic" flux tubes) with a finite energy cost  
per unit length. For $K < K_c$ the gauge theory confines
with area law Wilson loops, but there are
nevertheless string-like excitations in this phase as well.
These can be understood via duality,
which interchanges the area and perimeter law
phases - 
the string-like excitations in the area law phase
are simply flux tubes of the {\it dual} $Z_2$ gauge field.
Physically, these dual tubes are ``electric flux tubes"
responsible for the confinement
of ``electric" charge in the area law phase.
Specifically, when
two test $Z_2$ ``electric" charges separated by a distance $R$ are introduced into the system, the resulting
``electric" flux is concentrated in a tube that extends from one test charge to the other with an
energy cost proportional to $R$  - the linear confinement. Similarly, in the perimeter phase,
dual test charges ($Z_2$ ``monopoles'') that act as sources for the visons are confined.

\begin{figure}
\epsfxsize=3.5in
\centerline{\epsffile{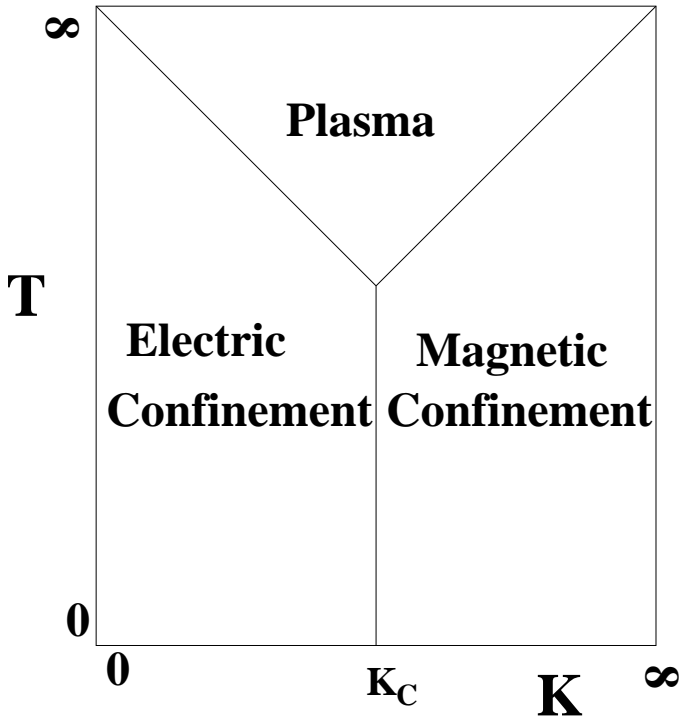}}
\vspace{0.15in}
\caption{Schematic finite temperature phase diagram for the pure $Z_2$ gauge theory in
three spatial dimensions. Upon including the coupling to the chargons and the spinons,
the finite temperature transition for $K < K_c$ is smeared and becomes a crossover only, while the one for $K > K_c$
continues to exist.}
\vspace{0.15in}
\label{z2g3dft}
\end{figure}

Now consider the properties of the gauge theory at finite temperature. The phase diagram
is well-known\cite{Svet} and is shown in Fig. \ref{z2g3dft}. There are three finite temperature phases.
For $K > K_c$, at small but non-zero temperatures, large 
(``magnetic") vison loops are bound as their 
energy cost is proportional to their length. Similarly, for $K < K_c$ at low temperatures, large ``electric" flux loops are bound. 
At high temperature, for any $K$, both kinds of loops are unbound. The transition from the low temperature
to the high temperature phase is therefore associated with the unbinding of (electric) magnetic vison loops
for $K$ (lesser) greater than $K_c$.

In the low temperature
phase for $K < K_c$, the free energy of an isolated static test 
``electric" charge diverges, so that test charges are confined.
In the high temperature deconfined phase, the free energy cost
is finite.
Formally, the pure $Z_2$ gauge theory has a {\it global} Ising symmetry at finite temperature which is broken in 
the {\it high temperature phase}. As shown by Polyakov\cite{Svet}, a convenient characterization
of this transition is through the following operator: 
\be
L_r = \prod_{n = 0}^{M -1} \sigma_{\vec{r} + n\vec{\tau}, \vec{r} + (n + 1)\vec{\tau}}
\ee
for each site $\vec{r}$ of the spatial lattice. Here $\vec{\tau}$ is a vector along the 
(imaginary) time direction of length the time-slice. The product is over all the temporal links at that 
site, and $M$ is the number of time slices.  
This operator $L_r$ is often referred to as the ``Polyakov loop''. 
The free energy $F(r, r')$ to introduce two test charges at sites $r$, $r'$ is directly related
\cite{Svet} to the correlator
of $L_r$ through
\be
e^{- F(r, r')/T} = < L_r L_{r'} >.
\ee
Thus, test charges will be confined if this correlator goes to zero at large distances - on the other 
hand, if this correlator goes to a constant, the test charges will be deconfined. Furthermore,
consider the following transformation on the gauge fields
\be
\label{center}
\sigma_{\vec{r} +  n_0 \vec{\tau}, \vec{r} +  (n_0 + 1)\vec{\tau}} \ra \epsilon 
\sigma_{\vec{r} +  n_0 \vec{\tau}, \vec{r} +  (n_0 + 1)\vec{\tau}}
\ee
where $\epsilon = \pm 1$ {\it independent} of $r$, and $n_0$ is fixed. The action of the pure gauge theory
is invariant under this transformation - implying a global Ising symmetry
of the theory. The operator $L_r$, however, transforms as
\be
L_r \ra \epsilon L_r  .
\ee
Thus $L_r$ is an order parameter for this global Ising symmetry. In the low temperature phase for $K < K_c$,
$L_r$ has no expectation value, the global Ising symmetry is unbroken, and test charges are confined.
At high temperatures, however, $L_r$ acquires an expectation value breaking the global Ising symmetry,
and the test charges are deconfined.

For $K > K_c$, the self-duality of the $Z_2$ gauge theory implies the existence of a dual global Ising
symmetry, with an order parameter that is the dual analog of the Polyakov loop. In the low temperature phase, this 
dual global symetry is unbroken - in this phase dual test charges ({\em i.e} $Z_2$ monopoles) are 
confined. At high temperatures this dual global symmetry is broken and the dual test charges are 
deconfined. 

Consider next the effects of coupling matter fields (the chargons and the spinons) to the
$Z_2$ gauge field. As these carry $Z_2$ gauge ``electric" charge, it is easy to see that the action is no 
longer invariant under the transformation in Eqn. \ref{center}.
Indeed, this transformation is equivalent to changing
the boundary conditions on the chargon fields from ($\beta-$)periodic
to anti-periodic, and vice versa for the spinons.
Moreover,  
if the matter coupling is weak, the matter fields may formally be integrated 
out\cite{Suss} to leave behind a ``magnetic field'' term that couples linearly to the Polyakov loop order parameter 
of the global Ising symmetry. There is then 
no longer any transition separating the low and high temperature
regimes. Physically, this is exactly as expected - for $K < K_c$, the electronic system is in a {\it conventional}
confined insulating phase at zero temperature. 

On the other hand, since
the chargons and spinons do {\it not} carry any {\it dual} $Z_2$
``magnetic" charge, the dual global Ising symmetry remains
even in their presence.  The 
finite temperature transition for $K > K_c$ should thus remain in tact.   
Consequently, we arrive at the striking conclusion that the three dimensional fractionalized 
insulator undergoes a finite temperature transition associated with the unbinding of vison 
loops. This conclusion will not be affected by the Berry's
phase term $S_B$, which is quite innocuous
in the fractionalized insulator.

\subsection{Spin-rotation non-invariant systems}
The $Z_2$ gauge theory formulation (in either the particle or vortex 
representations) works equally well in the absence of spin rotation invariance. 
In particular, fractionalized phases continue to exist even when spin is not a good quantum number.
(Spinless fermion systems can also be handled with no fundamental modifications). For these
reasons, we have avoided the term ``spin-charge separation'', in favour of the more
general term ``electron fractionalization''. 

\subsection{Analogies with nematics}
Certain aspects of our formulation might be familiar from
the {\em classical} statistical mechanics of nematics. The order parameter for a nematic
is a headless three component vector. Lattice models of nematics are usually formulated in terms 
of an ordinary three component vector - the headless nature 
being incorporated through a local $Z_2$ gauge symmetry which inverts the local vector order
parameter.  Here,
we briefly explore the analogies between the {\em classical} 
phases of nematic systems, and the {\em quantum} phases discussed in this paper.

The analogy is closest if we consider $s$-wave pairing with an even number of 
electrons per unit cell, and further, integrate out the spinons to work with just the 
chargons and the $\sigma$ field. The action describing the system is then,
\be
S = -2t_c \sum_{<ij>} \sigma_{ij} cos(\phi_i - \phi_j) - K\sum_{\Box} \prod_{\Box} \sigma_{ij}  .
\ee
As formulated, this describes a {\em quantum} problem of chargons coupled to a fluctuating 
$Z_2$ gauge field in two spatial dimensions. 
But alternately, we may view it as a {\em classical} Hamiltonian for a
three dimensional $XY$ nematic. Indeed, an $O(3)$ version of the same model was 
introduced a few years ago by Lammert, Rokhsar, and Toner\cite{LRT} to
describe nematic ordering in three dimensions. Further, they argued 
that their lattice gauge nematic model admits three distinct phases - an ordered
nematic phase, and {\em two} isotropic phases. 
The nematic phase breaks the rotational symmetry, and the $Z_2$ gauge symmetry. For an $XY$
system, this is the direct analog of the superconducting phase. Moreover, 
the physical $hc/2e$ vortices
of the superconductor correspond directly to the ``disclinations'' in the nematic fluid.

The two isotropic phases in the nematic are distinguished\cite{LRT} 
by the free energy cost per 
unit length to 
externally impose a disclination line through the system. 
In particular, in the conventional isotropic phase, the free energy cost per unit length 
is zero (as the length goes to infinity). The disclinations are condensed. But, in the 
unconventional isotropic phase\cite{LRT}, the free energy cost per unit length is a constant 
(as the length goes to infinity).  In the context of this paper, the isotropic phases
correspond to insulating phases. As we have elaborated at length, there are 
two insulating phases ${\cal I}$ and ${\cal I}^*$ which are distinguished 
by whether or not the visons (which are the relics of the $hc/2e$ vortices in the insulating
phases) are condensed. Thus, the conventional insulator corresponds, in the nematic 
analogy, to the conventional isotropic phase. Note that the energy cost of a vison (which is the action
cost per unit length of the worldline) is zero in this phase. Similarly, the fractionalized
insulator ${\cal I}^*$ corresponds to the unconventional isotropic phase of the $XY$ nematic.
In ${\cal I}^*$ the visons have finite energy cost, again just like the 
disclination lines in the unconventional isotropic fluid.

The phase transition between ${\cal SC}$ and either insulating phase is second order.
In contrast, for the $O(3)$ nematic system considered in Ref. \cite{LRT}, the transition
between the nematically ordered phase and the conventional isotropic phase is first order, 
while that to the other isotropic phase is second order. This difference is due to the $XY$
symmetry of the superconducting system, as opposed to the $O(3)$ symmetry of the nematic.

For the more general situation, with coupling to the spinons or with an 
odd number of electrons per unit cell, a direct correspondence with the nematic system 
no longer holds. Nevertheless, we believe that the discussion in this subsection
may help (some) readers get further intuition and insight into our formulation.

\section{Relation to previous approaches}
\label{prevapp}
We now comment on the connection between the
$Z_2$ gauge theory and earlier approaches to electron fractionalization.
We begin by making contact with earlier papers on the ``nodal liquid".
Earlier formulations of the nodal
liquid (in Ref. \cite{NLII} and \cite{NLIII}) focussed on the importance
of ``vortex-pairing" as a means to describe
charge fractionalization in two-dimensions.
In Ref. \cite{NLII} a theory was formulated
in terms of vortices in a local superconducting pair field,
and shares many features with the approach
taken here, particularly the dual formulation detailed
in Section \ref{DVT}.  In Ref. \cite{NLIII}, Chern-Simons theory
was used to convert spinful electrons into bosons,
and a dual formulation was developed in terms of
vortices in these bosonic fields.     
The $Z_2$ gauge theory and its dual Ising Chern-Simons vortex theory
developed in this paper, not only ties together both earlier approaches
into a unified framework, but allows for a more direct quantitative
analysis of ``microscopic" models.  We now describe this
connection in a bit more detail.

In Ref. \cite{NLII} a spinon operator was defined as an electron
with its charge screened by ``one-half" of a Cooper pair.
The latter coincides precisely with the ``chargon"
introduced in Eqn.\ref{chargon}, showing the equivalence
of the spinons as well.  The importance of the
long-ranged interaction between the spinon and $hc/2e$ vortex
was emphasized in Ref. \cite{NLII}.  It was suggested that this interaction
could be implemented by employing a $U(1)$ Chern-Simons term
to attach flux to both species of particles.  But since the
spinon number is not conserved, it was suggested that the flux
could be attached to the (conserved) z-component of the spin.
Moreover, it was argued in Ref. \cite{NLII} that due to the statistical
interactions, condensation of $hc/2e$ vortices
should lead to confinement of spinons.
In the dual vortex formulation presented in this paper the statistical
interaction between vortex and spinon is described in terms
of a novel Ising-like Chern-Simons term.  It is important to stress
that this {\it does not} require the spin of the spinon
to be conserved, in contrast to the $U(1)$ approach,
since the ``Ising-flux" is attached to the conserved $Z_2$ charge
of the spinons.  Moreover, the Ising formulation clearly
shows that condensation of the $hc/2e$ vortices - or
the visons - leads to confinement of spinons and chargons.
In the global Ising model for the visons with $\langle v_i \rangle \ne 0$,
the linear confinement is due to the required line
of negative Ising couplings connecting the two spinons.
In the $Z_2$ gauge theory formulation, it follows
from the area law for the Wilson loop.

In Ref. \cite{NLIII}, a theory was developed by converting
spinful electrons into spinful bosons - using Chern-Simons
to attach flux to the electrons {\it spin} - 
and then passing to a dual representation
of vortices in these bosonic fields,
denoted $\Phi_\alpha$ with spin label $\alpha = \uparrow, \downarrow$.
A lattice version of this theory can be written
in terms of the phases, $\theta_\alpha$,
of the vortex field operators, $\Phi_\alpha = e^{i\theta_\alpha}$,
with effective Euclidian action,
\be
S = -t_v \sum_{\langle ij \rangle}
cos(\theta_{i \alpha} - \theta_{j \alpha} + a^\alpha_{ij}) + S_{cs}(a^\sigma) .
\ee
Here, $i,j$ label sites of the 2+1 space-time lattice,
$t_v$ is a dimensionless vortex ``hopping" term
and $S_{cs}$ is a Chern-Simons terms involving
the field $a^\sigma = a^\uparrow - a^\downarrow$.
The curl of $a^\alpha$ corresponds to the conserved
electrical current of the electrons with spin $\alpha$. 
In Ref. \cite{NLII}, two different composite ``pair" vortex operators
were considered;  
\be
\Phi_\rho = \Phi_\uparrow \Phi_\downarrow = e^{i\theta_\rho}    ; \hskip0.4cm
\Phi_\sigma = \Phi_\uparrow \Phi^\dagger_\downarrow = e^{i\theta_\sigma}  ,
\ee
which are minimally coupled to $a^{\rho/\sigma} = a^\uparrow \pm
a^\downarrow$, respectively.  The action can be re-expressed
in terms of these composite phase fields
using the relation
\be 
\theta_{\uparrow \downarrow} = {1 \over 2} (\theta^\rho \pm \theta^\sigma )
+ {\pi \over 2} v   ,
\ee
giving,
\be
\label{aras}
S =  -t_v \sum_{\langle ij \rangle} v_i v_j cos[(\theta^\rho_i - \theta^\rho_j
+ a^\rho_{ij} )/2] cos[(\theta^\sigma_i - \theta^\sigma_j
+ a^\sigma_{ij} )/2] .
\ee
Here, the Ising spins $v_i = \pm 1$ are the ``visons". 
The primary emphasis of Ref. \cite{NLIII} was an analysis
of fractionalized phases, such as the nodal liquid.
It was emphasized that fractionalization occurs
when $\langle v_i \rangle = 0$, and breaking
the Ising symmetry with $\langle v_i \rangle \ne 0$
corresponds to confinement.  Deep within the 
deconfined phase it is possible to integrate
out the massive visons, which generates
local terms such as,
\be
S_{hc/e} = -t_{2v} cos(\theta^\rho_i - \theta^\rho_j + a^\rho_{ij}) ,
\ee
which describes the hopping of the $hc/e$ vortex pair, $\Phi_\rho$,
and
\be
S_{spinon} = -t_s cos(\theta^\sigma_i - \theta^\sigma_j + a^\sigma_{ij}) .
\ee
Due to the Chern-Simons terms above, this corresponds to the
hopping of fermionic spinons which
carry $S_z = 1/2$.

The relationship between this formulation,
in terms of ``electron" vortices,
and the dual vortex theory of Section \ref{DVT} constructed
in terms of BCS $hc/2e$ vortices is at first not apparent.
But consider introducing a vortex operator, $\Phi=e^{i\theta}$,
whose {\it square} equals the $hc/e$ vortex pair operator:
$\Phi^2 = \Phi_\rho$.  This requires that,
\be 
\theta = {1 \over 2} \theta_\rho + {\pi \over 2} (1-v)   ,
\ee
which implies,
\be
\Phi = v e^{i\theta_\rho/2}  .
\end{equation}
As defined $\Phi$ carries vorticity $hc/2e$, and can tentatively
be identified as the BCS vortex.  To complete this identification
it is necessary to show that there is a long ranged statistical
interaction between this $hc/2e$ vortex and the spinon.  
Evidence for this is provided by the following argument. We first imagine
explicitly adding the vortex hopping term $S_{spinon}$ to the action
in Eqn. \ref{aras}.
We then absorb the field $\theta^\sigma$ into $a^\sigma$. 
We may now re-express the action Eqn. \ref{aras} in terms
of $\theta_i$:
\be 
S = -t_v 
\sum_{\langle ij \rangle} \mu_{ij}cos(\theta_i - \theta_j + {1 \over 2} a_{ij}) + S_{spinon} ,
\ee
with 
\be 
\mu_{ij} = cos({1 \over 2} a^\sigma_{ij} )  .
\ee
Here, we have defined $a_{ij} = a^\rho_{ij}$.
In the presence of the vortex hopping term $S_{spinon}$ above, if we specialize to the 
limit of large $t_s$, it is
legitimate to restrict $a^\sigma_{ij}$ to be
$2\pi$ times an integer.  With that
restriction the gauge field $\mu_{ij} = \pm 1$, reducing
to an Ising $Z_2$ gauge field. 
   Now, imagine putting a stationary spinon
on one site of the original spatial lattice.  In this dual vortex representation
this corresponds to a plaquette with $\Delta \times a^\sigma = 2\pi$,
or equivalently to a product $\prod_{\Box} \mu_{ij} = -1$ for all plaquettes
pierced by the spinon ``world-line".  Since 
the $hc/2e$ vortex is minimally
coupled to $\mu_{ij}$, this establishes
that it does indeed acquire a minus sign upon
being transported around a spinon.
In the dual vortex formulation in Section \ref{DVT},
a $\pi$-flux tube in $\mu_{ij}$ is attached to each spinon
by the Ising-like Chern-Simons term.  To complete the mapping
between these two formulations
requires, finally, to re-fermionize the spinon
creation operator, $e^{i\theta^\sigma_i}$,
(fermionic due to the Chern-Simons term $S_{cs}[a^\sigma]$)
effectively replacing it with spinful fermions $f_{i\alpha}$.

Finally we comment briefly on the relationship with theories based on 
slave boson/fermion approaches to electron fractionalization. 
A number of authors have examined insulating Heisenberg antiferromagnetic
spin models in the hope of finding phases with deconfined spinon
excitations through these approaches. However this program has generally been 
quite unsuccesful - the $U(1)$ or $SU(2)$ gauge symmetry introduced in the
slave boson or fermion representations ultimately leads only to confined 
phases. A notable exception however is the work of Read and Sachdev\cite{RSSpN } on large-$N$
$Sp(2N)$ frustrated antiferromagnets, and related quantum dimer models\cite{subir_pc}. 
Under certain special conditions, these authors demonstrated
the existence of quantum disordered phases with deconfined spinons in their theory. 
It is worth pointing out that fractionalization is achieved when the 
$U(1)$ gauge symmetry (introduced by the Schwinger boson representation of the 
$Sp(2N)$ spins) is broken down to $Z_2$ by condensation of pairs of bosons. 
The fully frustrated transverse field Ising model appears in that description 
as well\cite{subir_pc}.

Slave boson representations of electron operators have been used extensively
to discuss spin-charge separation issues in doped $t-J$ models. However, the 
resultant compact $U(1)$ or $SU(2)$ gauge theories presumably always lead to confinement,
unless the gauge symmetry is broken down to $Z_2$. This may be be achieved by 
pairing the spinons\cite{Wen}. Indeed, the slave-boson mean field treatments of the $t-J$
model do find pairing of spinons below a finite temperature at low doping. 
As we have emphasized in this paper though, even in the undoped limit and without frustration, 
the Heisenberg spin model may be rewritten 
in terms of fermionic spinon operators coupled to a fluctuating $Z_2$ gauge field.
Equivalently spinon pairing terms may be added to the Hamiltonian
describing the Heisenberg magnet without altering any of the physical symmetries. 
We have shown that electron fractionalization is definitely possible
once charge fluctuations are incorporated into the description.  

\section{Conclusion and Discussion} 
\label{disc}

\subsection{Summary} 

The primary focus of this paper was to explore the possibility of
electron fractionalization in strongly correlated electron systems in 
spatial dimension greater than one, and in the presence of time reversal symmetry. We based our discussion on a particular class of microscopic
models designed to capture the physics essential to the cuprates,
although our decription of fractionalization is more general.
Starting from these models, we 
developed a new gauge theory of strongly correlated systems
consisting of  
charge $e$, spin-zero bosons (the ``chargons'')
and charge zero, spin $1/2$ fermions (the ``spinons''), both 
minimally coupled to a fluctuating $Z_2$ gauge field. 
Remarkably, the spin-sector of the theory at half-filling and 
in the absence of charge fluctuations, is formally
{\it identical} to a spin one-half Heisenberg antiferromagnet.
In this limit the $Z_2$ gauge field enforces the constraint
that the spinon number on each site is {\it odd} - physically
equivalent to the single occupancy constraint,
imposed with additional unneeded redundancy in
earlier $U(1)$ gauge theory formulations of
the Heisenberg model.  

Charge fluctuations, however,
are naturally incorporated into our $Z_2$ gauge theory,
and when they become large the theory describes
a $d_{x^2 - y^2}$
superconductor.  Analysis of the theory in the intermediate
region reveals that there are two 
qualitatively different routes for the evolution from the antiferromagnet
to the superconductor.  One route is through conventional insulating phases in which 
fluctuations of the $Z_2$ gauge field confines together the chargon and the spinon, leaving only the electron in the spectrum.
But a more interesting possibility
takes one through phases in which the 
electron is fractionalized, and the chargons and spinons exist as deconfined excitations. 
With $d_{x^2 - y^2}$ pairing, this fractionalized insulator is
the nodal liquid\cite{NLI,NLII}, with
gapless spinon
excitations at four points of the Brillouin zone.
It seems likely that the ultimate transition from
the insulating phases to the $d_{x^2 - y^2}$ superconductor occurs close to 
the boundary between the confined and deconfined insulating phases. Thus,
which of these two qualitatively different 
routes is realized in any particular experimental system could depend sensitively 
on microscopic details.

In addition to the chargons and spinons,
the $2d$ nodal liquid supports
Ising-like point excitations - the ``visons''-  which 
correspond to vortices in the $Z_2$ gauge field.  
These gapped vison excitations play a central role
in our analysis of fractionalization,
as becomes clear upon passing to
a dual description in terms of 
$hc/2e$ BCS vortices 
(of a conventional superconductor) and the spinons.
In this dual framework, 
the nodal liquid can be accessed by a pairing and condensation
of the $hc/2e$ vortices,
as emphasized in ealier work\cite{NLII,NLIII}.  This reveals
that the vison excitations are simply the remnant of the unpaired $hc/2e$ vortices
which survive in the insulating nodal liquid.

The utility of the vison excitations goes far beyond giving a simple
description of the nodal liquid.  Indeed,  
the pure $Z_2$ gauge theory in $2+1$ space-time dimensions 
is dual to the global $2+1$ dimensional Ising model - and
the Ising spins are simply the vison creation operators.
Remarkably, an unusual Berry's phase term in the 
gauge theory corresponds simply to frustration
in the dual Ising model, with full frustration at half-filling.
The fully frustrated quantum Ising model
arose in earlier work by Sachdev and coworkers\cite{FTFIM2,subir_pc} in their
analysis of frustrated magnets.
Ordering the dual Ising model by condensation of the
visons, generally will break translational symmetry
and lead to conventional confined insulating phases
such as the spin-Peierls phase.
In three spatial dimensions ($3d$), the visons become loop-like
excitations, and are closely
related to vortex-line excitations
which occur in a conventional superconductor.
Surprisingly, this implies that a $3d$ fractionalized
insulator ``survives" at finite temperature,
being separated from the high temperature regime
by a finite temperature phase transition.
As in a conventional superconductor,
the $3d$ fractionalized insulator
confines $hc/2e$ monopole excitations even at
non-zero temperature.

Within the $Z_2$ gauge theory approach,
a conventional superconductor
is described as a condensate of charge $e$ chargons. 
A superconducting phase involving condensation of chargon pairs ({\em i.e} Cooper pairs)
without condensation of single chargons was shown to exist - this has several exotic
properties distinguishing it from the conventional superconductor.

\subsection{Experiments}

We close with a very brief
discussion of some of the experimental signatures of 
electron fractionalization.
As we will see,
experimental detection of fractionalization may be quite subtle. Further 
theoretical understanding of fractionalized phases leading to detailed 
experimental predictions are clearly called for. Our discussion will
necessarily be brief.

\subsubsection{Two dimensional nodal liquid}
Earlier work on the nodal liquid\cite{NLI,NLII} outlined a number of experimental signatures
of the two-dimensional nodal liquid,
and we have little to add here.
As pointed out in the earliler papers, perhaps the most telling indication will
be in angle resolved photoemission (ARPES) which directly measures the 
electron spectral function as a function of the momentum $k$, and frequency 
$\omega$. As the electron is fractionalized into the chargon and the spinon
in the nodal liquid, its spectral function will not have a sharp
quasiparticle peak even at zero temperature. Note that 
bound states of the chargon and the spinon (which could lead to sharp 
spectral features) are not expected here at low energies as the spinons are gapless.  

\subsubsection{${\cal SC}^*$}
We have discussed the basic physics of the exotic superconductor $SC^*$
obtained by condensing chargon pairs in Section \ref{SCstar}. 
There are several 
qualitative experimental distinctions between this phase and the conventional superconductor
which we now briefly discuss.
The most striking is again in the electron spectral
function as measured in ARPES. As discussed in Section \ref{SCstar}, the
electron decays into a spinon and an Ising part of the charge - the ``ison''
excitation. Thus, we expect that the electron spectral function does {\em not} have a 
sharp quasiparticle peak
in the ${\cal SC}^*$ phase. Again, since the isons are massive excitations
while the spinons are gapless, bound states of the two are generally 
not expected at low energies. 
The presence of gapped ison excitations would also 
affect the thermodynamics,
and 
contribute to the
thermal conductivity at some intermediate temperatures. However, these
signatures are likely to be quite subtle. 
A striking theoretical feature of ${\cal SC}^*$ is that the conventional BCS $hc/2e$ vortices are splintered into pieces - the $U(1)$ ``vorton"
carrying the circulating electrical currents, and
the $Z_2$ 
vison.  Since the spinons do not have a 
long-ranged statistical interaction with the $hc/2e$ vorton,
it is tempting to 
speculate that the structure of the core states in such a vorton would be qualitatively
different from that of an $hc/2e$ vortex in a conventional superconductor. 

\subsubsection{Three dimensional effects}
In striking contrast to a two dimensional nodal liquid, a genuinely three dimensional nodal
liquid has a finite temperature phase transition associated with the 
unbinding of vison loops. This phase transition could lead to observable singularities
in the measured properties of the system.  But due to the
highly anisotropic nature of the cuprates,
it is perhaps more natural to speculate
that a fractionalized phase would consist
of decoupled $2d$ systems, with a confinement of spinons
within each layer.  Clarification of such interlayer
confinement physics will be necessary in order to disentngle
the subtle interlayer behavior of the cuprate materials,
both in the normal and superconducting phases.

We are grateful to Phil Anderson,
Leon Balents, Tom Banks, Eduardo Fradkin, Doug Scalapino
and R. Shankar for insightful discussions, and V. Ashvin for useful 
comments on the manuscript.
We would particularly like to thank Subir Sachdev
for many helpful comments and his explanations of $Z_2$ gauge
theories in the context of frustrated large $N$ magnets.
Thanks are due the Aspen Center of Physics where some of this
work was carried out.
This research was generously supported by the NSF 
under Grants DMR-97-04005,
DMR95-28578
and PHY94-07194.

\appendix
\section{Path Integral}

In this Appendix we derive a path
integral expression for the partition function 
of the spinon-chargon Hamiltonian.
A crucial role is played by the constraint on the Hilbert
space, which naturally introduces an $Z_2$ gauge field.

To this end, we work with fermionic coherent states
built from the spinon operators, $\hat{f}_\alpha$ and $\hat{f}^\dagger_\alpha$,
which are defined in the standard fashion:
\begin{equation}
|f_\alpha \rangle = e^{-f_\alpha \hat{f}^\dagger_\alpha} | 0 \rangle   ,
\end{equation}
\begin{equation}
\langle \bar{f}_\alpha | = \langle 0 | e^{\bar{f}_\alpha \hat{f}_\alpha  }  ,
\end{equation}
where the spinon {\it operators} are denoted with ``hats",
and $\bar{f}_\alpha$ and $f_\alpha$ are Grassman numbers.
The bra and ket states denoted with a ``0", are fermionic fock states
with no spinons present.  Here we have suppressed the dependence
of the fermion operators and Grassman fields on the spatial
coordinate, $r$.
In the charge sector of the theory we choose a basis of states
diagonal in the phase $\phi$ of the chargon field,
denoted $|\phi \rangle$. 

The partition function in Eqn. \ref{cspf} can then be expressed
as
\begin{equation}
Z = \int d\bar{f}_\alpha d f_\alpha \int_0^{2\pi} d\phi e^{-\bar{f}_\alpha f_\alpha} \langle
-\bar{f}_\alpha; \phi| (e^{-\epsilon H} {\cal P})^M |f_\alpha;\phi \rangle   ,
\end{equation}
with $\epsilon = \beta/M$ and ${\cal P}$ the projection operator
defined in Eqn. \ref{proj}.  Inserting the resolution of the identity
between each time slice gives,
\begin{equation}
Z = \prod_{\tau = 1}^M \int d\bar{f}_{\tau\alpha} df_{\tau\alpha} d\phi_\tau
e^{-\bar{f}_{\tau} f_{\tau -1}} {\cal M}_{\tau}  ,
\end{equation}
with matrix elements
\begin{equation}
{\cal M}_\tau = \langle \bar{f}_\tau; \phi_\tau |
e^{-\epsilon H} {\cal P} | f_{\tau}; \phi_{\tau -1} \rangle   ,
\end{equation}
and appropriate boundary conditions
on the fields:  $f_{M+1} \equiv - f_1$ and $\phi_0 \equiv \phi_M$. 

The matrix elements can be readily evaluated for small $\epsilon$
by inserting a complete set of states diagonal in the
chargon number, $N$.  Using the definition of the projection operator
in Eqn. \ref{proj} gives,
\begin{equation}
{\cal M}_\tau = {1 \over 2} \sum_{\sigma_\tau = \pm 1} \sum_{N_\tau = -\infty}^\infty e^{i N_\tau[\phi_\tau - \phi_{\tau -1} + {\pi \over 2}
(1-\sigma_\tau)]} e^{\bar{f}_\tau \sigma_\tau f_\tau} E_\tau  ,
\end{equation}
with
\begin{equation}
E_\tau = e^{-\epsilon H(N_\tau,\phi_\tau,\bar{f}_\tau,\sigma_\tau f_\tau)} .
\end{equation}

Upon making the change of variables in the Grassman functional integral,
\begin{equation}
\sigma_\tau f_\tau \rightarrow f_\tau   ,
\end{equation}
the full partition function can finally be re-expressed as,
\begin{equation}
Z = \int \prod_{\tau =1}^M d\bar{f}_{\tau} 
df_{\tau} d\phi_\tau \sum_{N_\tau = - \infty}^\infty
\sum_{\sigma_\tau = \pm 1}   e^{-S}    ,
\end{equation}
with,
\begin{equation}
S = S^f_\tau + S^\phi_\tau + \epsilon \sum_{\tau =1}^M H(N_\tau,\phi_\tau,
\bar{f}_\tau f_\tau )   .
\end{equation} 
with
\begin{equation}
S^f_\tau = \sum_{\tau=1}^M  [ \bar{f}_{\tau} (\sigma_{\tau+1} f_{\tau+1}
-f_{\tau}  )]  ,
\end{equation}
and
\begin{equation}
S^\phi_\tau = -i \sum_{\tau=1}^M  N_{\tau} [\phi_{\tau} - \phi_{\tau -1} + {\pi \over 2}
(1 - \sigma_{\tau}) ] .
\end{equation} 
Throughout, we have suppressed the explicit $r$ and $\alpha$ subscripts
on the fields, displaying only the time-slice dependences.

\section{$Z_2$ gauge theory with $d_{x^2 - y^2}$ pairing}
\label{dwaveigt}
In this Appendix, we will provide the outline of a microscopic derivation
of the $Z_2$ gauge theory in the presence of $d_{x^2 - y^2}$ pairing correlations.
We begin with the Hubbard-type Hamiltonian Eqn. \ref{dHam} discussed in Section \ref{intro}:
\be
H = H_0 + H_J + H_{\Delta} + H_u  .
\ee
The crucial difference with the $s$-wave case is in the structure of the ``pairing'' term $H_{\Delta}$.

We now follow exactly the same strategy as in the $s$-wave case, defining chargon and spinon 
operators. A path integral representation of the partition function
is readily set up with the main difference 
being  
in the pairing term which becomes
\bea
S_{\Delta} & = & \epsilon \sum_{<rr'>, \tau}\Delta_{rr'}(b^*_{r}b_{r'} + c.c) B_{rr'} , \\
B_{rr'} & \equiv & \Delta_{rr'}(\bar{f}_{r\ua}\bar{f}_{r'\da}
- (\ua \ra \da) + c.c ) .
\eea
We have suppressed the $\tau$ index on all fields. It will be convenient to use a slightly
different decoupling of the $H_J$ term. We write
\bea
e^{-S_J} & = & \int[d\chi_{rr'}d\chi^*_{rr'}d\eta_{rr'}d\eta^*_{rr'}] e^{-S_{hs}}  ,\\
S_{hs} & = & S_{hs}[\chi] + S_{hs}[\eta]  ,\\
S_{hs}[\chi] & = & \epsilon \sum_{<rr'>, \tau}J[2|\chi_{rr'}|^2 - (\chi_{rr'}\bar{f}_{r\alpha}f_{r'\alpha}
+ c.c)] , \\
S_{hs}[\eta] & = & \epsilon \sum_{<rr'>, \tau}J[2\eta_{rr'}|^2 \\
& + & (\eta_{rr'}a_{rr'}(f_{r\ua}f_{r'\da} - 
f_{r\da}f_{r'\ua}) + c.c) .
\eea
Here $a_{rr'} = +1$ for bonds along the $x$-direction, and equals $-1$ for bonds along the $y$-direction.
Note that $S_{hs}[\chi]$ is the same as before. This 
decoupling of the spin-spin interaction is standardly used in the $SU(2)$ gauge theory formulations of the 
$t-J$ model. We emphasize though that our formulation has, as we will show, only an $Z_2$ gauge symmetry.
We now shift the two Hubbard-Stratonovich terms:
\bea
\chi_{rr'} & \ra & \chi_{rr'} - \frac{t}{J}b^*_rb_{r'}  ,\\
\eta_{rr'} & \ra & \eta_{rr'} + \frac{\Delta}{J}(b^*_{r}b_{r'} + c.c) .
\eea
The shift of $\chi$ is as before, and eliminates the spinon-chargon interaction coming from
rewriting the electron hopping term. The shift of $\eta$ eliminates the pairing term.
The net spatial part of the action is then,
\bea
S_r & = & \epsilon\sum_{<rr'>} 2J(|\chi_{rr'}|^2 + |\eta_{rr'}|^2) + S_{cr} + S^1_{sr} + S^2_{sr}  ,\\ 
S_{cr} & = & -\epsilon\sum_{<rr'>}[(2t\chi_{rr'} + 2\Delta(\eta_{rr'} + \eta^*_{rr'}))b^*_r b_{r'} + c.c]  ,\\
S^1_{sr} & = & -\epsilon \sum_{<rr'>}J\chi_{rr'}\bar{f}_{r\alpha}f_{r'\alpha} + c.c. , \\
S^2_{sr} & = & \eta_{rr'}\Delta_{rr'}(f_{r\ua}f_{r'\da} - f_{r\da}f_{r'\ua}) + c.c. . \\
\eea
The shift in $\eta$ also generates a ``Cooper pair'' hopping term $cos(2\phi_r - 2\phi_{r'})$
with a negative hopping amplitude of order $\Delta^2/J$. This is not expected to be important
for the issues of fractionalization that we primarily wish to discuss. So we will for the most 
part drop it. 

The $\chi, \eta$ integrals may be done by saddle point - a uniform, real saddle point solution
$<\chi_{rr'}> = \chi_0$, $<\eta_{rr'}> = \eta_0$ breaks the $Z_2$ gauge symmetry. Parametrizing the fluctuations 
about it by $\chi_{rr'} = \chi_0 \sigma_{ij}$, $\eta_{rr'} = \eta_0 \sigma_{ij}$ 
as before, we arrive at the Ising 
gauge theory appropriate for the $d_{x^2 - y^2}$ superconductor.

\section{Ising self-duality}
\label{ISD}
In this Appendix, we will review the self-duality of the $Z_2$ gauge theory with matter fields in 
$2+1$ dimensions. As a limiting case, we recover the duality of the pure $Z_2$ gauge theory
to the global Ising model. The theory is defined by the lattice action
\bea
\label{igtm}
S[s, \sigma] & = & S_{s} + S_{\sigma}  ,\\
S_{s} & = & -J\sum_{ij} s_i \sigma_{ij} s_j  ,\\
S_{\sigma} & = & - K\sum_{\Box} \prod_{\Box} \sigma_{ij} .
\eea
The constants $J,K$ are assumed to be positive. The indices $i,j$ label the sites of
a three dimensional cubic lattice. It is convenient to first rewrite the $s_i\sigma_{ij}s_j$
term on each bond using the following identity:
\bea
\label{iiden}
e^{Js_i \sigma_{ij}s_j} & = & A \sum_{n_{ij} = 0,1} \exp{\left[2J_d n_{ij} \right.} \\
 & & + \left. i\frac{\pi}{2}n_{ij} 
\left(s_i -s_j + 1-\sigma_{ij} \right)\right]  .
\eea
 Here $tanh(J_d) = e^{-2J}$, and $A = \frac{1}{\sqrt{1 - e^{-2J}}}$. From now on, we will
drop the constant $A$ as it just contributes to an overall 
multiplicative constant to the partition function. 
The $n_{ij}$ take the values $0,1$. Upon using this 
identity for every bond, and 
doing the sum over $s_i$, we get
\bea
\EXP{-S_s}&  = & Tr_{\sigma_{ij}} Tr_{n_{ij}}
\left(\prod_{i} cos\left(\frac{\pi}{2}(\vec \Delta . \vec n) \right)\right) \\
&  & \EXP{2J_d \sum_{ij} n_{ij} + \sum_i i\frac{\pi}{2} n_{ij} (1 - \sigma_{ij})}  .
\eea
Here $\vec \Delta . \vec n$ is the lattice divergence of the link variable $n$. We now notice that
the cosine can be written as
\be
cos\left(\frac{\pi}{2}(\vec \Delta . \vec n) \right) = 
(-1)^{\frac{\vec \Delta . \vec n}{2}}\delta\left( (-1)^{\vec \Delta . \vec n}, 1 \right)  ,
\ee
where $\delta(m,n)$ is the Kronecker delta function for two integers $m,n$. 
The term multiplying the delta function
is a total derivative that contributes zero on summing over all sites - we will therefore drop it.
Note that the delta function imposes conservation modulo $2$ of the link variable $n_{ij}$ at every site.
This conservation can be made more explicit by defining a $Z_2$ current $\alpha$:
\be
\alpha_{ij} = (-1)^{n_{ij}}  .
\ee
We now solve the current conservation condition by
writing the $Z_2$ current $\alpha$ on any link as the flux of a dual $Z_2$ gauge field 
$\mu$ through the plaquette of the dual lattice pierced by this link:
\be
\alpha_{ij} = (-1)^{n_{ij}} =  \prod_{\Box} \mu_{ij} .
\ee
The $\mu_{ij}$ are understood to be defined on the links of the dual lattice, and the plaquette 
product for the $\mu$ is around the appropriate plaquette of the dual lattice. 
Note that this is directly analogous to the standard duality transformation 
of the $XY$ model.

We next solve for the $n_{ij}$ in terms of the $\mu_{ij}$:
\be
n_{ij} = \frac{1 - \prod_{\Box} \mu_{ij}}{2} .
\ee
The $n_{ij}$ may now be eliminated from the action in favor of the $\mu_{ij}$. The result
(after dropping overall multiplicative constants) is the following identity
\bea
\label{idual}
\sum_{s_i}e^{J\sum_{ij} s_i \sigma_{ij} s_j} & = & \sum_{\mu} \EXP{- S_{\mu} - S_{CS}} ,\\
S_{\mu} & = & J_d \sum_{\Box} \prod_{\Box} \mu_{ij} ,\\
S_{CS} & = & \sum_{<ij>} i\frac{\pi}{4} \left(1 - \prod_{\Box}\mu \right)(1-\sigma_{ij}) .
\eea
 
The last term has a structure similar to a Chern-Simons term, but for the
group $Z_2$. It's exponential is actually invariant under $\sigma \leftrightarrow \mu$. This
can be seen as follows.  Write
\bea
e^{-S_{CS}} 
& = & \prod_{<ij>}\left(\prod_{\Box} \mu \right)^{\left(\frac{1-\sigma_{ij}}{2}\right)}  ,\\
& = & \prod_{<ij>}e^{\frac{i\pi}{4}\sum_{<ij>}\left(\Delta \times (1-\mu)\right)\left(1-\sigma_{ij}\right)}  .
\eea
In the last equation, $\Delta \times \mu$ is the lattice curl of $\mu$ on the plaquette of the dual 
lattice pierced by $<ij>$. If we now perform a lattice integration by parts, we get
\bea
& & \EXP{\sum_{<ij>}-i\frac{\pi}{4}(1-\mu_{ij})\left(\Delta \times (1 - \sigma)\right) } \\
& = & \EXP{-\sum_{<ij>} i\frac{\pi}{4} \left(1 - \prod_{\Box} \sigma \right) (1- \mu_{ij})}  ,
\eea
where now the sum is over links $<ij>$ of the dual lattice. 

The full partition function can then be written as
\be
\label{idualg}
Z =  Tr_{\sigma, \mu} \EXP{-S_{\sigma} - S_{\mu} - S_{CS}} .\\ 
\ee

The duality of the full action is now apparent. In particular, the action is invariant under the exchange
$\sigma \leftrightarrow \mu$, $J_d \leftrightarrow K$. To make the duality even more explicit,
we again use the identity Eqn. \ref{idual} to write 
\be
\label{igtmd}
\sum_{\sigma} \EXP{-S_{\sigma}- S_{CS}} 
= \sum_{v_i}\EXP{K_d\sum_{ij}v_i \mu_{ij} v_j} ,
\ee
where $v_i = \pm{1}$ and $tanh(K_d) = e^{-2K}$. The partition function now 
becomes
\be
\label{igtmD}
Z = Tr_{\tau, \mu}e^{K_d\sum_{ij}v_i \mu_{ij} v_j +  J_d \sum_{\Box} \prod_{\Box} \mu_{ij}} ,
\ee
which is exactly of the same form as in terms of the original variables $(s_i, \sigma_{ij})$,
but with the dual couplings $(J_d, K_d)$, thus establishing the self-duality of the theory.

As a special case, consider the limit when $J = 0$. Then the action in Eqn. \ref{igtm} is that of the 
pure $Z_2$ gauge theory. Under the duality transformation, we now get the form Eqn. \ref{igtmD}
but with the dual coupling $J_d = \infty$. This means that the fluctuations of the dual gauge field $\mu$
are frozen - we may choose a gauge in which $\mu_{ij} = 1$ on every link. The dual action then simply reduces 
to that of a global Ising model for the $v_i$ with the dual coupling $K_d$.

\section{Duality of the model with combined $U(1)$ and $Z_2$ invariances}
\label{u1z2d}
In this Appendix, we will perform a duality transformation on the chargon-spinon
action $S = S_{c} + S_{s} + S_B$ derived in Section \ref{Mod} to work instead with
vortex variables instead of the chargons. For simplicity, we will
restrict ourselves to situations with an integer number of electrons 
per unit cell. In this case, the Berry phase term $S_B$ is independent of
the chargon phase field $\phi_i$. In Section \ref{Dop}, we will 
provide the generalization necessary to handle non-integer number of electrons per unit cell. 
All of our transformations will focus
entirely on the term in the action involving the chargon variables. This is simply
a chargon hopping term:
\bea
S_{c} & = & -\sum_{<ij>} \sigma_{ij}(t_c b^{*}_ib_j + c.c.)  ,\\
& = &  -\sum_{<ij>}2t_c cos\left(\phi_i - \phi_j + \frac{\pi}{2}(1- \sigma_{ij})\right) .
\eea
Note that in the absence of $\sigma_{ij}$, this is just the action for the 
three dimensional $XY$ model. The duality transformation for the $3D XY$ model is
standard - here we will generalize it to include the $Z_2$ gauge field $\sigma_{ij}$. 

Consider the partition function obtained by integrating over the chargon fields in the above action:
\be
Z_{hol}[\sigma] = \int_0^{2\pi} \prod_i d\phi_i e^{-S_{c}} .
\ee
As with the duality transformation of the $XY$ model, it will be convenient
to work with the Villain form of the action
\be
\label{svill}
S[\phi, J, \sigma] = \sum_{<ij>}\kappa J_{ij}^2 /2 + i J_{ij} (\phi_i - \phi_j + \frac{\pi}{2}(1-\sigma_{ij})) ,
\ee
where $J_{ij}$ are integer valued fields that live on the links of the lattice, and are to
be summed over in the partition function. As usual, this is 
strictly justified only in the limit  $t_c << 1$ when $t_c = \exp(-\kappa/2)$, though 
we do not expect any modifications to the physics by relaxing this assumption.
The $J_{ij}$ have the interpretation of being the total conserved electrical current on  
any link. This can be made more explicit by performing the integrals over 
$\phi_i$ which imposes the current conservation condition
\be
 \Delta \cdot J = 0  .
\ee
The symbol on the left hand side is the lattice divergence of the link variable $J_{ij}$.
We proceed, as usual, by solving the current conservation condition by writing
\be
2\pi J_{ij} = \vec \Delta \times \vec a   .
\ee
The quantity $\vec a$  lives on the links of the dual lattice, and is constrained to be
$2\pi$ times an integer. 
The right hand side is the lattice curl of this variable $\vec a$ on the plaquette of the dual lattice
pierced by the link $<ij>$. The chargon action now takes the form
\be
S[a,\sigma] = \sum_{\Box} \frac{\kappa}{8\pi^2}(\Delta \times a)^2 + 
\frac{i}{4}\sum_{<ij>}(\Delta \times a)(1 - \sigma_{ij})   .
\ee
Here the first term is a sum over plaquettes of the dual lattice, and the lattice curl in the second term
is on the plaquette pierced by the link $<ij>$. Now note that as $ \sigma_{ij} = \pm 1$, 
the exponential of the second term can be written
\begin{displaymath}
\prod_{<ij>}(-1)^{\left(\frac{\Delta \times a}{2\pi}\right)\left(\frac{1- \sigma_{ij}}{2}\right)}   .
\end{displaymath} 
It is useful now to separate the integer $\frac{a}{2\pi}$ into its even and odd part by writing
\be
a = 2\pi(2A +s)  ,
\ee
where $A$ is an integer and $s = 0,1$. Then, we have
\be
\prod_{<ij>}(-1)^{\left(\frac{\Delta \times a}{2\pi}\right)\left(\frac{1- \sigma_{ij}}{2}\right)}
= \prod_{<ij>} \left(\prod_{\Box} (-1)^s \right)^{\left(\frac{1- \sigma_{ij}}{2}\right)}   ,
\ee
where the product inside the brackets denotes the product over the links of the plaquette
of the dual lattice pierced by $<ij>$.
We now define
\be
  \mu_{ij} \equiv (-1)^s = 1-2s  .
\ee
Note that $\mu_{ij}$ lives on the links of the dual lattice and takes values $\pm 1$. 
The product above can then be written
\be
\EXP{i\frac{\pi}{4} \left(1 - \prod_{\Box}\mu \right)(1-\sigma_{ij})} .
\ee
Note that $\mu$ satisfies
\be
\prod_{\Box} \mu = (-1)^{J_{ij}} ,
\ee
where the plaquette product on the left hand side is on the plaquette of the dual lattice
penetrated by the link $<ij>$. Thus, the conserved $Z_2$ charge current determines the flux 
of $\mu$.  

The action now becomes
\bea
S & = & \sum_{\Box}\frac{\kappa}{8\pi^2}\left(\Delta \times \left(2A + \frac{1-\mu}{2}\right)\right)^2 + S_{CS}  ,\\
S_{CS} & = & i\sum_{<ij>}\frac{\pi}{4} \left(1 - \prod_{\Box}\mu \right)(1-\sigma_{ij}) .
\eea 
At this stage, $A$ is constrained to be  integer-valued. We impose this integer 
constraint on $A$ softly by adding a term 
\be
- t_{v}\sum_{<ij>}cos(2\pi A_{ij})  .
\ee
Here the sum is over the links of the dual lattice.
The action can now be rewritten in terms of $a = 2\pi\left(2A + \frac{1-\mu}{2}\right)$:
\bea
S & = & S_{v} + S_{a} + S_{CS} ,\\
S_{v} & = & -t_{v}\sum_{<ij>}\mu_{ij}cos\left(\frac{a_{ij}}{2}\right) ,\\
S_{a} & = & \sum_{\Box}\frac{\kappa}{8\pi^2} (\Delta \times a)^2   . 
\eea
It is convenient to extract a ``matter field'' from the $a_{ij}$ by letting 
\be
a_{ij} \ra a_{ij} + 2(\theta_i - \theta_j)  .
\ee
This changes $S_{v}$ to
\be
S_{v} = -t_{v}\sum_{<ij>}\mu_{ij} cos \left(\theta_i -\theta_j + \frac{a_{ij}}{2}\right)  ,
\ee
but leaves all the other terms unchanged. 
The field $e^{i\theta_i}$ may be interpreted as an $\frac{hc}{2e}$ vortex
creation operator. 
Several symmetries of the action above are apparent. It is invariant under a local
$U(1)$ gauge transformation
\bea
\theta_i & \ra & \theta_i + \Lambda_i, \\
a_{ij} & \ra & a_{ij} - \frac{\Lambda_i - \Lambda_j}{2} .
\eea
This is standard in the dual vortex description of $XY$ models in three dimensions.
However the action has an additional $Z_2$ gauge symmetry under which
\bea
e^{i\theta_i} & \ra & \epsilon_ie^{i\theta_i} ,\\
\mu_{ij} & \ra & \epsilon_i \epsilon_j \mu_{ij} ,
\eea
with $\epsilon_i = \pm 1$. This $Z_2$ gauge symmetry is actually dual to the
one in the chargon-spinon action.   
Note that the
action describes the vortices $e^{i\theta_i}$ {\em minimally coupled} 
to the fluctuating $U(1)$
gauge field $a$ , and also to the fluctuating $Z_2$ gauge field $\mu$. The field
$\mu$ is in turn coupled to the field $\sigma$ by the term $S_{CS}$.  

This completes the duality transformation to the vortex description. Adding together the spinon
action and the Berry phase term $S_B$ gives the full dual action of Section \ref{DVT}.

\end{multicols}

\begin{references}
\bibitem{note1} Integrating out the chargons will also generate other interactions between 
gauge fields on links that belong to 
different plaquettes. However, in the insulating phase, these will be short-ranged.
\end{references}
\end{document}